\documentstyle[12pt,epsfig]{ioplppt}          
\input feynman
\def\ee{e$^+$e$^-$}
\def\qbar{$\bar {\rm q}$}
\def\qqbar{q$\bar {\rm q}$}
\def\eetoqqbar{\ee$\rightarrow$~\qqbar}
\def\Ztoqqbar{${\rm Z}^0\rightarrow$~\qqbar}
\def\eetoZtoqqbar{\ee$\rightarrow {\rm Z}^0/\gamma^*\rightarrow$~\qqbar}
\def\eetoZtoh{\ee$\rightarrow {\rm Z}^0/\gamma^*\rightarrow$~hadrons}
\def\eetoWWtohadrons{\ee$\rightarrow {\rm W}^+{\rm W}^- \rightarrow$~hadrons}
\def\r00{$\rho_{00}$}
\def\Lambdab{$\Lambda_{\rm b}$}
\def\Lambdac{$\Lambda_{\rm c}$}
\def\ltap{\raisebox{-.4ex}{\rlap{$\sim$}} \raisebox{.4ex}{$<$}}
\def\gtap{\raisebox{-.4ex}{\rlap{$\sim$}} \raisebox{.4ex}{$>$}}
\def\ppair{p$^{\scriptsize (}\bar{\rm p}^{\tiny )}$}
\def\Lpair{$\Lambda^{\scriptsize (}\bar{\Lambda}^{\tiny )}$}
\begin{document}

{\large EUROPEAN LABORATORY FOR PARTICLE PHYSICS }

\begin{flushright}
CERN-PPE/97-040 \\
Edinburgh-PPE/97-01 \\
MAN-HEP/97-02 \\
16th April 1997
 \end{flushright}

\title{Hadronization in Z$^0$ decay}

\author{I.~G.~Knowles\dag\ and G.~D.~Lafferty\ddag\ftnote{4}{On leave
from Department of Physics and Astronomy, The University of 
Manchester, Manchester, M13 9PL, UK.} } 

\address{\dag\  Department of Physics and Astronomy, The University of
Edinburgh, Edinburgh, EH9 3JZ, UK.}
\address{\ddag\ PPE Division, CERN, 1211 Gen\`eve 23, Switzerland.}

\begin{abstract}
The confinement transition from the quark and gluon degrees of freedom
appropriate in perturbation theory to the hadrons observed by real world
experiments is poorly understood. In this strongly interacting transition
regime we presently rely on models, which to varying degrees reflect
possible scenarios for the QCD dynamics. Because of the absence
of beam and target remnants, and the clean experimental conditions 
and high event rates, \ee\ annihilation to hadrons at 
the Z$^0$ provides a unique laboratory, both experimentally
and theoretically, for the study of parton hadronization.
This review discusses current theoretical understanding
of the hadronization of partons, with particular emphasis on models 
of the non-perturbative phase, as implemented in Monte Carlo 
simulation programs. Experimental results 
at LEP and SLC are summarised and considered in the light of the 
models. Suggestions are given for further measurements which
could help to produce more progress in understanding hadronization.
\end{abstract}

\vskip 2cm
\noindent
Topical Review to be published in \hfil\break
Journal of Physics G: Nuclear and
Particle Physics 

\maketitle

\section{Introduction}
\label{sec-intro}

Hadronic systems produced in \ee\ annihilation
have their origin in a uniquely simple quark-antiquark state.   
While the standard model of particle physics provides a well tested
description of the reaction
\eetoZtoqqbar, the subsequent production of observable hadrons
is less well understood.  
A parton shower, described by perturbative quantum
chromodynamics (pQCD), is
normally invoked to describe the initial fragmentation phase.
In the subsequent non-perturbative
hadronization process, the partons become the hadrons
which are experimentally observed.    
Multihadronic \ee\ annihilation events, with no beam or
target fragments to confuse their experimental or theoretical
interpretation, provide the most powerful
system available for the study of the transition from the partons of
perturbative QCD to the hadrons of the
laboratory.   

Between 1989 and 1995, the Large Electron Positron collider
(LEP) at CERN delivered an integrated 
luminosity of about
170~pb$^{-1}$ per detector at and around the Z$^0$ peak, 
providing each of its four dedicated
experiments, ALEPH, DELPHI, L3 and OPAL,  
with some 6 million e$^+$e$^-$ annihilation events,
70\% of which were multihadronic.
These events have enabled the experiments to conduct
detailed studies of many 
aspects of parton hadronization.
Over the same period, the SLAC Linear Collider 
(SLC) 
delivered a more modest number of events, initially to the
MARK II detector and then to the SLD detector, which
has now accumulated about 200k Z$^0$ events.
Although its luminosity is lower than that
of LEP, the SLC is able to provide highly polarized electron
beams. While LEP has now entered phase 2, running at
higher energies, analysis of the Z$^0$ peak data 
will continue for some time. The SLC, with polarized beams,
is scheduled to run at the Z$^0$ for several more years.

In this review, we look at the present understanding of the 
non-perturbative hadronization process in the context of
the recent experimental results from LEP and SLC.  
A brief introduction to the electroweak theory of 
\ee\ annihilation is followed by a 
heuristic picture of the transition from perturbative partons
to final-state hadrons. After this, the role
of perturbative QCD in determining the structures 
and particle content of events is discussed.  
Models of the non-perturbative hadronization phase are then
covered --- independent fragmentation, string fragmentation 
and cluster fragmentation are discussed in detail and a 
comparison is given of the main features of the different 
models. The LEP and SLC machines and their 
associated detectors are then described, with
emphasis on the features relevant to measurements of 
individual hadrons from Z$^0$ decay.
Next the experimental data on single inclusive identified
particles are summarised, and the results interpreted in the
light of the different models. Results on spin phenomena, such as 
baryon polarization and vector meson spin alignment
are then covered, and this is followed by discussion of results 
on correlations, including Bose Einstein effects, strange particle rapidity 
and angular correlations, and intermittency; where possible 
the results are interpreted within the context
of theory and models. Differences between quark and
gluon initiated jets are then dealt with, and the review 
ends with a look forward to what may still be learned from the
available LEP and SLC data, and the forthcoming SLC data.

\subsection{Electroweak aspects of e$^+$e$^-$ annihilation}
\label{selwk}

The (initial) numbers, directions, polarizations and flavours of the quarks 
produced in \ee\ annihilation are determined by the electroweak couplings
of the exchanged vector bosons. The full, tree-level expression for the
differential cross section for \eetoqqbar\ in the centre of momentum (CoM)
frame is given as a function of quark's polar angle, $\theta^\star$,
measured relative to the electron beam direction, and total CoM energy
squared, $s$, by:
\begin{eqnarray}
\label{eeff}
\fl\frac{\d\sigma}{\d\Omega}(s,\theta^\star)=\frac{3N_c\alpha^2_{\rm em}}
{16s}\beta_{\rm q}\sum_{i,j}\chi_i(s)\chi^\star_j(s)(v^i_{\rm e}v^j_{\rm e}
+a^i_{\rm e}a^j_{\rm e})(v^i_{\rm q}v^j_{\rm q}
+\beta^2a^i_{\rm q}a^j_{\rm q})\times
\nonumber \\
\fl\Biggl\{\!\Bigl({\cal P}^{(1)}_{\rm e}-{\cal P}^{(2)}_{\rm e}A^{ij}_{\rm e}
\Bigr)\!\biggl[\!\Bigl(2-{\cal P}^{(1)}_{\rm q}\Bigr)B^{ij}_{\rm q}
(1-\beta^2_{\rm q})\sin^2\theta^\star\!
+\!\Bigl({\cal P}^{(1)}_{\rm q}+{\cal P}^{(2)}_{\rm q}A^{ij}_{\rm q}\Bigr)
(1+\cos^2\theta^\star)\biggr] \nonumber \\
\fl\hspace{3mm}+\Bigl({\cal P}^{(1)}_{\rm e}A^{ij}_{\rm e}
+{\cal P}^{(2)}_{\rm e}\Bigr)
\Bigl({\cal P}^{(1)}_{\rm q}A^{ij}_{\rm q}
+{\cal P}^{(2)}_{\rm q}\Bigr)2\cos\theta^\star\Biggr\}
\end{eqnarray}
where,
\begin{eqnarray}
\fl {\cal P}^{(1)}_{\rm f}=1-\rho_{\rm f}\rho_{\bar f}\hspace{10mm}
A^{ij}_{\rm f}=\frac{\beta_{\rm f}(v^i_{\rm f}a^j_{\rm f}+a^i_{\rm f}
v^j_{\rm f})}{v^i_{\rm f}v^j_{\rm f}+\beta^2_{\rm f}a^i_{\rm f}a^j_{\rm f}}
\hspace{10mm}\chi_i(s)=\frac{s}{(s-M^2_i)+\i M_i\Gamma_i} \nonumber \\
\fl {\cal P}^{(2)}_{\rm f}=\rho_{\rm f}-\rho_{\bar f}\hspace{12mm}
B^{ij}_{\rm f}=\frac{v^i_{\rm f}v^j_{\rm f}}{v^i_{\rm f}v^j_{\rm f}
+\beta^2_{\rm f}a^i_{\rm f}a^j_{\rm f}}\hspace{18mm}
\beta_{\rm q}=\sqrt{1-\frac{4m^2_{\rm q}}{s}}
\end{eqnarray}
The summation is over all exchanged bosons, $i=\gamma^\star,$Z$^0,\ldots$,
with mass and width $M_i$ and $\Gamma_i$, and vector and axial couplings 
$v^i_{\rm f}$ and $a^i_{\rm f}$. For example $v^\gamma_{\rm f}=Q_{\rm f},\;
a^\gamma_{\rm f}=0$, $v^{\rm Z}_{\rm f}=(T^{3L}_{\rm f}-2Q_{\rm f}s^2_W)
/2s_Wc_W$ and $a^{\rm Z}_{\rm f}=T^{3L}_{\rm f}/2s_Wc_W$ etc, with
$Q_{\rm f}$ the fermion's electric charge, normalized to that of the
electron, $T^{3L}_{\rm f}$ the fermion's third component of 
weak, SU(2)$_{\rm L}$,
isospin and $s_W(c_W)$ the sine(cosine) of the Weinberg angle. The number of
colours is $N_c=3$ and $\alpha_{\rm em}$ is the electromagnetic fine structure
constant. The dependence on the initial lepton and final quark longitudinal
polarizations are via $\rho_{\rm f}$, where $\rho_{\rm f}=+1$ for spin along
the CoM direction of travel (helicity basis). In the absence of transverse
beam polarization there is no azimuthal angular dependence.

\begin{figure}[htb]
\epsfig{file=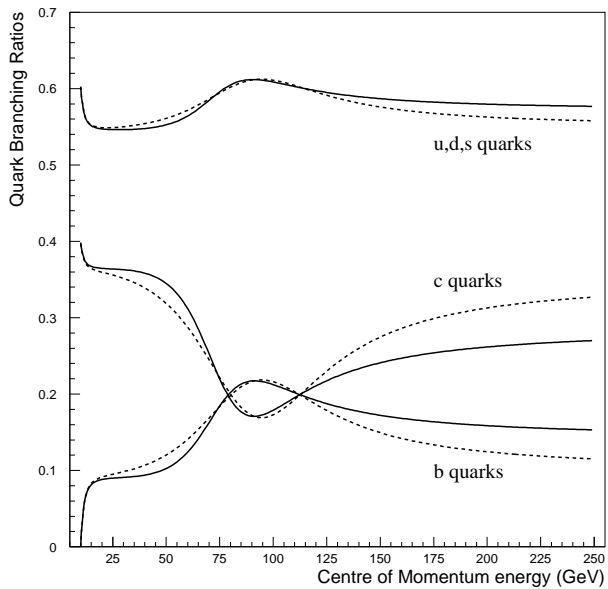,height=80mm}
\caption[]{The relative branching fractions of light (u,d,s), charm and
bottom quarks in \ee\ annihilation as a function of $\sqrt s$ for an
unpolarized (solid lines) and 100\% right polarized (dashed lines)
electron beam. \label{fbrs}}
\end{figure}

At $\sqrt s=M_{\rm Z}$ the Z$^0$ exchange term dominates and \eref{eeff}
simplifies considerably. In this limit the partial width for \Ztoqqbar\ is
given by:
\begin{equation}
\Gamma_{{\rm q}\bar {\rm q}}=
\frac{N_c\alpha_{\mathrm em}M_{\rm Z}}{6}\beta_{\rm q}
\Bigl[(3-\beta^2_{\rm q})v^2_{\rm q}+2\beta^2_{\rm q}a^2_{\rm q}\Bigr]
\end{equation}
This implies relative hadronic branching ratios of $\approx 17\%$ to each
up-type quark and $\approx 22\%$ to each down-type quark at the Z$^0$; the
relative fraction of b quarks peaks on resonance. \Fref{fbrs} shows the
relative fractions of light (u,d,s), charm and bottom quarks as a function
of $\sqrt s$ obtained using the full expression \eref{eeff}. The total
hadronic branching fraction is $\approx 70\%$.
The beam polarization has little influence on the relative rates of
produced quarks, although the total rate is proportional to $(v^2_{\rm e}+
a^2_{\rm e}){\cal P}^{(1)}_{\rm e}-2v_{\rm e}a_{\rm e}{\cal P}^{(2)}_{\rm e}$.
However only a single polarized beam can significantly alter the quark polar
angle distribution, as can be seen from the forward-backward
asymmetry (for pure Z$^0$ exchange and $\beta_{\rm q}=1$):
\begin{equation}
\frac{\frac{\d \sigma}{\d\Omega}(\cos\theta^\star)-\frac{\d \sigma}
{\d\Omega}(-\cos\theta^\star)}{\frac{\d \sigma}{\d\Omega}(\cos\theta^\star)
+\frac{\d \sigma}{\d\Omega}(-\cos\theta^\star)}
=\frac{{\cal P}^{(1)}_{\rm e}A_{\rm e}-{\cal P}^{(2)}_{\rm e}}
{{\cal P}^{(1)}_{\rm e}-{\cal P}^{(2)}_{\rm e}
A_{\rm e}}\times A_{\rm q}\frac{2\cos\theta^\star}{1+\cos^2\theta^\star}
\end{equation}
Since $A_e$ is only $\approx0.16\;(A_d\approx0.94$ and $A_u\approx0.69$)
then, given polarized beams, one has a powerful statistical way to
identify separately quark and antiquark jets based on their direction. The
quarks are naturally produced highly polarized:
\begin{equation}
\fl\frac{\frac{\d \sigma}{\d\Omega}(\rho_{\rm f}=+1)-\frac{\d \sigma}{\d
\Omega}(\rho_{\rm f}=-1)}{\frac{\d \sigma}{\d\Omega}(\rho_{\rm f}=+1)
+\frac{\d \sigma}{\d\Omega}
(\rho_{\rm f}=-1)}
=A_{\rm q}+\frac{A_{\rm e}(1-A^2_{\rm q})2\cos\theta^\star}
{1+\cos^2\theta^\star+A_{\rm e}A_{\rm q}2\cos\theta^\star}\approx A_{\rm q}
\end{equation}
This result is for unpolarized lepton beams, pure Z$^0$ exchange and
$\beta_{\rm q}=1$; the full expression depends little on the lepton beam
polarization.

For QCD studies at the Z$^0$~\cite{r1qcd} the effect of initial-state 
electromagnetic radiation (ISR) is of rather minor significance, unlike in
the case of electroweak studies~\cite{rlep1,rlepew}. The primary effect of
ISR is to lower the effective $\sqrt s$, leading to a distortion of the
Breit-Wigner lineshape given by \eref{eeff}. However the basic quantities of
relevance from a QCD perspective, quark flavour mix, polarization etc, are
only rather weak functions of $\sqrt s$. Further, because of the Z$^0$
resonance the cross section falls rapidly when ISR occurs, mitigating against
its effects (even so the peak cross section falls by $\approx25\%$ compared
to \eref{eeff}), in contrast to the situation at LEP 2 and higher energies
where ISR is very important~\cite{r2isr}. The effects of {\it final-state}
radiation, particularly of gluons but also of photons, form one theme of
this review.  The effect of final-state radiation on the total 
hadronic cross section can be
summarized in the multiplicative factor~\cite{rlepew,r3loop}:
\begin{equation}
1+\frac{3}{4}\left(Q^2_{\rm q}\frac{\alpha_{\mathrm em}}
{\pi}+C_F\frac{\alpha_s}
{\pi}\right)+\cdots
\end{equation}
where $\alpha_s$ is the strong coupling constant $(g^2_s/4\pi)$ and $C_F=4/3$
is a measure of the quark-gluon coupling strength. Flavour (mass)
dependent effects and electroweak radiative corrections are small --- both
occur at the 1\% level, and need not concern us here~\cite{rlepew}.

\subsection{A picture of the parton to hadron transition}
\label{sxtpic}

The description of a multihadronic event commences by specifying a set of
primary partons (q/\qbar/g) distributed according to an exact Z$^0$ decay
matrix element, such as \eref{eeff}. It is customary to identify three
basic stages in the transition of these partons into detected 
hadrons~\cite{rlep1mc}. First a parton (or, equivalently, dipole) cascade, formulated
according to pQCD, evolves the primary partons from the hard scattering 
scale $Q\approx M_{\rm Z}$ into secondary partons at a (fixed) cut-off
scale $Q_0\approx 1$ GeV. It is during the calculable stages, hard
subprocess and shower, that the event's global features are determined:
energy dependences, event shapes, multiplicity etc. In a second
stage, carried out at the fixed, low virtuality scale $Q_0$, a model is
employed to convert the secondary partons into hadrons. The second stage is
essentially energy $(Q)$ independent, up to power corrections, and assumed
to be local in nature. Finally unstable primary particles are decayed into
stable hadrons and leptons according to decay tables~\cite{rpdg}.
Schematically the fragmentation function is given by:
\begin{eqnarray}
\label{econv}
\fl D^a_{\rm h}(x,Q^2)=(\mbox{pQCD~evolution: }Q^2\to Q^2_0)
                &\otimes(\mbox{model:}a\to {\rm H})\Big|_{Q^2_0} \\ \nonumber
                &\otimes(\mbox{tables:}{\rm H}\to {\rm h},{\rm h}',..)
\end{eqnarray}
Here, $D^a_{\rm h}(x,Q^2)$ is the probability to find a hadron of type h
carrying a fraction $x$ of the parton's momentum,
in a jet initiated by the parton $a$, whose maximum virtuality is $Q$.
The H represent possible intermediate hadrons. 

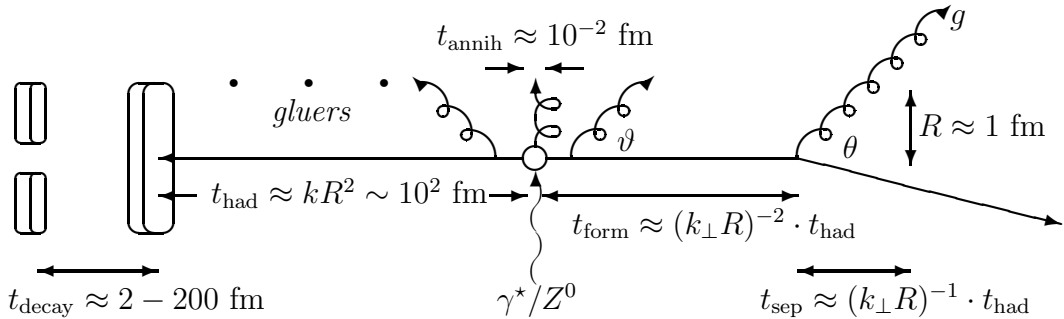
\begin{figure}
\unitlength 1mm
\begin{picture}(140,50)
\thicklines
\put(70,20){\circle{3}}
\put(68.5,20){\vector(-1,0){48.5}}
\put(71.5,20){\line(1,0){33.5}}
\put(105,20){\vector(4,-1){35}}
\put(111,20){$\theta$}
\put(20,20){\oval(4,20)}
\put(18,20){\oval(4,20)[l]}
\put(18,30){\line(1,0){2}}
\put(18,10){\line(1,0){2}}
\put(4,26){\oval(2,8)}
\put(2,26){\oval(2,8)[l]}
\put(2,30){\line(1,0){2}}
\put(2,22){\line(1,0){2}}
\put(4,14){\oval(2,8)}
\put(2,14){\oval(2,8)[l]}
\put(2,18){\line(1,0){2}}
\put(2,10){\line(1,0){2}}
\put(4,5){\vector(-1,0){0}}
\put(4,5){\vector(1,0){16}}
\put(0,0){$t_{\rm decay}\approx 2-200$~fm}
\multiput(30,30)(10,0){3}{\circle*{1}}
\put(35,25){\it gluers}
\put(27,14){$t_{\rm had}\approx kR^2\sim10^2$~fm}
\put(25,15){\vector(-1,0){5}}
\put(64,15){\vector(1,0){5}}
\put(75,10){$t_{\rm form}\approx (k_\perp R)^{-2}\cdot t_{\rm had}$}
\put(71,15){\vector(-1,0){0}}
\put(71,15){\vector(1,0){34}}
\put(121,23){$R\approx 1$~fm}
\put(120,19){\vector(0,-1){0}}
\put(120,19){\vector(0,1){10}}
\put(100,0){$t_{\rm sep}\approx (k_\perp R)^{-1}\cdot t_{\rm had}$}
\put(105,5){\vector(-1,0){0}}
\put(105,5){\vector(1,0){15}}
\put(57,35){$t_{\rm annih}\approx10^{-2}$~fm}
\put(63.5,32){\vector(1,0){5}}
\put(76.5,32){\vector(-1,0){5}}
\put(81,21){$\vartheta$}
\put(65,0){$\gamma^\star/Z^0$}
\put(68.75,20){
\Feynmanlength
\begin{picture}(1000,1000)
\drawline\gluon[\N\REG](0,400)[2]
\drawarrow[\N\ATBASE](\pbackx,\pbacky)
\drawline\photon[\S\REG](0,-800)[4]
\drawarrow[\N\ATBASE](\pfrontx,\pfronty)
\end{picture}}
\put(63.5,20){
\Feynmanlength
\begin{picture}(1000,1000)
\drawline\gluon[\NW\FLIPPED](0,0)[2]
\drawarrow[\NW\ATBASE](\pbackx,\pbacky)
\end{picture}}
\put(73.5,20){
\Feynmanlength
\begin{picture}(1000,1000)
\drawline\gluon[\NE\REG](0,0)[2]
\drawarrow[\NE\ATBASE](\pbackx,\pbacky)
\end{picture}}
\put(103.5,20){
\Feynmanlength
\begin{picture}(1000,1000)
\drawline\gluon[\NE\REG](0,0)[5]
\drawarrow[\NE\ATBASE](\pbackx,\pbacky)
\global\advance\pbackx by 200
\global\advance\pbacky by -500
\put(\pbackx,\pbacky){$g$}
\end{picture}}
\end{picture}
\caption{A schematic diagram of the spatial evolution of a hadronic Z$^0$
decay indicating the relevant time scales and distances associated with
gluon bremsstrahlung and hadron formation.\label{fxt}}
\end{figure}

We consider this process in terms of its space-time
structure: the discussion is based on~\cite{rbible} and illustrated in
figure \ref{fxt}. The hard subprocess, \eetoqqbar, may be viewed as the
production of a highly virtual photon, or a real Z$^0$, which impulsively
kicks a \qqbar\ pair out of the vacuum; the time scale is short,
$t_{\rm annih}\approx 1/\sqrt{Q^2}\sim 10^{-2}$~fm. In this non-adiabatic
process the quarks shake off most of their cloud of virtual particles, so
that any  structure they have is on a scale below $1/\sqrt{Q^2}$. That is
they behave as bare (more properly half-dressed) colour charges until the
gluon field has had time to regenerate out to a typical hadron size $R
\approx 1$~fm. Allowing for the boost, this takes a time $t_{\rm had}
\approx Q/m\times R\approx QR^2\sim 10^2$~fm, where the second approximation
is appropriate to light hadrons.

The fact that $t_{\rm had}\gg t_{\rm annih}$ raises the issue of how charges
are conserved over the space-like separated distances involved. Of course
the accelerated quarks will radiate gluons and here two new time scales are
relevant. First, from the off-shellness of the quark prior to emission, the
formation time of a real gluon of 3-momentum (=energy) $k$ is $t_{\rm form}
\approx k/k^2_\perp$ (the uncertainty relation gives the proper lifetime of
the virtual state as the reciprocal of its off-shellness, during which time
it will travel a 4-distance $q^\mu/q^2$ in the laboratory frame, where
$q^\mu$ is the virtual state's 4-momentum). Second, for
the gluon to reach a transverse separation of $R$ and become independent of
the quark takes a time $t_{\rm sep}=(k_\perp R)\cdot t_{\rm form}$, whilst 
the hadronization time may be written $t_{\rm had}(\approx kR^2)=(k_\perp
R)^2\cdot t_{\rm form}$.

For this quark-gluon picture to make sense we require $k_\perp>R^{-1}$ so
that $t_{\rm form}<t_{\rm sep}<t_{\rm had}$; this is natural as it implies
that $\alpha_s(k^2_\perp)<1$, making perturbation theory applicable. If
$k_\perp<R^{-1}$ then we can say nothing. On the borderline are quanta with
$k_\perp=R^{-1}$; these feel the strong interaction and are responsible for
holding hadrons together. We distinguish these from the essentially free
perturbative gluons by the name {\it gluers}. The first {\it gluers} form
after only 1~fm, having $k\approx k_\perp\approx R^{-1}$. Thus {\it gluers}
immediately form in the wake of the primary partons blanching their colour
field and leaving two fast separating charge neutral systems. Thus the
slowest hadrons form first, close to the interaction point, in what may be
called an `inside-out' pattern~\cite{rinout}.

On quite general grounds the distribution of {\it gluers} in QCD can be
estimated using:
\begin{equation}
\label{engluer}
d{\cal N}_{\mathrm gluers}\sim C_F\left[\int_{k_\perp\sim R}
\frac{\alpha_s(k^2_\perp)}{\pi}{\cal P}(k)\frac{\d k^2_\perp}{k^2_\perp}
\right]\times\frac{\d k}{k}\propto\frac{\d k}{k}
\end{equation}
Here the logarithmic $k^2_\perp$ dependence reflects the fact that the
coupling constant is dimensionless, the logarithmic $k$ dependence follows
from the gluon being massless and only the kernel function ${\cal P}$, which
is of order unity, depends on details of the quark-gluon vertex. The 
distribution
is thus governed by longitudinal phase space and leads to hadron production
on rapidity plateaux along the directions of the initial partons.

Not all quanta are emitted at low $k_\perp$ and in a significant range,
$R^{-1}\ll k_\perp<k\ll \sqrt{Q^2}$, a shower of perturbative gluons is
possible. A hard gluon emitted at an angle $\theta$ becomes a separate
colour source at $t_{\rm sep}=R/\theta(=(g_\perp R)\cdot t_{\rm form})$,
again raising the issue of charge conservation. Fortunately a {\it gluer}
emitted at the same angle $\theta$ would appear just in the right place
and at the right time to blanch the tail of the separating  gluon's colour
field. This gluon then acts as a new source of {\it gluers}, restricted
such that $k\gtap \theta^{-1}R^{-1}$. That is equivalent to a boosted
($\gamma=1/\theta$) quark jet of a reduced scale $Q=k_\perp$ and with the
substitution $C_F\mapsto C_A$ for the parton's charge 
in \eref{engluer}~\cite{rplumb}. It should be noted that {\it gluers} 
emitted at larger
angles $\vartheta>\theta$ have $R/\vartheta<R/\theta=t_{\rm sep}$ and
should therefore be associated with the parent q+g system. For example at
time $R$ the first {\it gluers} emitted actually see ensembles of partons
equivalent to the original q$\bar {\rm q}$ pair. The coherence of soft
emissions simply reflects colour charge conservation and is fundamental to
a gauge theory such as QCD.

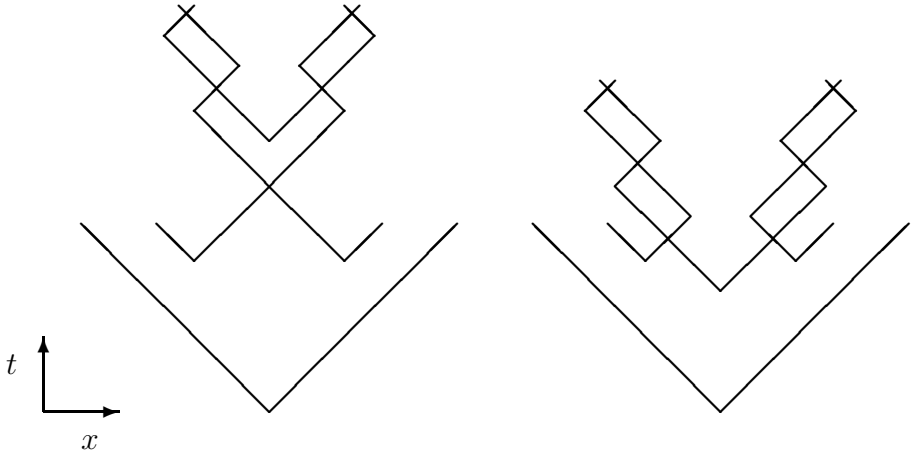
\begin{figure}
\unitlength 1mm
\begin{picture}(125,60)
\thicklines
\put(5,5){\vector(1,0){10}}
\put(0,10){$t$}
\put(5,5){\vector(0,1){10}}
\put(10,0){$x$}
\put(35,5){\line(-1,1){25}}
\put(35,5){\line(1,1){25}}
\put(25,25){\line(-1,1){5}}
\put(25,25){\line(1,1){20}}
\put(45,45){\line(-1,1){6}}
\put(39,51){\line(1,1){8}}
\put(35,41){\line(1,1){14}}
\put(49,55){\line(-1,1){4}}
\put(45,25){\line(1,1){5}}
\put(45,25){\line(-1,1){20}}
\put(25,45){\line(1,1){6}}
\put(31,51){\line(-1,1){8}}
\put(35,41){\line(-1,1){14}}
\put(21,55){\line(1,1){4}}
\put(95,5){\line(-1,1){25}}
\put(95,5){\line(1,1){25}}
\put(85,25){\line(-1,1){5}}
\put(85,25){\line(1,1){6}}
\put(91,31){\line(-1,1){14}}
\put(77,45){\line(1,1){4}}
\put(95,21){\line(-1,1){14}}
\put(81,35){\line(1,1){6}}
\put(87,41){\line(-1,1){8}}
\put(105,25){\line(1,1){5}}
\put(105,25){\line(-1,1){6}}
\put(99,31){\line(1,1){14}}
\put(113,45){\line(-1,1){4}}
\put(95,21){\line(1,1){14}}
\put(109,35){\line(-1,1){6}}
\put(103,41){\line(1,1){8}}
\end{picture}
\caption{Two possible views from the string model perspective of $\pi^+\pi^-$
production via an intermediate $\rho^0$, illustrating the practical
difficulty in separating strong resonance decays from the direct string
fragmentation. \label{frho}}
\end{figure}

Finally, the time scales for the decay of the primary particles are set
by the reciprocal of their widths: for the strong resonances typical values
of $\Gamma\sim 1-100$~MeV give $t_{\rm decay}\sim 200-2$~fm. Clearly these
scales are commensurate with those for primary hadron production so that any
distinction between hadronization and resonance decays may be only semantics.
This is illustrated in \fref{frho} which shows a string-inspired space-time
picture (see \sref{string}) of two equivalent ways to form a $\pi^+\pi^-$
pair differing only in the time sequence of the string breaks. In the first
figure the pions are directly produced whilst in the second they appear to
come from a $\rho^0$; however observationally they are indistinguishable.
The time scale for the weak decays of s, c and b hadrons is of order $t_{\rm
decay}\sim (M_{\rm W}/m_{\rm Q})^2k/m_{\rm Q}$ (where $k/m_{\rm Q}$ is
the hadron's $\gamma$ factor); this is $\approx 10^{12}$~fm
for c and b and much longer for s, and so these decays may safely be treated
as separate subprocesses.

\subsection{Monte Carlo event generators}

A typical hadronic final state in Z$^0$ decay contains about 20 stable 
charged particles and 10 neutral hadrons, mainly $\pi^0$'s. 
This is a complex, non-perturbative system
beyond direct, first principles calculation at the present time. However
Monte Carlo event generator programs~\cite{rlep1mc,rlep2mc} have been 
developed which
provide remarkably accurate, detailed descriptions of complete hadronic
events. These are based on a combination of judicious approximations to
pQCD, discussed in \sref{spqcd}, and per/in-spirational models for
hadronization, to be discussed more fully in \sref{shad}. Here we list
the main features of the major event generators used in Z$^0$ studies
and refer the reader to later sections and to the literature for more 
complete descriptions.

\begin{description}
\item[ARIADNE]~\cite{rariadne} implements a pQCD shower based on the dipole
cascade model~\cite{rdcm}, which is equivalent to a coherent parton shower.
The evolving chain of dipoles corresponds naturally to a string and indeed
the hadronization is performed using the JETSET implementation of the Lund
string model~\cite{rlund}.

\item[COJETS]~\cite{rcojets} is based on a virtuality-ordered parton shower,
that takes no account of colour coherence. The resulting jets are hadronized
using a refined version of the Field-Feynman independent fragmenation 
model~\cite{rff}.

\item[HERWIG]~\cite{rherwig} is based on a highly developed parton shower
algorithm~\cite{rhersh,rdepl}, that automatically takes into account colour
coherence, and a relatively simple cluster hadronization scheme~\cite{rclus}.

\item[JETSET]~\cite{rjetset} uses a virtuality-ordered parton shower with an
imposed angular ordering constraint to take into account colour coherence. A
relatively sophisticated hadronization scheme based on the Lund string model
is provided~\cite{rlund}. An option also exists which does not employ a
parton shower but tries to use instead higher-order hard-subprocess matrix
elements plus string hadronization.
\end{description}

\section{The role of perturbation theory in \ee\ annihilation to hadrons}
\label{spqcd}

As stated in \sref{sxtpic} the bulk properties of hadronic events
in Z$^0$ decay are established early in the fragmentation when virtualities
are large and QCD perturbation theory is valid. It is an important issue to
establish to what extent pQCD dominates and what contributions are made by
non-perturbative effects.

\subsection{Matrix elements and showers}

Two basic approaches are used to calculate hadronic event properties. The
use of fixed-order perturbation theory is justified by the smallness of the
running strong coupling constant~\cite{run} at $M_{\rm Z},\;\alpha_s(Q^2)=
4\pi/\beta_0\ln(Q^2/\Lambda^2)\approx0.12$. Known results include the total
hadronic cross section to three loops, order-$\alpha_s^3$~\cite{r3loop}.
Complete one-loop, order-$\alpha^2_s$ calculations are available~\cite{rert}
for planar (i.e. 3-jet dominated) event shapes, including contributions
due to the orientation of the event plane with respect to the beam
direction~\cite{rertor} and to quark mass effects~\cite{rertb}. Partial
order-$\alpha^3_s$ results are known for 4-jet distributions~\cite{r4jet}.
Tree-level calculations, up to order-$\alpha_s^3$, of 5-jet distributions
are also
available~\cite{r5jet}. However a complication arises in this approach
because tree-level diagrams diverge whenever external partons become soft or
collinear and related negative divergences arise in virtual (loop) diagrams
(in addition to ultraviolet divergences). Fortunately, in sufficiently
inclusive measurements such as the total hadronic cross section it is
guaranteed that the two sets of divergences cancel~\cite{rkln}.
Unfortunately in more exclusive quantities such as event shapes the
cancellation is no longer complete and large logarithmic terms remain,
generically of the form $L=\ln(Q^2/Q^2_0)$. Since $\alpha_sL$ is 
of order unity
this can spoil the convergence of naive perturbation theory.

In the second approach the perturbation series is rearranged in terms of
powers of $\alpha_sL$: $\sum_n a_n(\alpha_s L)^n+\alpha_s(Q)\sum_n b_n
(\alpha_s L)^n+\cdots$. The first infinite set of terms represent the
leading logarithm approximation (LLA), then the $\alpha_s$-suppressed
next-to-LLA and so on. Many of these terms have been identified~\cite{rlogs}
and summed using renormalization group techniques. They are conveniently
expressed via $Q^2$-dependent fragmentation functions whose evolution is
controlled by Altarelli-Parisi type equations~\cite{rap}:
\begin{equation}
\label{eap}
Q^2\frac{\partial D^a_{\rm h}(x,Q^2)}{\partial Q^2}=\int^1_x\frac{\d z}{z}
\frac{\alpha_s}{2\pi}\sum_{\rm b} 
P^a_{{\rm bb}'}(z)D^{\rm b}_{\rm h}\left(\frac{x}{z},Q^2\right)
\end{equation}
Here, the so-called splitting functions, $P^a_{{\rm bb}'}(z)$, 
may be thought of as giving the probability of finding parton
$b$ (and $b'$) inside $a$ and carrying a fraction $z$ of its
momentum. 
Monte Carlo event generators implement solutions to these equations as
parton showers~\cite{rlep1mc,rlep2mc,rmcsh}. More recently the resummed
results for a number of event shapes have been calculated~\cite{resum}
and combined with the fixed order approach~\cite{rmatch}.
The Durham jet clustering algorithm~\cite{rdur} for example was proposed 
to allow such a resummation for jet rates.

\subsection{Global event properties}

At very low $\sqrt s$ little structure is present in hadronic events, beyond
that due to the presence of hadronic resonances, and these can be described
essentially by isotropic phase space~\cite{rlows}. This is because the
momentum scales involved in the hadronization, typically 200~MeV, are
comparable with those of the proto-jets. At higher energy, due to the
preferred collinear nature of gluon radiation, the hadrons clearly form
collimated pairs of back-to-back jets with angular distributions compatible
with those of the primary q$\bar {\mathrm q}$ pairs~\cite{rmeds}. 
Around $\sqrt s
\approx30$~GeV a significant fraction of events deviate from a simple linear
configuration: this was interpreted as first evidence for gluons~\cite{r30s}.
At the Z$^0$, hard, non-collinear, gluon radiation is manifested in multi-jet
event structures~\cite{rjets}.

Event shape variables have been developed to be sensitive to the amount of
acollinearity and hence the presence of hard gluon radiation~\cite{r1qcd}.
Since theoretical predictions are at parton level~\cite{rertmc}, hadronization
corrections, derived from Monte Carlo models, must be applied before making
comparisons to data. Away from phase space boundaries the corrections are
modest, but not negligible (of order 10\%), and order-$\alpha^2_s$ calculations
generally describe the data well~\cite{revsh}, particularly so when 
resummed~\cite{rsfit}. In the case of planar variables such as thrust,
transverse momentum in the event plane, wide jet broadening etc, 
this allows $\alpha_s$ to be measured, and good consistency is seen
with other measurements~\cite{rbw}. Only a leading-order description of
aplanar, four-parton, event shapes such as the $D$-parameter, transverse
momentum out of the event plane $p^{\mathrm out}_\perp$, the four-jet rate
etc, are presently available, although partial results are
available~\cite{r4jet} in higher order. 
Here agreement between data and Monte Carlo is less
satisfactory, with 30\% discrepancies occurring for $p^{\mathrm out}_\perp$
and the 4-jet rate~\cite{rlep2mc,rAmega,rDtun2}. Hopefully the situation will
be improved by matching the showers to exact fixed-order results. This will
prove important because of a reliance on Monte Carlo models in establishing
variables to discriminate at higher energies between continuum QCD events
and other physics signals, such as W$^+$W$^-$ pairs. This is in addition to
their importance in QCD and in increasingly demanding electroweak
measurements such as those involving jet charge~\cite{rqfb}.

Establishing agreement between the predictions of event generator programs
and data for event shape variables has proved very important for confirming
our ability to model QCD as a whole and has led to a first generation tuning
of the Monte Carlo models~\cite{rtunxi,rtun1}. A second generation of tunings,
which also take into account data on identified particle production, are now
becoming available~\cite{rlep2mc,rAmega,rDtun2}. This tuning is not a trivial
exercise as
it is often the case that a distribution depends in a complex way on a
model's free parameters. However some specific sensitivities have been
identified. The 3-jet rate is very sensitive to the QCD scale parameter
$\Lambda$, indicating the dominance of the shower and pQCD. Event shapes and
inclusive momentum spectra are sensitive to the shower cut-off $Q_0$, and
also to parameters controlling the generation of transverse
momentum in the
hadronization. In programs that lack a shower stage it is necessary to retune
the hadronization model at each $\sqrt s$ since this must describe the whole
fragmentation process including the perturbative $\sqrt s$ dependence found
in a shower.

\subsection{Power corrections}

As noted above, hadronization corrections, which at present are inherently
model dependent, need to be applied to partonic predictions before comparison
can be made to hadronic event properties. Empirically the differences
between quantities measured at the hadron level and the partonic predictions
are found to be power behaved: $\delta X\equiv X_{\mathrm had}-X_{\mathrm par}
\sim 1/Q^n$. For example, in the case of massless quarks, $\delta R_{\mathrm
had}\sim 1/Q^4$~\cite{rren}, where $R_{\mathrm had}$ is the usual
ratio of hadronic to $\mu$ pair cross sections; in fact this result 
follows from the operator
product expansion (OPE)~\cite{rope} and can be related to the value of the
gluon condensate: $\delta R_{\mathrm had}\sim \langle\alpha_s G.G\rangle/Q^4$.
In the case of average event shape variables, a $1/Q$ behaviour is found (the
OPE does not apply in this case, due to the presence of multiple scales, and
indeed no corresponding dimension-1 local operator exists). This behaviour
may be established by going to the power-enhanced, low $Q$, region where a
$1/Q^n$ variation is easily distinguished from the slow logarithmic variation
of the perturbative expressions. Such a $1/Q$ behaviour is significant because
$\alpha^2_s(Q)\approx1/Q|_{Q=M_{{\rm Z}^0}}$; thus to be able to take
advantage of a future $\Or(\alpha^3_s)$ prediction will require a better
understanding of hadronization.

Three sources of power corrections are known: instantons, infrared (IR) 
renormalons
and ultraviolet renormalons~\cite{rren}. Only IR renormalons are 
believed to be relevant to the issue of hadronization corrections;
instanton effects are too highly power-suppressed. The coefficients of pQCD
series in $\alpha_s$ generically suffer from a factorial growth, $\sim
\beta^n_0n!$, making the series formally divergent. A
standard technique for dealing with this type of behaviour is to employ a
Borel transformation~\cite{rren}. A pole in the transform 
at $n.2\pi/\beta_0$ would
correspond to an $\exp(-n.2\pi/\beta_0\alpha_s)=(\Lambda/Q)^n$ term in the
Borel summed series: these poles are called renormalons. Unfortunately the
`residue' of the pole, equivalent to the coefficient of the $(\Lambda/Q)^n$
term, is not calculable by this purely mathematical technique.

Recently a relationship has been suggested between the positions of the
renormalons and the power behaviour of perturbative series in the presence
of an IR `regulator'~\cite{rpowid}. The basic idea is that the (resummed)
perturbative calculation probes regions of phase space involving 
low-virtuality partons where non-perturbative confinement effects should also
be important. If these regions are isolated by introducing a `cut-off'
into the perturbative calculation then requiring that any cut-off
dependence is compensated by the non-perturbative hadronization correction
allows the power behaviour to be determined. Using this idea, together with
a gluon mass~\cite{rpow}, a $1/Q$ correction was derived for the thrust $T$,
$C$-parameter and longitudinal cross section $\sigma_L$
with all the coefficients
proportional to a common scale: $\delta\langle T\rangle\propto -4C_F/\pi,\;
\delta\langle C\rangle\propto6C_F$ and $\delta\langle\sigma_L\rangle\propto
C_F$. Refinements of this calculation~\cite{rfine} have attempted to relate
the common scale to a fixed $\overline\alpha_s$, representing an effective
measure of long-range confinement forces at an inclusive level. A
dispersive approach indicates that the numerator of the $1/Q$ term may
involve several coefficients, for example proportional to $\ln(Q/\Lambda)$.

The subject of power corrections is an active theoretical area and still
subject to dispute. However there exists the exciting prospect of a
phenomenology of power corrections. This is particularly so if the corrections
to various processes can be related and so shown to be universal~\cite{runiv}
or could be developed for full distributions rather than average values. For
example, according to the calculations above, $\langle T+(2/3\pi)C\rangle$
would have no leading $1/Q$ power correction, enabling a more accurate 
high-$Q$ prediction~\cite{rpow}.

It is interesting to compare these speculations with the results of a simple
tube model~\cite{rbw,rtube} calculation. The idea is that after hadronization
a parton jet is equivalent to a tube of hadrons distributed uniformly in
rapidity, $y$, along the jet axis with transverse mass $\mu/$unit $y$ and
length $Y$. Thus the total energy and momentum of the jet are:
$E=\mu\sinh Y$ and $ P=\mu(\cosh Y-1)$. In a two-jet event the thrust is simply
given by $T=P/E$, so that $\delta\langle T\rangle=-2\mu/Q$; likewise $\delta
\langle C\rangle=3\pi\mu/Q$ and $\delta\langle\sigma_L/\sigma_{\mathrm tot}
\rangle=(\pi/2)\mu/Q$. Fitting to data, all these estimates are consistent
with $\mu=0.5$ GeV.

\subsection{Direct photons}

Direct photons, which are unaffected by hadronization, offer an alternative
way to probe the early perturbative event structure~\cite{rphoton}. In these
events, cross sections are not simply given in terms of the quark charges but
reflect the competition in showers between q$\to $qg and q$\to $q$\gamma$
branchings in a way that is sensitive to the choice of evolution 
variable~\cite{rmike}. Here agreement with experiment has been less 
satisfactory~\cite{rcart}, particularly for rather soft or 
very isolated photons; for 
example $n$-jet+$\gamma$ cross-section results~\cite{r1jgam} disfavour the
virtuality ordering employed in JETSET's parton shower~\cite{rjetset}. This
has led to some development of the Monte Carlo shower 
algorithms~\cite{rcart}. A more critical test will be the rate of 
soft, wide-angle
photons, but here one must be especially wary of non-prompt photons arising
from decays of particles produced in the hadronization~\cite{rwide}.

\subsection{Colour coherence phenomena}
\label{shump}

Long-wavelength quanta see event structures on larger scales and so are
sensitive to the presence of neighbouring charged partons~\cite{rbible}.
This results in the effective radiating units being charge-anticharge
dipoles~\cite{rstexp} and leads to a suppression of soft gauge quanta 
due to the
requirement of coherence. This is true of QED and QCD but in the latter,
because gluons carry colour charge, one must also allow at leading order
for changes in the colour antennae formed by the hard partons.

Effects due to colour coherence are expected in both the perturbative and
non-perturbative stages of an event~\cite{rdom}. In the parton shower they
can be simply incorporated by requiring angular ordering --- that is 
successive
branchings are nested~\cite{rbible,rao}; this is known as the modified 
leading logarithm approximation
(MLLA). As noted earlier, in JESTET angular ordering is imposed on an
initially incoherent shower~\cite{rjetset}, in HERWIG it is built into the
choice of evolution variable~\cite{rherwig}, it is intrinsic to the colour
dipole model used by ARIADNE~\cite{rariadne} and it is not included in
COJETS~\cite{rcojets}. Hadronization models also respect an event's colour
structure in accordance with Local Parton Hadron Duality (LPHD)~\cite{rlphd},
discussed below. Strings may be regarded as the natural limit of an evolving
chain of colour dipoles whilst clusters may be thought of as the final, colour
neutral, dipoles. In all Monte Carlo models the treatment of both stages is
based on the large $N_c$ approximation to the colour flow~\cite{rlnc} which
gives very good agreement with known full analytic results~\cite{rgood}.

\subsubsection{The string effect}

The classic direct test of inter-jet colour coherence is the 
`string effect' --- a comparison of the particle flows between the jets 
in \qqbar g events~\cite{rstrng1}. 
As the historic name suggests, the relative depletion between
the \qqbar\ and qg jets was expected to have a non-perturbative 
origin~\cite{rsteff}; later, assuming LPHD, a perturbative 
explanation was found~\cite{rstexp}. To describe 
the Z$^0$ data~\cite{rAmega,rstrng6,rstrng2} it is
necessary to
employ a coherent shower with an approximately equal contribution coming
from the non-perturbative hadronization model (but see~\cite{rdom}). A second
classic measurement~\cite{rstrng3}, proposed in~\cite{rstgam}, is the
relative depletion between the \qqbar\ pair in \qqbar g and \qqbar$\gamma$
events~\cite{rstrng2,rstrng4,rstrng8}. The azimuthal angular dependence of
particle flow in
\qqbar g events has also been studied~\cite{rstrng5}. In order to avoid
having to find jets, the energy-multiplicity-multiplicity correlation can
be used to measure the inter-jet coherence~\cite{rAmega}, though in this
case the pQCD calculation suffers large corrections~\cite{remmc}. (A related
quantity, the asymmetry in the particle-particle correlation, has even been
suggested~\cite{rappc} as a way to measure intra-jet 
coherence effects~\cite{rAmega,rstrng7}.) Again all of 
the above studies at the Z$^0$ confirm
that Monte Carlo models incorporating a coherent shower give the best
descriptions of the data. Unfortunately a test for the `negative \qqbar\
dipole', expected in \qqbar g events when the large $N_c$ approximation is
not used, does not appear practical~\cite{rstexp}.

\subsubsection{Fully inclusive momentum spectra}
\label{sfull}

The intra-jet effect of colour coherence is to limit the production of soft
particles in the parton shower, leading to a `hump-backed plateau' shape for
the particle spectrum in the variable $\xi=\ln(1/x)$~\cite{rbible,rhump},
where $x$ is the ratio of the hadron momentum to the beam energy.
By contrast, in the incoherent
tube model, based on \eref{engluer}, there is no gluon suppression and no
broad peak forms. The actual calculation is of the parton (essentially gluon)
spectrum in a jet at the cut-off scale: $\sigma^{-1}\d\sigma/\d\xi=f(\xi;Q,
Q_0,\Lambda)$, with $Q\gg Q_0\gtap\Lambda_{QCD}$, which it is then argued
also applies to the hadron spectrum. The result is a rather unwieldy
expression so that often LPHD is invoked, in the rather technical sense of
taking the limit $Q_0\to\Lambda$~\cite{rlphd}, to obtain the simpler, more
fully evolved, limiting spectrum. Finally in the peak region a distorted
(downward skewed, platykurtic) Gaussian function~\cite{rgaus} is applicable.
The Z$^0$ data~\cite{rtunxi,rxiexp,rmult2,rxiexp2,rscvioex1,rscvioex2} are 
qualitatively
well described; quantitatively the limiting spectrum is a little too narrow in the
peak region, where the distorted Gaussian fits better, although not in the
tails. The coherent Monte Carlo event generators~\cite{rariadne,rherwig,
rjetset} give the best fits (however see~\cite{rcohq}). 

A particularly interesting measurement on the $\xi$ spectrum is that of the
peak position, $\xi^\star$. This occurs at low momentum where the occurence
of successive parton branchings should lessen any dependence on the primary
quark flavour. (Measurements of the width, skewness, etc are statistics
limited and sensitive to primary quark flavour effects in the distribution's
high momentum tail). In pQCD the $Q$ dependence of $\xi^\star$ is predicted
to be linear in $\ln(Q/Q_0)$ whie the width grows as 
$\ln^{3/2}(Q/Q_0)$~\cite{rhump}. Specifically $\xi^\star=a+n\ln(Q/Q_0)$ (i.e. $x^\star=C
(Q_0/Q)^n$) with $n=1$ in the incoherent double logarithm approximation and
$n=1/2$ in the coherent MLLA, corresponding to a harder spectrum. The Z$^0$
data when combined with lower energy results (see~\cite{bib-datarev}) clearly
prefer $n=1/2$ especially when higher-order corrections 
are included~\cite{rgaus,rxi*cor}. Similar conclusions have 
been reached from Breit frame
analyses of deep inelastic scattering ep events at HERA~\cite{rbreit}.

A closely related measurement is that of two-particle momentum 
correlations~\cite{r2pcmea}. Qualitative agreement is seen with 
the expectations based
on the pQCD distribution of gluon pairs in a shower~\cite{r2pcpre}. However
to obtain quantitative agreement large corrections to the prediction should
be anticipated~\cite{r2pcorr}.

Finally, scaling violations are seen when comparing the fragmentation
function, $\sigma^{-1}\d\sigma/\d x$, measured at the Z$^0$ to those from
lower energies~\cite{rscvioex1,rscvioex2}, after making allowance for 
the varying
primary quark favour mix (see \fref{fbrs}). As in deep inelastic scattering
these variations are controlled~\cite{rap} by Altarelli-Parisi 
equations \eref{eap} which allows $\alpha_s$ to be determined~\cite{rscvioth}.

\subsubsection{Charged particle multiplicity}

The integral of the momentum distribution of all particles gives the event
multiplicity. Soft particles make a significant contribution to this total
so that again colour coherence is important. The $\sqrt s$ dependence of
the first few moments of the distribution of event multiplicities is
calculable in MLLA~\cite{rmmom}. The effect of coherence is to slow the
growth of the mean multiplicity with $\sqrt s$~\cite{rmmlla} as compared to
the LLA result~\cite{rmlla}. The MLLA prediction for the mean works well
when higher-order corrections are included~\cite{rmult2,rxi*cor,rmult,
bib-mult1,bib-mult2,bib-mult3}, as do
several more phenomenological functions of $\sqrt s$~\cite{rigk}, including
a simple statistical phase space model~\cite{rfermi}. However this is not
the case for the incoherent COJETS model. The width (or equivalently the
second binomial moment) is larger than the data~\cite{bib-mult1,rmdist}
although
further relatively large higher-order corrections might be anticipated from
the ratio of leading to next-to-leading order predictions~\cite{rmmom}. At
high energy the ratios of these moments are expected to become $\sqrt s$
independent, a feature already present in the data. This would imply KNO
scaling~\cite{rkno} which is indeed seen to work well. The ratios of
moments as given by QCD behave approximately like those of the negative
binomial distribution (NBD)~\cite{rnbd} (with parameter $1/k\approx0.4-0.9
\sqrt{\alpha_s}$), though empirically the discretized log normal distribution
(LND)~\cite{rlnd} is a better fit to the multiplicity distribution.

Multiplicity distributions have also been studied in restricted rapidity
intervals~\cite{bib-mult2,rmydist} where the influence of global conservation
laws
may be reduced. The shape of the distribution becomes narrower for small
rapidity ($y$) windows, and some structure, possibly attributable to multi-jet
events~\cite{rmjdist}, arises at intermediate $y$ ranges. These features
are described well by JETSET, but less well by HERWIG or the simple NBD and
LND.

In addition to studies of the natural flavour mix, light and heavy quark
initiated events have been investigated separately~\cite{rQmult,rbruce}. A
large
quark mass reduces phase space and shields the the collinear singularity
associated with forward gluon emission, causing a relative suppression of
forward hadrons~\cite{rdepl}. In practice observation of the reduction in
associated multiplicity due to this `dead cone'~\cite{rdead} is hampered
by the presence of the heavy hadron's decay tracks ($5.5\pm0.1$ 
per b-hadron~\cite{rbruce}) which actually ensure a higher multiplicity 
than in u,d,s
events. However the difference in multiplicity is predicted~\cite{rbdif}
and seen to be $\sqrt s$ independent~\cite{rQmult,rbruce,rchrin}. (A
variation of
0.4 tracks is expected from the change in the heavy quark fraction between
12 ($\Upsilon(4S)$) and 91 (Z$^0$) GeV~\cite{rbruce}; this is below the level
of the experimental errors).

\subsection{Local parton-hadron duality}

Implicit in the above calculations of event features, including those by
Monte Carlo methods, is the assumption that pQCD provides the dominant
contribution. This is especially true for infrared (soft gluon) sensitive
quantities. In other words hadronization and resonance decays causes
little disruption of the features already established by the cascade.
In the absence of a well developed theory of hadronization this is in fact
a minimum requirement for pQCD calculations to be worthwhile. In terms of
the earlier space-time picture, described in subsection \ref{sxtpic}, 
this idea of
`soft confinement' appears very natural and it is indeed in many instances
supported by data~\cite{rochs}. Since in this picture it is the relatively
soft {\it gluers}, following in the wake of the hard quarks and gluons,
which cause hadronization to occur, then except for a `collective' action
by the {\it gluers}, large momentum transfers during hadronization are
precluded.

An important observation which allows this picture to be taken further is
that after a coherent shower, hadronization should occur locally. This is
seen by considering two separating perturbative partons, with opening 
angle $\theta$, and two {\it gluers} each emitted at $\theta_\epsilon$ to
their parent partons~\cite{rbible}. At the time the two {\it gluers}
simultaneously hadronize, their transverse and longitudinal separations
are given by:
\begin{equation}
d_\perp\approx R.\frac{\theta}{\theta_\epsilon}\gg R\hspace{20mm}
d_{||}\approx R\theta.\frac{\theta}{\theta_\epsilon}\gg R\theta\sim R
\end{equation}
Here the inequalities follow from the strong angular ordering condition
$\theta_\epsilon\ll\theta$~\cite{rao}. This implies that the hadronizing
{\it gluers} form only a rather low density system in configuration space.
The large inter-{\it gluer} separations effectively limit the influence
which one hadronizing parton system can have on a neighbour. This locality
hypothesis also receives strong support from the pre-confinement property
of pQCD~\cite{rprecon} (see \sref{sclus}).

In this scenario the local, and essentially independent, nature of each
parton's hadronization leaves little scope for long-range effects. 
Therefore hadronization should not significantly alter an event's angular
structure or its energy and multiplicity distributions or their
correlations, but only provide a `correction'. It is beyond this approach
to predict actual production rates of specific hadrons; however it would
support the notion that these rates are essentially constants, dependent
on at most a few nearest neighbour partons. To a certain extent present
models do respect this idea and it is the quantum numbers of a few nearby
partons which determine the properties of a produced hadron. This in turn
gives rise to an approximate local conservation of flavour, baryon number
etc.

This idea has become known as local parton-hadron
duality and represents perhaps the simplest working hypothesis for the
effects of hadronization. Experimental results on `hard event properties'
clearly indicate that the concept contains a grain of truth but also that
it is manifestly untenable at the level of `one parton one hadron'. Its
failure should be interpreted as a need for a less trivial, non-perturbative
model of hadronization. A more specific, technical definition of LPHD in
terms of the limiting momentum spectrum ($Q_0\to\Lambda$)~\cite{rlphd} was
encountered in \sref{sfull} and will be discussed further in \sref{semi}.

\section{Models for hadronization}
\label{shad}

So far it has been argued that the parton to hadron transition should occur
locally in space-time, only involving a few neighbouring partons, and that
local parton-hadron duality provides a reasonable summary of its 
effects on perturbative event
structures. However this does not mean that hadronization is simple in
detail. In practice many factors influence the production of a specific
hadron and a realistic description is presently feasible only within the
context of stochastic, non-perturbative models.

\subsection{Independent fragmentation}

Independent fragmentation is perhaps the earliest framework for 
hadronization~\cite{rfif}, later
becoming synonymous with the work of Field and Feynman~\cite{rff}. As the
name suggests, the hadronization of each individual parton is treated in
isolation as a sequence of universal, iterative q$_1\to$q$_2$+h branchings
based on the excitation of (di)quark pairs. Unfortunately the scheme has
no strong theoretical underpinning and is rather arbitrary in its details,
leading to many variants. It is used by COJETS~\cite{rcojets} 
(and ISAJET~\cite{risajet} for hadron-hadron collisions) and is a 
available as a set of non-default options in JETSET~\cite{rjetset}.

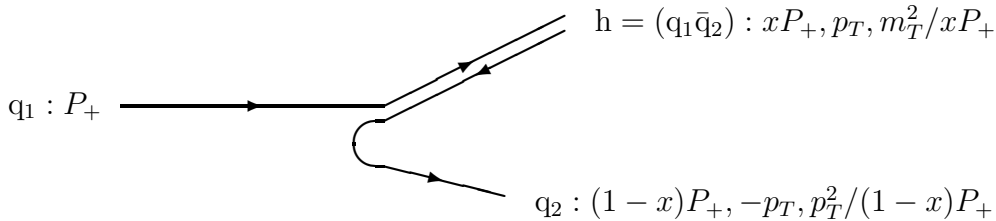
\begin{figure}
\unitlength 1mm
\begin{picture}(140,35)
\thicklines
\put(0,16){${\rm q}_1:P_+$}
\put(15,17){\line(1,0){35}}
\put(34,17){\vector(1,0){0}}
\put(50,17){\line(2,1){24}}
\put(62,23){\vector(2,1){0}}
\put(50,15){\line(2,1){24}}
\put(62,21){\vector(-2,-1){0}}
\put(78,27){${\rm h}=({\rm q}_1\bar {\rm q}_2):xP_+,p_T,m^2_T/xP_+$}
\put(50,12){\oval(8,6)[l]}
\put(50,9){\line(4,-1){16}}
\put(58,7){\vector(4,-1){0}}
\put(70,3){${\rm q}_2:(1-x)P_+,-p_T,p^2_T/(1-x)P_+$}
\end{picture}
\caption{The `unit cell' that is iterated in the independent fragmentation
scheme, showing the light-cone momentum fractions: $p_+,p_T,p_-$ where
$p_\pm = (E\pm p_z)$. Note that the parent quark acquires a mass $(m^2_h+
p^2_T)/x+p^2_T/(1-x)$.
\label{fif}}
\end{figure}

\Fref{fif} illustrates the `unit cell' used in the iterative implementation
of the scheme. The flavours of the (di)quarks generated are selected in
fixed ratios. Empirically it is found necessary to suppress both strange
quarks and diquarks, u:d:s = 1:1:$r_{\rm s}$ with $r_{\rm s}\approx 1/3$
and q:qq$'$ = 1:$r_{\rm qq}$ with $r_{\rm qq}\approx1/9$. Further rules
are required to choose between the various (low lying) hadron states of
a given flavour. The light-cone momentum fraction, 
$x=p^{\rm had}_+/p^{{\rm q}_1}_+=(E+p_z)_{\rm h}/(E+p_z)_{{\rm q}_1}$,
of the produced hadron, ${\rm h}={\rm q}_1\bar{\rm
q}_2$, is given by a longitudinal fragmentation function, such as
\begin{equation}
\label{eff}
f(x)=1-a+a(1+b)(1-x)^b
\end{equation}
In COJETS a dependence on the mass of an emitted light hadron is also built in.
Special forms are employed in the case of heavy quarks where harder momentum
spectra are expected (see \sref{sQfrag}). The transverse momentum is chosen
from a Gaussian distribution $\exp(-p^2_T/\sigma^2)$ possibly with a width
that narrows as $x\to 0$  and $x\to 1$, 
as expected from phase space considerations. The
iteration continues until a backward moving hadron would be produced ($p_z<
0$) or, in the case of COJETS~\cite{rcojets}, the sum of jet masses violates
an available energy bound~\cite{rjmb}. Diquarks are treated just as quarks,
whilst gluons are first split into a light q$\bar{\rm q}$ pair (the momentum
is shared equally in COJETS) and the above algorithm used but with retuned
parameters.

Typically a large cut-off value, $Q_0\approx 3$ GeV, is used, resulting in
only a few final-state partons. Since these partons are treated in strict
isolation, essentially ad hoc remedies must be used to ensure global
conservation of quantum numbers. Conserving four-momentum after the partons
have acquired masses proves particularly troublesome because event shape
variables, and hence $\alpha_s$ determinations, are sensitive to the
nature of the chosen solution~\cite{rasdep}. A fully Lorentz-invariant
scheme has been proposed in~\cite{rmont} but it is hard to
implement in general and only JETSET~\cite{rjetset} contains a simplified
version, as an option.

In addition to the independent hadronization of the final-state partons,
independent fragmentation models also naturally employ incoherent parton
showers, the combination of the two features offering, by today's 
standards, a mediocre
description of the exacting Z$^0$ data. However when applied to the
hadronization of a back-to-back q$\bar{\rm q}$ pair there is little
practical difference between independent and string-based models. (It is
only after gluon jets occur that differences become apparent). One can
therefore speculate that a possible way to improve the situation is, after
the forced gluon splitting, to apply the model to each neighbouring, colour
singlet q$\bar{\rm q}'$ pair in its own CoM frame just as to a pair of
back-to-back jets.

\subsection{String models}
\label{string}

When a pair of oppositely coloured quarks move apart it is thought that the
self-interacting colour field between them collapses into a long, narrow
flux tube/vortex line, called a string. Neglecting a short-range `Coulomb'
term the energy of this system appears to grow linearly with the separation.
That is, the string has a uniform (rest) energy density or constant tension,
estimated to be $\kappa\approx 1$ Gev/fm. This is equivalent to a linear
confining potential as expected from Regge phenomenology ($\kappa=1/2\pi
\alpha'$)~\cite{regge}, bag model calculations~\cite{rbag}, lattice 
studies~\cite{rwilson} and quarkonium spectroscopy~\cite{ronia}. 
This picture of a
collapsed field is analogous to a chain of magnets~\cite{rnambu} and the
behaviour of magnetic fields in (type I/II) superconductors~\cite{rsuper}.

The transverse size of a string, $\langle r^2\rangle=\pi/(2\kappa)$, is small
compared to a typical length. Therefore it is reasonable to try to model the
string dynamics on those of a `massless', relativistic string possessing no
transverse excitations. The classical equations of motion, derived from a
covariant area action, imply that in the CoM frame the two end quarks simply
oscillate backwards and forwards along a line in what are known 
as yo-yo modes~\cite{rartru1}, seen as diamonds in \fref{fxt}.
The equations of motion also admit solutions that include
localized energy-momentum carrying `kinks' which have successfully been
identified with hard gluons~\cite{rgstr}. At the end of a perturbative shower,
string segments develop between neighbouring partons, the ends terminating on
quarks. The full three-dimensional treatment of such a string system is
rather complex, being characterized by the appearance of new string segments
as intermediate gluons lose all their momentum to the system. However, robust,
covariant evolution algorithms are available~\cite{rgsyst,rmorris}, and whilst
ambiguities exist they are largely confined to the matching at gluon 
`corners'. 
The classical equations of motion are also more complex, becoming non-linear
when the end quarks are massive~\cite{rQstr}. It is worth noting at this point
that the quantized theory of the idealized string has spawned a rich and still
growing subject of its own~\cite{rgsw}, which may still prove of relevance to
the hadronization and confinement problems~\cite{rduality}.

An alternative scenario for a string description of gluons arises when the
possibility of an octet colour flux tube is admitted. For example, in a
q$\bar{\rm q}$g event the quarks may be attached to triplet strings and the
gluon to an octet string, all three of which join at a central node. A bag
model calculation~\cite{rbag} suggests the ratio of octet to singlet string
tensions is $r\equiv\kappa_8/\kappa_3=\sqrt{C_A/C_F}$. If however $r>2$ (as
suggested by lattice calculations) then it becomes favourable for the octet
string to split into two triplet strings and the above picture is recovered.
The Montevay independent fragmentation model~\cite{rmont} requires choosing
a frame in which the central junction is at rest.

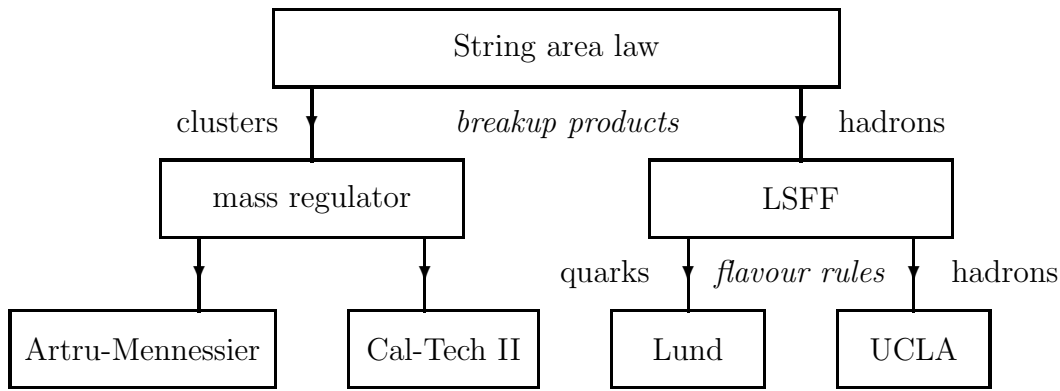
\begin{figure}
\unitlength 1mm
\begin{picture}(140,50)
\thicklines
\put(35,40){\framebox(75,10){String area law}}
\put(40,40){\vector(0,-1){6}}
\put(40,34){\line(0,-1){4}}
\put(22,34){clusters}
\put(59,34){\it breakup products}
\put(20,20){\framebox(40,10){mass regulator}}
\put(25,20){\vector(0,-1){6}}
\put(25,14){\line(0,-1){4}}
\put(55,20){\vector(0,-1){6}}
\put(55,14){\line(0,-1){4}}
\put(0,0){\framebox(35,10){Artru-Mennessier}}
\put(45,0){\framebox(25,10){Cal-Tech II}}
\put(105,40){\vector(0,-1){6}}
\put(105,34){\line(0,-1){4}}
\put(110,34){hadrons}
\put(85,20){\framebox(40,10){LSFF}}
\put(90,20){\vector(0,-1){6}}
\put(90,14){\line(0,-1){4}}
\put(73,14){quarks}
\put(93.5,14){\it flavour rules}
\put(120,20){\vector(0,-1){6}}
\put(120,14){\line(0,-1){4}}
\put(125,14){hadrons}
\put(80,0){\framebox(20,10){Lund}}
\put(110,0){\framebox(20,10){UCLA}}
\end{picture}
\caption{A family tree for string models.\label{fstree}}
\end{figure}

The above describes the motion of an idealized classical string due to the
exchange of energy between the end quarks and the string; in reality a second
process is also believed to contribute. Quantum mechanical effects allow the
the creation of q$\bar{\rm q}$ or q${\rm q}'\bar{\rm q}\bar {\rm q}'$ 
pairs in the
colour field of a stretched string, causing it to break in two {\it \`a la} the
snapping of a magnet. This picture is the basis for the Lund group's familiar
string hadronization model~\cite{rlund,rgsyst,rpop}. However the Lund version
is only one of several possible~\cite{rartru2,rctech2,rucla}, as illustrated
in \fref{fstree}.

The starting point for all string-inspired hadronization models is Wilson's
exponential area decay law, $\d{\cal P}/\d A=P_0\exp(-P_0A)$~\cite{rwilson}.
This describes the probability of a string break occurring due to the
creation of a q$\bar{\rm q}$ or q${\rm q}'\bar{\rm q}\bar{\rm q}'$ pair in
the colour field at a point containing the space-time area $A$ within its
backward light cone. As strings are believed to be uniform along their
length, the probability of pair creation, $P_0$, is a constant per unit
area. Now the momentum of a string fragment is proportional to its spatial
extent ($E=\kappa\Delta x$ and $p_Z=\kappa\Delta t$) such that the string
fragment mass is given by $m^2_{\rm string}=2\kappa^2A$, so that the decay
law can be reformulated as $\d{\cal P}/\d m^2=b\exp(-bm^2)$ with $b=P_0/(2
\kappa^2)$. Since $\langle A\rangle=P_0^{-1}$ this implies that on average
the string break-up points, and hence the hadron formation points, lie
scattered about a hyperbola: $\tau^2=t^2-x^2=4/P_0\sim2\langle m^2_{\rm
string}\rangle/\kappa^2$. This in turn implies that the slowest moving
fragments form first near the centre of the string (this is true in any
frame as the break-up points are space-like separated) --- an inside-out
pattern is assured~\cite{rinout}.

Before constructing an actual model one must decide what the string
fragments are to be identified with: either continuous mass 
fragments~\cite{rartru2,rctech} --- clusters
--- which then decay into hadrons, or actual discrete-mass 
hadrons~\cite{rlufrg}. The first choice leads to the original
Artru-Mennessier~\cite{rartru2} or CalTech-II~\cite{rctech2} schemes and
the latter to the 
Lund~\cite{rlund} and UCLA~\cite{rucla} schemes.

\subsubsection{The Artru-Mennessier/CalTech-II schemes}

Repeatedly applying the area decay law alone will result in an infinite
sequence of ever smaller string fragments~\cite{rartru2,rbowl}. One way to
see this is to note that the area law is equivalent to a joint 
distribution in light-cone
momentum fraction $x$ and transverse mass $m^2_\perp$, which
reduces to a divergent $m^2_\perp$ distribution:
\begin{equation}
\label{ezmperp}
\fl\frac{\d^2 {\cal P}}{\d x\d m^2_\perp}=\frac{b}{x}\exp\left(
-b\frac{m^2_\perp}{x}\right)\hspace{5mm}\Longrightarrow\hspace{5mm}
\frac{\d{\cal P}}{\d m^2_\perp}=bE_1(bm^2_\perp)\stackrel{m^2_\perp\to 0}
{\sim}\ln(bm^2_\perp)
\end{equation}
($E_1$ is the first exponential integral function.) Since a physical
interpretation of string fragments with very low mass is implausible, in
practice a regulator is required.

In the more fully developed CalTech-II model~\cite{rctech2} this regulator
is supplied by introducing a probability to allow a given string fragment
to undergo any further splitting:
\begin{equation}
\fl{\cal P}(\mbox{further break})=\Theta(m_{\rm string}-m_0)\left[1-
\exp\left(-P_0\frac{(m_{\rm string}-m_0)^2}{2\kappa^2}\right)\right]
\end{equation}
where $P_0$ and $m_0$ are free parameters, the latter related to the threshold
mass for a string's decay to two hadrons. The function $\Theta$ is the
Heaviside step function: $\Theta(x) = 0$ for $x<0$ and $\Theta(x)=1$
for $x \ge 1$.
If allowed, a break-up point is
selected according to the area law with the q$\bar{\rm q}$ or 
q${\rm q}'\bar{\rm q}\bar{\rm q}'$ flavour chosen according to
fixed probabilities from those kinematically allowed; at this 
point no transverse momentum is introduced.
Occasionally a final string fragment is below a second cut-off and this is
replaced by a single hadron. Otherwise the fragments undergo a comparatively
complex sequence of cluster decays in which phase space determines the
produced flavours and momenta.

Whilst the CalTech-II model has attempted to combine the desirable features
of both string and cluster models, its success in confronting data has been
at best mixed. Consequently the Monte Carlo program has fallen out of 
favour and is no longer being actively developed.

\subsubsection{The Lund scheme}

The alternative to a continuous mass spectrum for the string fragments is a
discrete spectrum, the allowed values of which are identified with the masses
of known hadrons. This approach is followed by the Lund~\cite{rlufrg} 
and UCLA~\cite{rucla} models. Requiring a 
string to fragment into hadrons of given
(transverse) mass leaves only the choice of the hadrons' light-cone momentum
fractions, $x$, free. These $x$ values may be iteratively chosen according to
several possible distributions~\cite{rjetset} and still remain true to the
area decay law. However a set of plausible assumptions greatly reduces the
number of allowed fragmentation functions~\cite{rlsff,rlsff2}. The assumptions
are: the equations of motion are those of a classical, relativistic, constant
tension string with no transverse excitations; a statistical left-right
symmetry; a central rapidity plateau; and negligible end effects. The resulting
$x$-distributions are known as the Lund symmetric fragmentation functions
(LSSF).
\begin{equation}
\label{elsff}
\fl f(x)=\frac{N_{\alpha\beta}}{x}x^{a_\alpha}\left(\frac{1-x}{x}
\right)^{a_\beta}\exp\left(-b\frac{m^2_\perp}{x}\right)
\hspace{5mm}\Biggr|_{a_\alpha=a}\hspace{-3mm}\equiv\hspace{2mm}
\frac{\d^2 {\cal P}}{\d y\d A}=C_0C_aA^ae^{-b\kappa^2A}
\end{equation}
The coefficient $a_\alpha$ relates to the parent quark flavour and $a_\beta$ 
to that of the quark or diquark produced in the colour field: in practice 
only diquarks are allowed a different value of $a$~\cite{rjetset}. 
Taking every $a_\alpha=a$ the
LSFF simplifies (compare to \eref{ezmperp} where $a=0$) and is 
equivalent~\cite{rlsff} to a flat rapidity, $y=1/2\ln[(E+P_z)/(E-P_z)]$, 
distribution and
the Wilson area law modified by a perimeter (Coulombic) term. 

An important issue now is to prescribe how the actual hadrons are chosen,
and here again the models diverge. The Lund approach~\cite{rschwin1} is based
on an attempt to model, principally through flavour and spin selection rules,
the supposed quark dynamics in the strong colour field that is a string. The
idea is a development of the concept of fermion pair production in a strong
electromagnetic field~\cite{rschwin2,rschwin3}. To supply the energy for a
\qqbar\ pair, each of transverse mass $m^2_\perp=m^2_{\rm q}+p^2_\perp$, it is
necessary to consume a finite length of string ($2m_\perp/\kappa$). The quarks
have equal and opposite $p_\perp$ since no transverse string excitations are
permitted in the model. If the \qqbar\ pair are produced locally at a point
then they must tunnel out to this classically required separation. Using the
WKB approximation for the matching of the quark wavefunctions at the classical
turning points suggests a suppression factor~\cite{rschwin2,rschwin3}:
\begin{equation}
\label{etun}
\exp\left(-\frac{\pi}{\kappa}(m^2_{\rm q}+p^2_\perp)\right)
=\exp\left(-\frac{\pi}{\kappa}m^2_{\rm q}\right)\times
\exp\left(-\frac{\pi}{\kappa}p^2_\perp\right)
\end{equation}
This would be a crude approximation to the known full QED expression for the
production of a single \qqbar\ pair, $\sum_nn^2\exp(-n\pi m^2_\perp/\kappa)$.
Unfortunately it is not known what to use for the quark masses in \eref{etun}
(see reference~\cite{rpdg} for some discussion of the range spanned by current
 and constituent quark masses) and so only qualitative conclusions can be
drawn. Among these are:
\begin{itemize}
\item
The transverse momentum suppression is the same for all quark flavours
\item
Charm and bottom quarks will not be produced from the string
\item
Since $m_{\rm s}>m_{\rm u}\approx m_{\rm d}$ (SU(3)$_{\rm F}$ is broken), 
strange quarks will be suppressed relative to up and down quarks:
\begin{equation}
\frac{{\cal P}({\rm s})}{{\cal P}({\rm u})}\equiv\gamma_{\rm s}<1
\end{equation}
\item
Production of diquarks will be suppressed:
\begin{equation}
\frac{{\cal P}({\rm uu})}{{\cal P}({\rm u})}<1\hspace{10mm}
\frac{{\cal P}({\rm us})}{{\cal P}({\rm uu})}\ltap\gamma_{\rm s}\hspace{10mm}
\frac{{\cal P}({\rm ss})}{{\cal P}({\rm uu})}\approx\gamma^4_{\rm s}
\end{equation}
the power 4 arising in the last term because of the quadratic dependence
of the tunnelling probability on the diquark mass.
\end{itemize}
Also the Gaussian $p_\perp$-distribution in \eref{etun} is best regarded as a
first approximation, leaving open the possibility of long tails originating
from unresolved gluon emission~\cite{rjetset,rbo}. This may account for the
fact that
the prediction for the width of the hadron $p_\perp$-distribution, $\sqrt{
\kappa/\pi}=0.25$ GeV, proves too 
narrow compared with measurements of $\approx0.40$ GeV~\cite{rAmega,rDtun2}.

As yet the model does not supply as much guidance on how to account for any
possible spin dynamics. Two factors influence the relative production ratios
of same flavour mesons~\cite{rlund,rvp}. First are the ($2J+1$) spin counting
factors. Second is the need for the quark produced in the string to match
onto the wavefunction of the produced meson. At the classical boundary to
the tunnelling region the meson wavefunction is expected to behave as
$1/\sqrt{m_\perp}$~\cite{rlund}. The result for the vector to pseudoscalar
meson production ratio is therefore:
\begin{equation}
\frac{V}{P}=3\times\left(\frac{m_\perp(V)}{m_\perp(P)}\right)^{\alpha
\approx 1}
\end{equation}
As $(m_V-m_P)/(m_V+m_P)$ is $0.69$ for $\pi/\rho$ and 0.004 for 
B/B$^\star$, the primary $\rho/\pi$ ratio is taken to be 1 whilst the 
B$^\star/$B ratio is fixed at 3. Radially and orbitally excited mesons 
are expected to be 
suppressed~\cite{rjetset}, although $L=1$ mesons may be included at the
expense of a new parameter for each of the four additional families of
states.

The fact that the transverse mass is involved in the suppression factor
proves unimportant except for pions, since $\langle p^2_\perp
\rangle\approx (0.3\mbox{ GeV})^2\gtap m^2_\pi$. Hence it is anticipated
that (directly produced) pions will be enhanced at low $p_\perp$ and have
a tighter $p_\perp$ distribution than, say, $\rho$ mesons or kaons. 
Also, because of their small mass, neighbouring-pion correlations are 
anticipated~\cite{rpicor}.

To summarize, the Lund prescription for an iterative string fragmentation
scheme is: first choose a quark flavour; then choose a produced hadron species;
select a quark transverse momentum; and finally select an $x$ value. The
normalization, $N$, of the LSFF in \eref{elsff} for this case is thus:
\begin{equation}
\label{enorm}
N^{-1}=\int\!\d x \frac{(1-x)^a}{x}\exp\left(-b\frac{m^2_\perp}{x}\right)
\equiv F(m^2_\perp)
\end{equation}
The resulting model requires the specification of a relatively large number
of free parameters~\cite{rjetset}. Much of this can be traced back to 
the unknown
properties and dynamics of (di)quarks. The UCLA model~\cite{rucla} attempts
to finesse these problems by formulating an iterative scheme only in terms
of known hadron properties, thereby trying to avoid the issue of quark
production in a string

\subsubsection{The UCLA scheme}

At the heart of the difference between the Lund and UCLA models is a
reinterpretation of the LSFF~\cite{rucla2}. In the UCLA model~\cite{rucla}
the LSFF is used to choose both $x$ and the species of produced hadron.
This means that $N$ \eref{enorm} becomes an {\it absolute} normalization:
\begin{equation}
N^{-1}=\sum_{{\rm h}}({\mathrm CG})^2\int\!dp^2_\perp 
F(m^2_{\rm h}+p^2_\perp)
\end{equation}
The sum runs over all hadrons containing the parent quark, with CG the
appropriate Clebsch-Gordon coefficients for the hadron wavefunction. Since
$N$ is now a common constant the hadron mass dependence appearing in the
exponential term in \eref{elsff} immediately implies a suppression of heavy
hadrons and a stiffening of their fragmentation function~\cite{rucla2}.

\begin{figure}
\unitlength=1mm
\begin{picture}(100,80)
\thicklines
\put(5,5){\vector(0,1){15}}
\put(0,15){$t$}
\put(5,5){\vector(1,0){15}}
\put(15,0){$x$}
\put(50,0){\line( 1,1){40}}
\put(50,0){\line(-1,1){40}}
\put(48,10){$A$}
\put(30,10){$\bar {\rm q}_N$}
\put(65,10){q$_0$}
\put(10,40){\line(1,1){10}}
\put(17.5,43){$N_N$}
\put(20,50){\line(-1,1){20}}
\put(0,70){\line(1,1){5}}
\put(25,35){\line( 1,1){5}}
\put(25,35){\line(-1,1){20}}
\put(21,30){$S_{N-1}$}
\put(32.5,38){$\bar {\rm q}_{N-1}$}
\put(5,55){\line(1,1){10}}
\put(15,65){\line(-1,1){10}}
\put(11,51){h$_N$}
\multiput(30,52)(2,0){5}{\circle*{1}}
\put(90,40){\line(-1,1){15}}
\put(75,55){\line(1,1){17.5}}
\put(85,45.5){$N_1$}
\put(83,57){h$_1$}
\put(70,35){\line(1,1){25}}
\put(95,60){\line(-1,1){12.5}}
\put(68,30){$S_1$}
\put(60,38){q$_1$}
\put(77.5,38){$\bar {\rm q}_1$}
\put(70,35){\line(-1,1){20}}
\put(62.5,43){$N_2$}
\put(62,58){h$_2$}
\put(50,55){\line(1,1){22.5}}
\put(42.5,27.5){\line(-1,1){5}}
\put(42.5,27.5){\line( 1,1){35}}
\put(40.5,22.5){$S_2$}
\put(50,30.5){$\bar {\rm q}_2$}
\put(77.5,62.5){\line(-1,1){15}}
\end{picture}
\caption{A schematic diagram of a fragmenting string: the $S_i$ control
the production of quark flavours; the ``knitting'' factors $N_i$ control
which hadrons form from the quarks.\label{fstring}}
\end{figure}
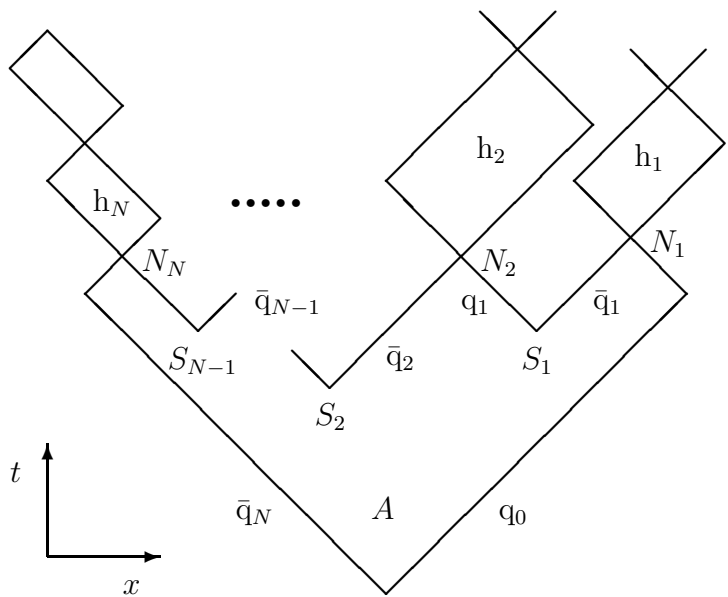

An instructive way to view the difference between the Lund and UCLA approaches
is to consider the complete weight for the production of a set of $N$ hadrons
$\{{\rm h}_i\}$ in a string of mass $s$ stretched between two quarks q$_0$ and
$\bar{\rm q}_N$, as illustrated in \fref{fstring}.

The `master' equation for this process's weight is given by:
\begin{eqnarray}
\label{emaster}
\fl\d W^{{\rm q}_0\bar {\rm q}_N}_{\{{\rm h}_i\}}(s)=
\delta^{(4)}(P_0-\sum^N_{i=1}p_{{\rm h}_i})
\exp\{-bA\}\times\prod^{N-1}_{i=1}
S_{{\rm q}_i}\exp\{-\kappa(p^2_\perp)_{{\rm q}_i}\}
\nonumber \\
\fl\times\prod^N_{i=1}N_{{\rm h}_i}[\mathrm{CG}
({\rm q}_{i-1},\bar {\rm q}_i;{\rm h}_i)]^2
\exp\{-\kappa'(p^2_\perp)_{{\rm h}_i}\}\d^4p_{{\rm h}_i}
\delta(p^2_{{\rm h}_i}-m^2_{{\rm h}_i})
\end{eqnarray}
The first two terms impose overall four-momentum conservation and the area
decay law. The second set of terms is associated with the production of the
$N-1$ intervening \qqbar\ pairs: $S_{{\rm q}_i}$ is a possible quark flavour
suppression factor and $\kappa$ controls the quark's Gaussian
$p_\perp$-distribution. The third set of terms is associated with the $N$
hadrons $\{{\rm h}_i\}$ formed out of the quark-antiquark pairs
$\{({\rm q}_{i-1}\bar {\rm q}_i)\}$: CG is the Clebsch-Gordon coefficient
for the hadron's SU(6) wavefunction, $N_{{\rm h}_i}$ is an additional
``Knitting factor'' and $\kappa'$ controls the hadron's assumed Gaussian
$p_\perp$-distribution. The approaches to these various terms in the UCLA
and Lund models are summarized in table~1.

\Table{The two sets of factors appearing in (21) and figure 6
in the UCLA and Lund string models indicating the typical values assigned.}
\br
Factor & UCLA~\cite{rucla} & Lund~\cite{rjetset} & Controls \\
\mr
$S_{\rm q}$ & 1 & $\gamma_{\rm s},\;($ud$)_1/($ud$)_0,\ldots$ 
& quark flavour suppression \\
$\kappa$ & 0 & ``$\pi/\kappa$'' & quark $p^2_\perp$ suppression \\
$\kappa'$ & $(n-2)b/(n+1)$ & 0 & hadron $p^2_\perp$ suppression \\
$N_{\rm h}$ & $N$ (const.) & $F(m^2_\perp)^{-1}\{V/P,\ldots\}$ & hadron
``knitting'' factor
\\
\br
\endtab

A number of points in the iterative implementation of the UCLA scheme are
noteworthy. First, the original model was designed to look ahead to the next
iterate. For example, if the first hadron leaves behind a u or an s quark
then the next hadron is most likely to be a pion or a kaon, and the latter
choice is (doubly) suppressed because of the higher masses of strange
hadrons. This was an attempt to mimic the quantum mechanical projection of
the string onto a set of hadrons $\{\rm h\}$. In fact, using \eref{emaster}
and the UCLA ansatz for an event's overall weight it is possible to
derive an iterative scheme that automatically generates chains of hadrons
according to this overall distribution. That is, the projection from partons
to hadrons is automatically taken into account. The required fragmentation
function is remarkably similar to the original LSFF:
\begin{equation}
\frac{(1-x)^a}{x}\left(1-\frac{m^2_\perp}{xS}\right)^a\exp\left(
-b\frac{(m^2_\perp+p^2_\perp/(n-1))}{x}\right)
\end{equation}
Compared to \eref{elsff} a finite mass correction term appears 
together with a term,
$p^2_\perp/(n-1)$, coming from local transverse momentum compensation, ($n
\approx 2$). In the case of hadrons containing a heavy quark, $x$ is replaced
by $x_{\rm eff}>x$ which softens the momentum spectrum and favours heavier
states~\cite{rucla}. Second, `multiple' so-called popcorn (discussed below)
baryon production  B$\overline{\rm B},\;$BM$\overline{\rm B},\;$BMM$
\overline{\rm B}\ldots$ is included (here B represents a baryon and M a
meson),  but due to its slow `convergence' an ad hoc suppression
needs to be introduced: $\exp(-\eta\sum m_M)$ with $\eta$ free ($\approx
10$ GeV$^{-1}$). Third, the mechanism for the local conservation of hadron
transverse momentum proves awkward due to an ambiguity between the quark
and hadron levels. Finally, but perhaps most significantly, the model only
contains four (+two) free parameters: $a,\;b,\;n$ and $\eta\;(+\Lambda$
and $Q_0$).

\subsubsection{Consequences of the string's space-time structure}

A further interesting aspect of string models is that inferences can
be drawn from their associated space-time picture. These include 
predictions on spin correlations and effects due to quantum statistics.

If a \qqbar\ pair are produced with some (equal and opposite) transverse
momentum, $p_\perp$, with respect to the string, then because they are
separated by a distance $2m_\perp/\kappa$ a non-zero angular momentum
$L=2m_\perp p_\perp/\kappa$ is necessarily introduced. Since total angular
momentum $J=L\otimes S_{\rm q}\otimes S_{\bar {\rm q}}$ must be conserved 
and $\langle L\rangle\approx 1\hbar$ the \qqbar\ pair typically form in a
$^3P_0$ state, particularly so at higher $p_\perp$. This is expected to lead
to $p_\perp-$transverse-spin correlations, and spin correlations between
neighbouring hadrons~\cite{rspcor}. Transverse polarizations are only
possible because the string introduces a preferred axis. (Such an axis is
implicit in the chain-like structure found in cluster models, though it is
not presently utilized in them). Conservation of angular momentum is also
expected to lead to a suppression of orbitally excited hadrons.

\begin{figure}
\unitlength 1mm
\begin{picture}(100,65)
\put(5,5){\vector(1,0){10}}
\put(7,14){$t$}
\put(5,5){\vector(0,1){10}}
\put(17,4){$x$}
\put(42.5,45){\line(1,1){5}}
\put(47.5,50){\line(1,-1){5}}
\put(52.5,45){\line(1,1){7.5}}
\put(60,52.5){\line(1,-1){5}}
\put(65,47.5){\line(1,1){5}}
\put(70,52.5){\line(1,-1){7.5}}
\thicklines
\put(60,5){\line(1,1){40}}
\put(60,5){\line(-1,1){40}}
\put(55,32.5){\line(1,1){7.5}}
\put(62.5,40){\line(1,-1){5}}
\put(67.5,35){\line(1,1){10}}
\put(77.5,45){\line(1,-1){5}}
\put(82.5,40){\line(1,1){5}}
\put(87.5,45){\line(1,-1){5}}
\put(92.5,40){\line(1,1){5}}
\put(55,32.5){\line(-1,1){12.5}}
\put(42.5,45){\line(-1,-1){5}}
\put(37.5,40){\line(-1,1){5}}
\put(32.5,45){\line(-1,-1){5}}
\put(27.5,40){\line(-1,1){5}}
\put(57.5,42.5){$\Delta A$}
\put(73,37){1}
\put(46,36){2}
\put(45.5,45){1}
\put(71,46){2}
\end{picture}
\caption{The two sequences of string breakings possible when two identical
particles are present; all other particles are the same. This results in a
difference $\Delta A$ between the enclosed total string areas.\label{fbe}}
\end{figure}
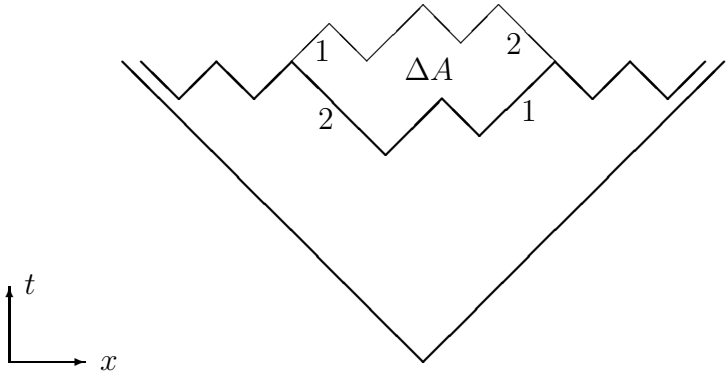

If two identical hadrons are produced from a string there exists an ambiguity
in the possible rank ordering of the particles, unless they have exactly equal
momentum. These two string drawings, illustrated in \fref{fbe}, enclose
different areas. Now if, as is believed, the string area law derives from the
modulus squared of a matrix element ${\cal M}=\exp[\i A(\kappa+\i P_0/2)]$
then quantum mechanical interference will occur~\cite{rstbe}. Dependent on
the Bose-Einstein or Fermi-Dirac nature of the particles the expected joint
production probability becomes
\begin{equation}
\label{ebe}
(P_{12}+P_{21})\left[1\pm\frac{\cos(\kappa\Delta A)}{\cosh(P_0\Delta A/2)}
\right]
\end{equation}
which is clearly modified from the naive sum of the weights. (A quark
transverse momentum correction, $(\pi/2\kappa)\Delta p^2_\perp$, is not
shown in \eref{ebe}.) As the momentum difference squared between the
identical hadrons vanishes, $\Delta A\rightarrow 0$ and any enhancement
or cancellation becomes maximal. The area difference, $\Delta A$, is
related to the size of the emitting volume, likened to the Hanbury-Brown
Twiss effect in optical astronomy~\cite{rhanb}, but not related directly
to the total string size. This approach therefore has close parallels to
a more geometric picture based on the Fourier transform of the source
distribution~\cite{rbegeom}. Only identical neutral pions can be produced
side by side from a string --- identical charged pions must have at least
one intervening hadron. Therefore $\Delta A$ will be larger for charged
pions, leading to smaller correlations than for neutral pions~\cite{rbo}.
The effects of correlations on short lived resonance decays can also be
included~\cite{rberes1,rberes2}.

Two basic algorithms are currently available for including a Bose-Einstein
event `weight'~\cite{rbealg}. The standard scheme available 
in JETSET~\cite{rjetset,WWBEC} involves rearranging identical 
boson momenta so that
they are distributed according to the pairwise correlation function. Full
multiboson correlations may be included, but at the cost of additional
computing time.
The effect of including Bose-Einstein correlations may be likened to adding
an attractive inter-boson force, leading to `lumpier' distributions. The
experimental situation is discussed in \sref{sbe}.

\subsubsection{Baryons and the popcorn mechanism}
\label{spop}

In the Lund string model, baryon production poses particular problems of
principle. The basic difficulty appears to stem from incomplete knowledge
of the internal structure of a baryon~\cite{rdiq}. Is it a quark-diquark
system or a three quark system? This ignorance poses less of a problem for
HERWIG and to some extent the UCLA string model because these essentially
only need to know a baryon's mass and spin. However the Lund string works
directly with the (di)quarks themselves and so in the absence of a guiding
principle it therefore allows for two options, the diquark~\cite{rstdiq}
and popcorn~\cite{rpop} baryon production mechanisms.

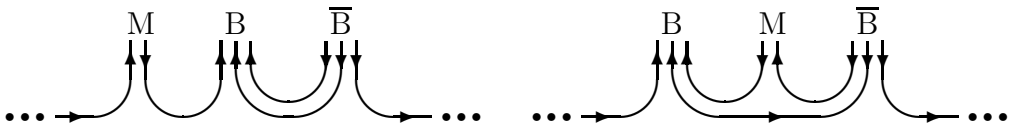
\begin{figure}
\unitlength 1mm
\begin{picture}(140,15)
\thicklines
\multiput(4,0)(2,0){3}{\circle*{1}}
\put(10,0){\line(1,0){5}}
\put(14,0){\vector(1,0){0}}
\put(15,5){\oval(10,10)[br]}
\put(20,5){\line(0,1){5}}
\put(20,9){\vector(0,1){0}}
\put(19.5,11){M}
\put(22,5){\line(0,1){5}}
\put(22,6){\vector(0,-1){0}}
\put(27,5){\oval(10,10)[b]}
\put(32,5){\line(0,1){5}}
\put(32,9){\vector(0,1){0}}
\put(34,7){\line(0,1){3}}
\put(34,9){\vector(0,1){0}}
\put(32.5,11){B}
\put(36,7){\line(0,1){3}}
\put(36,9){\vector(0,1){0}}
\put(41,7){\oval(10,10)[b]}
\put(41,7){\oval(14,14)[b]}
\put(46,7){\line(0,1){3}}
\put(46,6){\vector(0,-1){0}}
\put(48,7){\line(0,1){3}}
\put(48,6){\vector(0,-1){0}}
\put(46.5,11){$\overline{\rm B}$}
\put(50,5){\line(0,1){5}}
\put(50,6){\vector(0,-1){0}}
\put(55,5){\oval(10,10)[bl]}
\put(55,0){\line(1,0){5}}
\put(58,0){\vector(1,0){0}}
\multiput(62,0)(2,0){3}{\circle*{1}}
\multiput(74,0)(2,0){3}{\circle*{1}}
\put(80,0){\line(1,0){5}}
\put(84,0){\vector(1,0){0}}
\put(85,5){\oval(10,10)[br]}
\put(90,5){\line(0,1){5}}
\put(90,9){\vector(0,1){0}}
\put(92,7){\line(0,1){3}}
\put(92,9){\vector(0,1){0}}
\put(90.5,11){B}
\put(94,7){\line(0,1){3}}
\put(94,9){\vector(0,1){0}}
\put(99,7){\oval(14,14)[bl]}
\put(99,7){\oval(10,10)[b]}
\put(104,7){\line(0,1){3}}
\put(104,6){\vector(0,-1){0}}
\put(103.5,11){M}
\put(106,7){\line(0,1){3}}
\put(106,9){\vector(0,1){0}}
\put(111,7){\oval(10,10)[b]}
\put(116,7){\line(0,1){3}}
\put(116,6){\vector(0,-1){0}}
\put(99,0){\line(1,0){12}}
\put(107,0){\vector(1,0){0}}
\put(111,7){\oval(14,14)[br]}
\put(118,7){\line(0,1){3}}
\put(118,6){\vector(0,-1){0}}
\put(116.5,11){$\overline{\rm B}$}
\put(120,5){\line(0,1){5}}
\put(120,6){\vector(0,-1){0}}
\put(125,5){\oval(10,10)[bl]}
\put(125,0){\line(1,0){5}}
\put(128,0){\vector(1,0){0}}
\multiput(132,0)(2,0){3}{\circle*{1}}
\end{picture}
\caption{A schematic of baryon production in the diquark model (left) and
`popcorn' model (right) leading to MB$\overline{\rm B}$ and 
BM$\overline{\rm B}$ configurations respectively. \label{fpop}}
\end{figure}

The diquark mechanism is a straightforward generalization of the quark
meson production model and was the first to be fully developed~\cite{rstdiq}. 
However a stepwise quark model for the production of baryons was the first to
be proposed~\cite{rschwin3} and implemented~\cite{rbpop}. Whilst this only
evolved later~\cite{rpop} (and is continuing to evolve~\cite{rpop2}) into the
popcorn mechanism it would be a misconception to regard it as especially
contrived or unnatural. When a qq$'\bar{\rm q}\bar{\rm q}'$ (or \qqbar) pair
is produced in a string's colour field with the same colour as the end quarks
they precipitate a string break: the diquark mechanism~\cite{rstdiq}. When a
(virtual) \qqbar\ pair is produced the possibility that they have a different
colour to that of the end quarks allows a non-zero colour field to exist
between them in which further real \qqbar\ pairs could form, leading to the
sequence B$\overline{\rm B},$BM$\overline{\rm B}$ etc: 
popcorn production~\cite{rpop}. Perhaps not unsurprisingly 
the popcorn mechanism requires a
(modest) number of new free parameters. 

The main practical consequence, so far, of introducing popcorn production
appears to be a lessening of the phase space correlations between
baryon-antibaryon pairs (see~\sref{sbcor}). The fragmentation function for
baryons is also expected to be softened~\cite{rbpfrg} (compared to a
meson with the same transverse mass) due to popcorn 
production: in particular BM$\overline {\rm
B}$ sequences cause a suppression of leading baryons~\cite{rpop2}. (This would
suggest using $a_{\rm qq}>a_{\rm q}$ in the LSFF \eref{elsff}.) The actual
level of popcorn production required to describe data is still the subject
of debate but it may be related to the $a$ parameter of the LSFF~\cite{rbpfrg}.
Interestingly a search for the expected chain like structures, such as
correlated $\Lambda $K$^+\bar{\rm p}$ systems, has failed to see any positive
evidence~\cite{rpopex}. Allowing three body decays of clusters is expected
to have similar effects to introducing popcorn production in string models.

\subsection{Cluster models}
\label{sclus}

\begin{figure}[t]
\epsfig{file=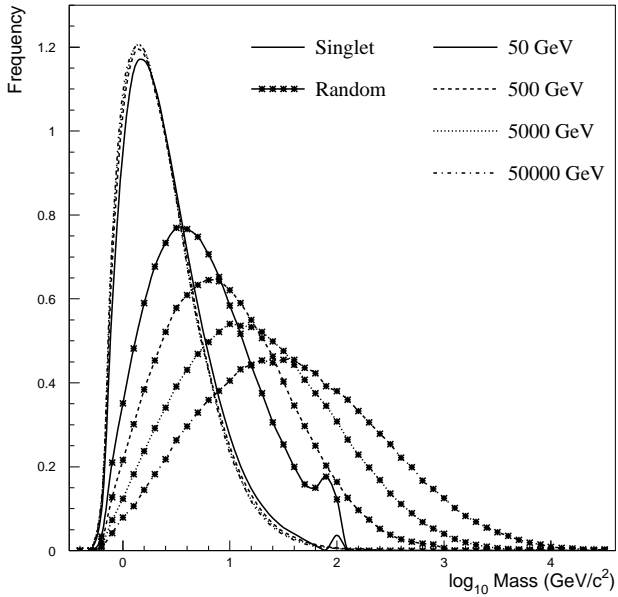,height=80mm}
\caption{The mass spectrum for colour singlet and random q$_1\bar{\rm q}_2$
clusters in a u-quark jet of four initial energies. In the random sample
\qqbar\ pairs from gluon splittings are excluded.\label{fclus}}
\end{figure}

Whilst clusters initially appeared as intermediate states in the string
model of Artru and Mennessier~\cite{rartru2} the first fully fledged
cluster hadronization models originated at CalTech~\cite{rctech,rctech1}.
(Cluster-like structures also naturally arose in the earlier statistical
bootstrap model~\cite{rboot} and multiperipheral models~\cite{rmpm}.)
Today the scheme is best known through the HERWIG implementation 
of Webber~\cite{rherwig,rclus}. This is based on 
the pre-confinement property proved
for pQCD~\cite{rprecon}: at the end of the perturbative shower the mass
and spatial distributions of colour singlet clusters of partons spanned by
quark-antiquark pairs have a universal distribution. In practice gluons
remaining at the cut-off scale $Q_0$ are forcibly split into  light \qqbar\
pairs, the Wolfram ansatz~\cite{rwolf}, so that in the planar 
approximation~\cite{rlnc} 
neighbouring quark-antiquark pairs form colour singlets. 
\Fref{fclus} illustrates the resulting cluster mass distributions for
u-quark initiated jets at four energies, showing clearly the universality
of the colour singlet cluster mass distribution, in contrast to the
distributions for random clusters. 

\begin{figure}
\unitlength 1mm
\begin{picture}(140,50)
\thicklines
\put(5,45){Perturbative}
\put(0,20){\framebox(15,10){parton}}
\put(15,25){\vector(1,0){4}}
\put(19,25){\line(1,0){1}}
\put(20,20){\framebox(15,10){shower}}
\put(35,25){\vector(1,0){4}}
\put(39,25){\line(1,0){1}}
\put(40,20){\framebox(15,10){clusters}}
\multiput(55,0)(0,2){10}{\line(0,1){1}}
\multiput(55,31)(0,2){10}{\line(0,1){1}}
\put(55,30){\vector(1,1){8.5}}
\put(63.5,38.5){\line(1,1){6.5}}
\put(65,34){heavy}
\put(70,40){\framebox(25,10){splitting}}
\put(82.5,40){\vector(0,-1){6}}
\put(82.5,34){\line(0,-1){4}}
\put(55,25){\vector(1,0){8.5}}
\put(63.5,25){\line(1,0){6.5}}
\put(70,20){\framebox(25,10){2-bdy decay}}
\put(55,20){\vector(1,-1){8.5}}
\put(63.5,11.5){\line(1,-1){6.5}}
\put(65,14){light}
\put(70,0){\framebox(25,10){1-bdy decay}}
\put(95,28){\vector(1,0){6}}
\put(101,28){\line(1,0){4}}
\put(95,22){\vector(1,0){6}}
\put(101,22){\line(1,0){4}}
\put(95,5){\vector(1,0){6}}
\put(101,5){\line(1,0){4}}
\put(105,0){\framebox(15,30)
{\shortstack{\mbox{r\hspace{4mm}}\\ \mbox{e\hspace{4mm}}\\[-.5mm]s d\\o e\\
n c\\a a\\n y\\[-.5mm] \mbox{c\hspace{4mm}}\\ \mbox{e\hspace{4mm}} }}}
\put(120,15){\vector(1,0){4}}
\put(124,15){\line(1,0){1}}
\put(125,10){\framebox(15,10){hadrons}}
\put(105,45){Non-perturbative}
\end{picture}
\caption{A schematic diagram for fragmentation via cluster
hadronization.\label{fcldk}}
\end{figure}
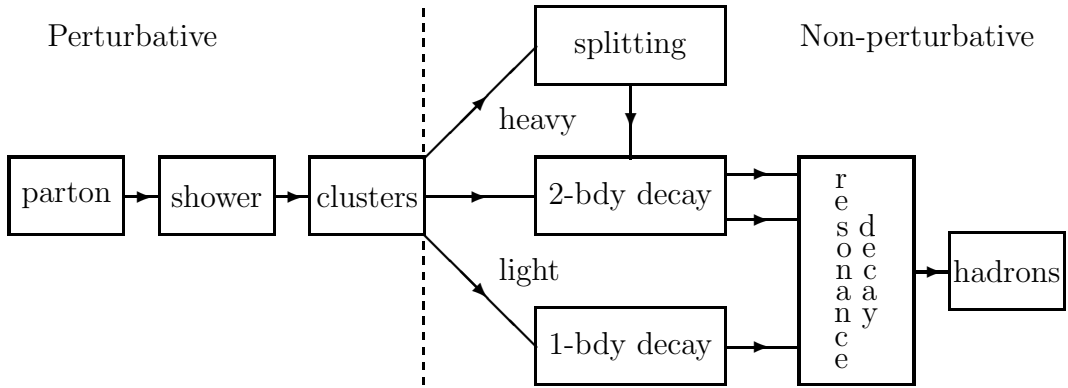

The mean cluster mass is of order $Q_0$, a few GeV, and for a colour-coherent
parton shower has a spectrum falling faster than any power~\cite{rfast}. The 
distribution is independent of the initial parton type (q or g) and virtuality:
this is in contrast to random quark-antiquark pairs. This universality is
suggestive of the formation of intermediate `super-resonances' which 
independently decay into the familiar resonances. \Fref{fcldk} illustrates the
stages in the fragmentation of a parton into hadrons via cluster 
hadronization. 

Three masses of cluster need to be treated:
\begin{description}
\item[Average] Most clusters have masses, $m_{\rm cl}$, close to $Q_0$ and
undergo two-body decays~\cite{rclus}. A quark or light diquark pair (here the
diquarks serve only as a mnemonic device for quantum number conservation),
$\bar{\rm q}_3$q$_3/$q$_3$q$_4\bar{\rm q}_3\bar{\rm q}_4$, is selected at
random, and for a
cluster of flavours q$_1\bar{\rm q}_2$ the putative hadrons h$_1=$q$_1\bar{
\rm q}_3/$q$_1$q$_3$q$_4$ and h$_2=\bar{\rm q}_2$q$_3/\bar{\rm q}_2\bar{\rm
q}_3\bar{\rm q}_4$ are formed. The selection is accepted or rejected according
to its phase space weight: $(2J_{{\rm h}_1}+1)(2J_{{\rm h}_2}+1)\hat p(m_{
\rm cl},m_{{\rm h}_1},m_{{\rm h}_2})\Theta(m_{\rm cl}-m_{{\rm h}_1}
+m_{{\rm h}_2})$ where $J$ and $m$ are the hadron spins and masses, and 
$\hat p$ is the CoM momentum of the kinematically allowed two-body
decays. If accepted, the decay momenta are selected isotropically in the
cluster
rest frame. This prescription is in accord with the OZI rule~\cite{rozi}.
\item[Heavy] A number of clusters in the tail of the distribution are
too massive for an isotropic decay to be plausible. The criterion used
is: $m_{\rm cl}\equiv m_{{\rm q}_1\bar {\rm q}_2}\geq m_{\rm f}$ 
where $m^n_{\rm f}=m^n_{\rm max}+(m_{{\rm q}_1}+m_{\bar {\rm q}_2})^n$
and constituent quark masses are used. Here the label f denotes
the flavour type of the cluster and $m_{\rm max}$ and $n$ are variable
parameters of the model. By the (repeated) device of introducing a
light quark pair, heavy clusters are forcibly split into two daughter
clusters whose masses are chosen according to a power law spectrum and
whose directions are aligned with the q$_1-\bar {\rm q}_2$ axis in the
cluster rest frame~\cite{rherwig}. This appearance of a preferred, colour
field, axis is reminiscent of a string splitting.
\item[Light] Occasionally it is kinematically impossible for a cluster
to decay into two hadrons. Such a cluster is decayed into the lightest
hadron h$_0=$q$_1\bar {\rm q}_2$ and four-momentum transferred to a 
neighbouring cluster to satisfy the mass-shell constraint~\cite{rherwig}.
\end{description}

In the absence of a theory of cluster decay the simplest viable approach,
pure phase space, is adopted (Okham's razor)~\cite{rclus,rctech} resulting
in very few free parameters~\cite{rherwig,rctech2}. Here two-body dominance
of the decays is one such simplification~\cite{rctech}. However non-trivial
matrix elements are not precluded and parameters controlling the relative
production rates of (di)quark pairs in the cluster decays and also weights
for the hadron representations (knitting factors) are made available to
users~\cite{rherwig,rctech}. The details of heavy cluster splitting prove
important: $m_{\rm max}$ (related to the available kinetic energy)
influences the light baryon yield, $n$ the heavy hadron yield and both 
influence the
momentum spectra~\cite{rDtun2}. The one-body cluster decay mechanism is
important for the production of rare, heavy states such as the $\Upsilon$
or B$_{\rm c}$, and also for describing leading particles, to which end a
number of extra parameters have been added~\cite{rherwig}.

Despite the fact that its initial purity has been somewhat corrupted by the
need to handle exceptional mass clusters, the model remains a simple, compact
and predictive scheme. Since clusters are typically light, limited transverse
momentum is automatic, hadrons with non-zero strangeness and baryon number
are suppressed because they are heavier, and the spin ratios of iso-flavour
hadrons follow partly from the ($2J+1$) factor and partly from the larger
masses of higher spin states. The model represents a well founded attempt to
go as far as possible with as little as possible.

\subsection{A comparison of the main hadronization models}

All the major hadronization models, cluster, independent and string, are
local, universal, stochastic models. They are based on a small number of
recursively applied branchings, where at each iterative step probabilistic
rules are applied to select flavours, spins and momenta. The main features 
the models are summarized in table~2. 

\fulltable{A comparison of the main hadronization model approaches.}
\br
Feature & \centre{4}{Hadronization Model} \\
\ns
&\crule{4} \\
& Cluster & Independent & \centre{2}{String} \\
\ns
&&&\crule{2} \\
&&& Lund & UCLA \\
\mr
Principle & very simple & simple & complex & less complex \\
Lorentz invariant & yes & no & \centre{2}{yes} \\
Flavour, charge & automatic & ad hoc & \centre{2}{automatic} \\
etc, conservation &&&& \\
Mass dep. via & hadrons & quarks & quarks & hadrons \\
Strangeness supp. & predicted & free param. & restricted params. & predicted \\
Baryon supp. & predicted & free param. & restricted params. & predicted \\
$J^P$ ratios & predicted & free params. & restricted params. & predicted \\
Limited $P_T$ & natural & built in & built in & natural \\
Frag. fn. & N/A & free & \centre{2}{restricted by L--R symm.} \\
Cut-off $(Q_0)$ dep. & significant & very strong & \centre{2}{modest} \\
Stability & infrared prob. & collinear prob. & \centre{2}{stable} \\
Limitations & massive clusters & requires & \centre{2}{light strings treated}
\\
& treated like & large $Q_0$ & \centre{2}{like clusters} \\
& `strings' &&& \\
\br
\endtab

Lorentz invariance and quantum number conservation are not troublesome for
strings or clusters but are an issue for independent hadronization models
requiring post facto adjustments. These essentially ad hoc
remedies are not always implemented (particularly in application to dirtier
hadronic collisions~\cite{risajet}), and, worse, physical observables are
known to be sensitive to details of the chosen solution~\cite{rasdep}.

Only the cluster and UCLA models, with their emphasis on hadron properties,
provide succinct basic algorithms for flavour, spin and momentum selection.
In their initial formulations the lack of free parameters gives these models
a laudable predictive power, which is in marked contrast to the Lund string
and independent hadronization models. Of course, poor fits to data could
mandate elaborations of the models which in turn may then dilute their
predictive power. The Lund model ought to have an advantage because it is based
on a semi-quantitative picture of an underlying dynamics but this is
undermined by the indirect measurability of the basic (di)quark parameters.
Thus whilst the model's parameters possess a large degree of internal
coherence, their must be disappointment at their large number and seemingly
Byzantine complexity. To emphasize this point, 13 inputs are needed in the
basic Lund model~\cite{rjetset} to describe the 13 (assuming
u$\leftrightarrow$d isospin) $L=0$ light (u,d,s) hadrons (the number of
hadrons increases  to 27 if u and d are distinguished). Thus the flavour and
spin aspects of the Lund model show little practical difference to the rather
similar, but assumed, rules found in an empirical independent hadronization
model. However a significant difference is the way in which a string's causal
structure restricts the longitudinal fragmentation function to a unique
family of left-right symmetric functions~\cite{rlsff}. No such restriction
applies to independent models.

The issue of stability with respect to collinear and soft gluons arises in
cluster and independent hadronization models. Specifically, the spectrum of
clusters is sensitive to the emission of isolated, soft gluons. This infrared
instability may be regarded either as a serious problem or, perhaps, as a
warning that it is important to treat the theory correctly. Perturbative QCD
does not like isolated colour charges, which are strongly Sudakov suppressed;
this is responsible for the fast falling tail of the cluster 
mass spectrum~\cite{rfast}. 
(Recall also that many other observables, such as the mean multiplicity, 
are known to be infrared sensitive.) In independent fragmentation a
similar problem arises when one final-state parton is replaced by two parallel
partons of equal net energy, giving a different multiplicity. This
collinear instability occurs essentially because the two partons are oblivious
to each other's presence.

To recap, the motivation for the rules used in these models varies from 
the QCD-inspired, complex dynamics of strings through the minimalism 
of clusters to
the simple expediency of independent fragmentation. In the subsequent sections
we shall see how well these hadronization models compare to the various Z$^0$
data. However since independent hadronization makes no claim to be based on
QCD, a fact reflected in the extreme arbitrariness of its parameters, we drop
it from further discussion. While it is not considered a viable scheme for
describing Z$^0$ physics it does however survive in the ISAJET~\cite{risajet}
Monte Carlo program for hadron hadron collisions.

\subsection{Colour rearrangement}

Recently interest has arisen in the possibility that soft, long wavelength,
gluons may cause non-perturbative rearrangements in the colour structures
of events developed during the showering stage, particularly in relation to
W$^+$W$^-$ pair production~\cite{rlep2mc,rlep2ww}. Several phenomenological
models are available based on the three main \ee\ event generators, and are
summarized in table~3.

\fulltable{The main features of colour rearrangement models.}
\br
Feature & \centre{5}{Colour Rearrangement Model} \\
\ns
\mr
Basic MC & \centre{2}{ARIADNE} & HERWIG & \centre{2}{JETSET} \\
Author(s) & J. H\"akkinen & L. L\"onnblad & B.R. Webber & T. Sj\"ostrand &
\u{S}. Todorova \\
& G Gustafson & & & V A Khoze & \\
Reference(s) & \cite{rjhreco} & \cite{rllreco} & \cite{rherwig} &
\cite{rjetset,rtsreco} & \cite{rlep2mc} \\
\ns
\mr
rearrangement: & & & decrease cluster &
\centre{2}{I space-time overlap} \\
- criterion & \centre{2}{decrease $\lambda$ measure} & spatial size & \centre{2}{II string crossing} \\
- in shower        & no  & yes & no  & no  & yes \\
- in hadron.       & yes & yes & yes & yes & yes \\
- mult. re-arr.    & yes & yes & yes & no  & yes \\
- inter-singlet    & yes & yes & yes & yes & yes \\ 
- intra-singlet    & no  & yes & yes & no  & yes \\
\br
\endtab

The rearrangement criterion in the ARIADNE-based models is a decrease in the
$\lambda$-measure~\cite{rlamb} which quantifies the momentum-space size of a
string system. Only the second model~\cite{rllreco} should be considered as
realistic for typical events. In HERWIG at the end of the perturbative shower
pairs of \qqbar$'$ singlet clusters may be rearranged, with fixed probability
($\approx1/9$), if this results in a reduction in the quadratic sum of their
space-time sizes. In JETSET the rearrangement criterion is based on the
space-time evolution of the strings. In model I the spatial overlap of `wide'
flux tubes is used to assign probabilities for rearrangement, whilst in model
II rearrangement occurs at the first crossing of the `narrow' vortex lines. 
Todorova's model~\cite{rlep2mc} is an elaboration of 
the original model~\cite{rtsreco}, allowing for
multiple reconnections, including within a single colour singlet
system, and self-interactions; this leads to string loops --- glueballs.

As yet these models have not been thoroughly investigated, nor their
consequences for Z$^0$ decays found. Note that when including colour
rearrangement the Monte Carlos must first be retuned: for example reducing
the $\lambda$-measure also lowers the average multiplicity, which must be
compensated. Since the physics of colour rearrangement is universal in
nature, effects should also be anticipated in all other types of hadronic
interactions, including for example B-hadron decays (B$\to J/\psi X_s$)
and rapidity gap events~\cite{rllrap}, where important constraints may be
found.

\section{The colliders and experiments}

\subsection{The colliders}

\subsubsection{LEP}

The LEP machine, a 27~km circumference storage ring, was 
conceived and constructed primarily as a Z$^0$ factory. 
In the LEP 1 mode, running at and near the Z$^0$ peak, a 
luminosity of $2.2 \times 10^{31}$~cm$^{-2}$s$^{-1}$ could be obtained,
with beam lifetimes of up to 20 hours. Until 1992, 
four bunches of electrons and positrons crossed 
every 22~$\mu$s at the experiments. During 1993 and
1994 LEP ran in a mode allowing eight bunches, with a
crossing time reduced to 11~$\mu$s. In 1995, so-called bunch 
trains were introduced, with
four trains, each of three bunches of particles,
providing a further increase in luminosity.
The beam spot in 1995, the last year of extensive running
at the Z$^0$, was 250~$\mu$m in
the vertical and 5~$\mu$m in the horizontal directions.
In the years 1990, 1992 and 1994, LEP ran on the Z$^0$ peak
to produce the largest possible number of events. In
1991, 1993 and 1995, the centre-of-mass energy was
scanned across the Z$^0$ peak to allow studies of the
Z$^0$ line shape and other tests of the electroweak theory.

\subsubsection{SLC}

Unlike LEP, the SLC is a single pass collider, with the electron
and positron bunches lost after each pass. The bunch crossing
frequency, 120~Hz, is therefore very much lower than at LEP.
The SLC is however capable of providing 
electron beams with large polarization. In order to increase 
the luminosity, the beams are tightly squeezed before the
collisions to reduce their cross sectional area. 
This technically difficult
procedure results in a beam spot of diameter 0.5~$\mu$m 
in the vertical plane and 2.3$\mu$m horizontally. The
experiment, SLD, is able to exploit this by placing
precision microvertex detectors only 3~cm from the 
interaction point. 
Since 1989 the SLC luminosity has steadily
improved to around $6 \times 10^{29}$~cm$^2$s$^{-1}$,
and electron
beam polarization values of 80\% are now routinely achieved.  

\subsubsection{Experimental conditions}

Both LEP 1 and SLC could provide their host experiments with 
clean, low background, experimental conditions. In addition,
the relatively low beam crossing rates allowed 
the experiments to implement bias-free triggers and to collect 
data with essentially no dead time. For studies of multihadron
production at the Z$^0$ peak, initial-state radiation is 
negligible.
Multihadronic Z$^0$ decay events could be triggered
with efficiencies greater than 99\% and selected offline
with backgrounds (mainly due to $\tau$ lepton pair production)
of less than 1\%.  

\subsection{The detectors}

The four LEP detectors, ALEPH, DELPHI, L3 and OPAL, and
the SLD detector at the SLC are typical large, multipurpose,
particle physics detectors designed to allow measurement
of events over a large part of the solid angle.  
For studies of hadronization,
certain features of the detectors are
particularly relevant. A large, active tracking volume 
within a strong magnetic field allows reconstruction of jets of
charged particles and
of secondary vertices from strange particle decays; 
good track momentum and direction measurement give 
accurate reconstruction of systems 
of two or more particles; charged particle identification
permits the study of inclusive identified hadrons;
electromagnetic calorimetry
allows measurement of inclusive $\pi^0$ and $\eta$ mesons;
and precision secondary vertex reconstruction and inclusive 
electron and muon identification may be used to identify charm
and beauty particles.

The coordinate systems used by the experiments define $z$ to be
along the beam direction, so that the $xy$ plane is
perpendicular to the beams. Then $r$ and $\phi$ are the usual
cylindrical polar coordinates, and the $xy$ plane is also
called the $r\phi$ plane. The angle $\theta$ is normally the
polar angle to the $z$ (beam) axis. 

\subsubsection{ALEPH}

The ALEPH detector~\cite{bib-ALEPH} was designed to provide
high three-dimensional granularity with large solid angle
coverage for charged particle tracking and for calorimetry. 
Its silicon vertex detector, drift chamber and large
time projection chamber (TPC), in a 1.5~T magnetic field, give
a resolution on momentum transverse to the beam directions
of $\sigma(1/p_T) = 0.6 \times 10^{-3}$~(GeV/c)$^{-1}$
for 45~GeV/c tracks. At low momentum, the resolution is
dominated by multiple scattering which contributes $0.5\%$
to $\sigma(p_T)/p_T$. The silicon vertex detector
permits the measurement of track impact parameters with
an accuracy of $25~\mu$m$+95~\mu$m$/p$ (with $p$ in GeV/c) 
in both the $r\phi$ and $rz$ planes, allowing excellent
reconstruction of charm and beauty particles.  
The TPC also measures ionization energy loss, $dE/dx$, 
giving good electron identification to high momenta, 
$\pi$/K separation at two standard deviations (2$\sigma$) 
in the relativistic rise region above 2~GeV/c, and
K/p separation at 1$\sigma$ for momenta over 5~GeV/c.
The efficiency to measure K$^0_{\rm s}$ and $\Lambda$ particles
is about 50\% at maximum, which occurs at about 8~GeV/$c$ momentum.
The lead/wire-chamber electromagnetic calorimeter has 
an energy resolution of $\sigma(E)/E = 0.18/\sqrt{E} + 0.009$
and an angular resolution of $(2.5/\sqrt{E} + 0.25)$~mrad 
($E$ in GeV). Along with the $dE/dx$ information, the
calorimeter gives an average electron identification 
efficiency of 65\% in hadronic jets. The efficiency for
reconstruction of $\pi^0$ mesons peaks at about 50\% at 
10~GeV, falling to 20\% at 2~GeV and 10\% at 30~GeV. 
The $\pi^0$ energy resolution is around 7\%, independent of 
energy. Muons are identified using the hadron calorimeter
and muon chambers; only muons above 3~GeV/$c$ momentum
penetrate the system, and these are detected with an
average efficiency of 86\%. 

\subsubsection{DELPHI}

A pivotal feature of the DELPHI~\cite{bib-DELPHI} detector
is the particle identification capability of its ring imaging
Cherenkov detectors (RICH). These cover both the barrel and
end cap regions, and have both liquid and gas radiators. When combined 
with ionization energy loss information from the tracking detectors, 
the system gives clear identification of charged particles 
over the whole momentum range 
at LEP 1. The tracking detectors of DELPHI operate in a magnetic 
field of 1.23~T and consist of a silicon vertex detector,
an inner drift chamber, a TPC and an outer detector of drift
tubes. In addition there are two forward chambers, containing
planes of drift tubes, to improve reconstruction at low
polar angles.   
The system gives a momentum resolution $\sigma(p)/p$ 
of 0.7\% to 1.4\% in the barrel region, varying with
track momentum.
Measurement of K$^0_{\rm s}$ and $\Lambda$ in multihadron
events has an efficiency of 30--40\% over a wide momentum range. 
The silicon vertex detector allows track impact parameters
to be measured with an accuracy which depends on momentum
and polar angle, and which 
varies between 30 and 130~$\mu$m in $r\phi$
and between 40 and 200~$\mu$m in $z$.  
DELPHI's electromagnetic
calorimeters consist of high-density projection chambers 
in the barrel 
region and lead glass blocks in the endcap regions. They are preceded
by $0.8/\sin \theta$ radiation lengths of material in the 
barrel region, and more in the endcap regions, so that resolution
is somewhat degraded. When combined with $dE/dx$ information, the
system allows electron identification with efficiency and 
purity values both around 50\% over a wide momentum range. Photons
are identified both in the calorimeters and by measuring
conversion e$^+$e$^-$ pairs in the TPC, allowing 
reconstruction of $\pi^0$ mesons. 
Muons above 3~GeV/$c$ are detected using a hadron calorimeter
and muon chambers with an efficiency between 75 and 85\%.

\subsubsection{L3}

The L3 detector~\cite{bib-L3} was designed with the primary aim
of reconstructing electrons, muons and photons. It is therefore 
more limited than the other detectors in its capabilities for 
studying hadronization. L3 has a low magnetic field of 0.5~T 
in a large cylindrical volume of diameter 12~m.
The central time expansion chamber measures 
tracks out to a radius of 31.7~cm 
with a high spatial resolution
in the $r\phi$ plane. It is supplemented by a $z$-chamber to
measure the polar angle of tracks. 
The L3 arrangement gives optimized momentum resolution 
for penetrating muons, with $\sigma(p)/p \approx 2.5\%$
for 45~GeV/c muons. 
Its bismuth germanium oxide (BGO) electromagnetic calorimeter
is preceded by less than 10\% of a radiation length
in the barrel region, and has a
spatial resolution better 2~mm above 2~GeV. The energy
resolution is about 5\% at 100~MeV and 1.4\% at 45~GeV.
The calorimeter permits electron
identification, with only 0.1\% probability of misidentifying 
a hadron. It is also well suited for
measurements of inclusive $\pi^0$, $\eta$ and $\eta'$ 
production. 
  
\subsubsection{OPAL}

The OPAL~\cite{bib-OPAL} detector has a warm solenoid providing
a magnetic field of 0.435~T. The  
main central tracking jet chamber lies outside of
a silicon microvertex detector and a precision 
vertex drift chamber; a set of $z$-chambers around the 
jet chamber give precise measurement of track polar angles.
The combination of the tracking chambers gives a 
momentum resolution of $\sigma(p_T)/p_T \approx 
\sqrt {0.02^2+(0.0015 p_T)^2}$ with $p_T$ in GeV/$c$.
Efficiency for reconstruction of K$^0_{\rm s}$ and $\Lambda$
particles varies with momentum, having a maximum value
of 30\% at about 5~GeV/$c$.
The silicon detector, orginally an $r\phi$ device but
improved in 1993 to also measure $z$, gives an impact
parameter resolution of 15~$\mu$m in $r\phi$ for high
momentum tracks. 
The jet chamber
measures ionization energy loss, $dE/dx$, of 
tracks in multihadronic events with a resolution of 3.8\%,
allowing excellent identification efficiency and high 
purity for electrons, pions, kaons and protons 
over almost the whole momentum range at LEP 1. 
OPAL's lead glass electromagnetic calorimeter, which
is preceded by some two radiation lengths of material, 
has an energy resolution varying with energy from 1 to 5\%.
The muon chambers, together with the instrumented flux 
return of the hadron calorimeter,
are highly efficient for identification
of muons with momentum greater than 3~GeV/$c$.

\subsubsection{SLD}

The SLD~\cite{bib-SLD} experiment, like DELPHI, makes use of
Cherenkov ring imaging to identify charged particles, with a 
detector which uses both liquid and gas radiators to allow
coverage of a wide momentum range. Tracking is done by a 
central drift chamber within a 0.6~T magnetic field which
gives a momentum resolution, 
$\sigma(p_T)/p_T^2 = \sqrt {0.005^2+(0.01/p_T)^2}$, with
$p_T$ in GeV/$c$.
For electromagnetic
calorimetry, the SLD uses a liquid argon device, inside
the magnet coil, which gives a
resolution of around 15\%/$\sqrt {E}$ (with $E$ in GeV). The SLD has 
silicon charge-coupled-device pixel detectors for 
microvertex measurements. The pixels are 22 micron square, and 
the setup covers radii from 3 to 4~cm from the interaction vertex. 
Resolution on track impact parameter in $r\phi$ 
is in the range 11 to 20~$\mu$m, depending on track momentum
and polar angle. Muons are identified by layers of streamer
tubes between the slabs of iron which make up SLD's 
warm iron calorimeter.

\section{Measurements of inclusive single identified particles}

Inclusive single identified particles are usually studied in 
terms of their fractional energy ($x_E$) or momentum ($x_p$)
relative to that of the beams, with the fragmentation functions
being reported as 
$(1/\sigma_{\rm h})\d \sigma / \d x$. Here $\sigma_{\rm h}$ is the total
cross section for \eetoZtoh. 
Its inclusion is experimentally advantageous since it 
obviates the need to measure absolute cross sections, so
reducing systematic errors: 
$\d \sigma / \sigma_{\rm h}$
is simply calculated as the number of particles, $\Delta N$, in 
bin $\d x$ 
relative to the total number, $N_{\rm tot}$, of Z$^0$ 
hadronic decays. 
In a real measurement the bin width is finite, $\Delta x$,
and the differential cross section is taken as 
$1/N_{\rm tot}\times\Delta N / \Delta x$. Because fragmentation
functions vary rapidly with $x$, care has to be
taken for large values of $\Delta x$ in
interpreting a measurement as a differential cross section at
some particular value of $x$~\cite{bib-xlw}. 

Total inclusive particle yields, or average multiplicities
per hadronic Z$^0$ decay, are obtained by
integrating the measured fragmentation functions 
and extrapolating
into any unmeasured regions of $x$ with the aid of one or more 
models or interpolation functions. Systematic errors are included
to account for uncertainties in this procedure.  

Fragmentation functions and total inclusive yields have been
measured for an impressively large number of particle species
at LEP. The string and cluster models, as implemented in JETSET and HERWIG, 
are usually confronted with the data. 
A comprehensive compilation of inclusive
particle production data in \ee\ annihilation at all available
CoM energies above the $\Upsilon$ mass, as of mid-1995, 
may be found in~\cite{bib-datarev} where measured fragmentation
functions are plotted along with curves obtained from 
JETSET version 7.4. A more recent review~\cite{bib-Boehrer} contains
a good summary of measurements published after~\cite{bib-datarev}.

\subsection{Overall inclusive rates}
\label{smult}

Tables~4 and~5 list
of all the measured inclusive yields of mesons and baryons
published to date.
Where an experiment has reported more than one measurement, 
only the most recent is taken.
For each measurement of a particular particle, statistical and systematic
errors have been combined in quadrature; then the weighted mean 
of the available measurements has been calculated to give the
results shown in the tables (no attempt has been made to take into account
systematic errors correlated between experiments).
Yields reported over a restricted $x$ range are given separately.
 
In general there is very good agreement among the 
measurements of the different experiments. In only two cases,
where the measurements are listed separately in the tables, is there evidence 
of disagreement: the DELPHI and OPAL measurements of 
the $\Delta(1232)^{\pm \pm}$ 
are possibly inconsistent, and the 
$\Xi(1530)^0$ yield reported by DELPHI does not agree well
with the numbers given by ALEPH and OPAL.

\fulltable{Average measured charged particle and identified meson 
multiplicities in Z$^0$ decay together with the rates from 
Monte Carlo models. The letters ADLMO indicate the contributing  
experiments. Where appropriate, the rates always include both
particle and antiparticle.}
\br
Particle& Multiplicity& HERWIG59& JETSET74& UCLA74& Comments \\ 
\mr
Charged       &$20.96\pm0.18$       & 20.40 & 20.95 & 20.88 & 
ADLMO~\cite{bib-mult1,bib-mult2,rmult2,rxiexp2,bib-mult3} \\
$\pi^+$       &$17.05\0\0\pm0.43$   & 16.62 & 16.95 & 17.04 &
O~\cite{bib-Ocharged} \\
$\pi^0$         &$\09.39\0\0\pm0.44$    & 10.15 & \09.59 & \09.61 &  
ADL~\cite{rAxpi0, bib-DELpi0, bib-L3pi0} \\
$\eta$          &$\00.282\0\pm0.022$  & \00.246 & \00.286 & \00.232 &
A~\cite{rAmega} $x_E>0.1$  \\
                &$\00.93\0\0\pm0.09$    & \00.92 & \01.00 & \00.78 &
L~\cite{bib-L3eta} \\
$\rho(770)^0$   &$\01.29\0\0\pm0.12$    & \01.12 & \01.50 & \01.17 &
AD~\cite{bib-AVM, bib-Dlight} \\
$\omega(782)$   &$\01.11\0\0\pm0.11$    & \01.05 & \01.35 & \01.01 &
AL~\cite{bib-AVM, bib-L3ometa}  \\
$\eta'(958)$    &$\00.064\0\pm0.014$    & \00.071 & {\bf\00.127} & \00.061 &
A~\cite{rAmega} $x_E>0.1$  \\
                &$\00.25\0\0\pm0.04\0$  & \00.143 & \00.297 & {\bf\00.121} &
L~\cite{bib-L3ometa} \\
f$_0(980)$      &$\00.098\0\pm0.016$  & \00.068 & \0--- & \0--- &
D~\cite{bib-Dlight} $x_E>0.06$   \\
$\phi(1020)$    &$\00.108\0\pm0.005$  & {\bf\00.181} & {\bf\00.194} &
{\bf\00.132} & ADO~\cite{bib-AVM, bib-DEL96, bib-OVTM}  \\
f$_2(1270)$     &$\00.170\0\pm0.043$  & \00.137 & \0--- & \0--- &
D~\cite{bib-Dlight} $x_E>0.05$ \\
f$_2'(1525)$    &$\00.020\0\pm0.008$  & \00.021 & \0--- & \0--- &
D~\cite{bib-f2'} \\
\mr
K$^+$         &$\02.37\0\0\pm0.11$    & \02.08 & \02.30 & \02.24 & 
DO~\cite{bib-Dcharged, bib-Ocharged}    \\
K$^0$           &$\02.010\0\pm0.029$  & {\bf\01.87} & \02.07 & \02.06 & 
ADLO~\cite{bib-AK0, bib-Dlight, bib-L3pi0, bib-OK0} \\
K$^*(892)^+$  &$\00.714\0\pm0.044$  & {\bf\00.524} & {\bf\01.10} & \00.779 & 
ADO~\cite{rAmega, bib-Dlight, bib-OK*}   \\
K$^*(892)^0$  &$\00.759\0\pm0.032$  & {\bf\00.530} & {\bf\01.10} & \00.760 & 
ADO~\cite{bib-AVM, bib-DEL96, bib-OVTM} \\
K$_2^*(1430)^0$ &$\00.079\0\pm0.040$  & \00.067 & \0--- & \0--- &
D~\cite{bib-DEL96} \\ 
                &$\00.19\0\0\pm0.07$  & {\bf\00.054} & \0--- & \0--- &
O~\cite{bib-OVTM} $x_E<0.3$ \\
\mr
D$^+$         &$\00.187\0\pm0.014$  & \00.190 & \00.174 & \00.196 & 
ADO~\cite{bib-AD, bib-DELD, bib-Ocharm} \\
D$^0$           &$\00.462\0\pm0.026$  & \00.406 & \00.490 & \00.497 &
ADO~\cite{bib-AD, bib-DELD, bib-Ocharm} \\
D$^*(2010)^+$ &$\00.181\0\pm0.010$  & \00.151 & {\bf\00.242} & {\bf\00.227} & 
ADO~\cite{bib-AD, bib-DELD, bib-Ocharm}             \\
\mr
D$^0_{\rm s}$   &$\00.131\0\pm0.020$  & \00.087 & \00.129 & \00.130 &
O~\cite{bib-Ocharm}  \\
\mr
B$^*$           &$\00.28\0\0\pm0.03$  & {\bf\00.182} & \00.260 & \00.254 &
D~\cite{bib-DELB*}  \\
\mr
B$^{**}_{\rm u,d}$ &$\00.118\0\pm0.024$& {\bf\00.032} & \0---   & \0--- &
D~\cite{bib-DELB**} \\
\mr
J/$\psi$           &$\00.0054\pm0.0004$ & {\bf\00.0018} & \00.0050 & \00.0050 & 
ADLO~\cite{bib-AJpsi, bib-DJpsi, bib-LJpsi, bib-OJpsi} \\
$\psi(3685)$       &$\00.0023\pm0.0005$ & \00.00097 & \00.0019 & \00.0019 &
DO~\cite{bib-DJpsi, bib-OJpsi}   \\
$\chi_{{\mathrm c}1}$ &$\00.0086\pm0.0027$ & \00.00088 & \0--- & \0--- & 
DL~\cite{bib-DJpsi, bib-LJpsi}         \\
\mr
$\Upsilon$       &$\01.4\0\0\0\pm0.7\!\times\!10^{-4}$ &
{\boldmath$<1.0\!\times\!10^{-7}$} & {\boldmath$\02.2\!\times\!10^{-6}$} &
{\boldmath$\01.8\!\times\!10^{-6}$} &
O~\cite{bib-Oupsilon} $\Sigma$(3 lightest $\Upsilon$) \\
\br
\endtab

\fulltable{Measured baryon multiplicities in Z$^0$ decay together with the
rates from Monte Carlo models. The letters ADLO indicate the contributing
LEP experiments. Where appropriate, the rates always include both particle
and antiparticle.}
\br
Particle &Multiplicity &HERWIG59 &JETSET74 &UCLA74 & Comments \\
\mr
p                       &$0.98\0\0\pm0.09$  & {\bf1.41} & 1.19 & 1.09 & 
DO~\cite{bib-Dcharged, bib-Ocharged}              \\
\mr
$\Delta(1232)^{++}$     &$0.079\0\pm0.015$  & {\bf0.278} & {\bf0.189} &
{\bf0.139} & D~\cite{bib-DDELTA}  \\
                        &$0.22\0\0\pm0.06$  & 0.278 & 0.189 & 0.139 &
O~\cite{bib-ODELTA} \\
\mr
$\Lambda$               &$0.373\0\pm0.007$  & {\bf0.605} & 0.385 & {\bf0.332} & 
ADLO~\cite{bib-AK0, bib-Dlam, bib-L3pi0, bib-OPSB96} \\
$\Lambda(1520)$         &$0.0213\pm0.0028$& \0--- & \0--- & \0--- &
O~\cite{bib-OPSB96} \\
\mr
$\Sigma^+$            &$0.092\0\pm0.017$  & 0.123 & 0.073 & 0.061 & 
O~\cite{bib-OPsigma96} \\
$\Sigma^-$            &$0.084\0\pm0.017$  & 0.102 & 0.068 & 0.056 & 
O~\cite{bib-OPsigma96} \\
$\Sigma^++\Sigma^-$   &$0.174\0\pm0.021$  & 0.225 & 0.140 & 0.118 & 
DO~\cite{bib-DELB95, bib-OPsigma96} \\
$\Sigma^0$              &$0.074\0\pm0.009$  & 0.093  & 0.073 & 0.074 & 
ADO~\cite{rAmega, bib-DELB96, bib-OPsigma96} \\
$\Sigma^{\star+}+\Sigma^{\star-}$ &$0.0474\pm0.0024$ & {\bf0.202} &
{\bf0.074} & {\bf0.074} & ADO~\cite{rAmega, bib-DELB95, bib-OPSB96} \\
\mr
$\Xi^-$                 &$0.0265\pm0.0009$  & {\bf0.0746} & 0.0271 &
{\bf0.0220} & ADO~\cite{rAmega, bib-DELB95, bib-OPSB96} \\
$\Xi(1530)^0$           &$0.0072\pm0.0007$  & {\bf0.0352} & 0.0053 &
0.0081 & A~\cite{rAmega} \\                        
                        &$0.0041\pm0.0006$  & {\bf0.0352} & 0.0053 &
{\bf0.0081} & D~\cite{bib-DELB95} \\ 
                        &$0.0068\pm0.0007$  & {\bf0.0352} & 0.0053 &
0.0081 & O~\cite{bib-OPSB96} \\
\mr
$\Omega^-$              &$0.0012\pm0.0002$  & {\bf0.0093} & 0.00072 & 0.0011 & 
ADO~\cite{rAmega, bib-DELB96, bib-OPSB96}    \\
\mr
$\Lambda_{\rm c}^+$     &$0.078\0\pm0.017$  & 0.0129      & 0.059 &
{\bf0.026} & O~\cite{bib-Ocharm} \\
\br
\endtab

\subsection{Conclusions for Monte Carlo models}

In tables~4 and~5 the measured rates
are compared with the outcome of the three major Monte Carlo
models which attempt a full simulation of particle production.
The numbers in bold font show results which are more than three
standard deviations from the experimental measurements. 
In each case, the default versions of the programs have been
used, although for JETSET version 7.4 various sets of alternative
parameters have been suggested which improve the agreement with the 
overall rates. The most
recent HERWIG version 5.9 does not fit as well as version 5.8,
but a new default set of parameters will no doubt follow careful
comparisons with data. 

It is clear from the various JETSET tunings suggested by the 
four LEP experiments in~\cite{rlep2mc} that
there are strong correlations among the program's parameters, 
possibly such that there is no unique best set. To this extent 
JETSET may be underconstrained despite the large number of 
experimental measurements. DELPHI have 
published~\cite{rDtun2} comprehensive
sets of tuned parameters for various Monte Carlo models
(including ARIADNE and JETSET with matrix elements as well as with
parton showers)
in which they take account of event shape variables as well
as inclusive identified particle rates. This exercise is useful
but probably premature. Some of the recent
measurements differ significantly from those used in the tuning, and 
many have much reduced errors.
For example the $\Omega^-$ baryon is now known to be 
produced at a much lower rate than previously measured, and
it turns out the DELPHI tuned rate fits better with 
this new rate than with the one used as input; the same is true of
the $\phi(1020)$ meson rate. All of the models considered in~\cite{rDtun2}
describe the inclusive rates reasonably well, with the
exception of the performance of HERWIG in the baryon
sector. 

\subsubsection{Production of L=1 mesons}

Although the available meson measurements are predominantly of the 
$L=0$ pseudoscalar and vector states, the presence of the 
$L=1$ mesons shows clearly their importance 
in the hadronization; this is confirmed also
in the baryon sector with the observation by OPAL 
of the $\Lambda(1520)$.
Other scalar, axial vector and tensor mesons,
together with orbitally excited baryons, must presumably also 
be produced, although large widths and small branching
ratios will make them difficult to measure. 
Many of the lighter particles are therefore decay products
of other hadrons, 
and care has to be taken in any interpretation of the 
data using only relative light particle rates. 
So far only HERWIG 5.9
includes by default the P and some D wave meson states, and as the 
tables show
their inclusion does not mar the agreement with data in the
u,d sector. However HERWIG does poorly with mesons 
containing s quarks as well as with baryons. 

The production of the light tensor mesons is discussed by DELPHI
in~\cite{bib-Dlight} where a comparison
is made of relative rates of tensor to
corresponding vector mesons. While the production 
ratio f$_2/\rho^0$ is
similar to that for f$'_2/\phi$, at about 20\%, there
is evidence for a lower K$_2^*$/K$^*$ ratio, in agreement
with results from hadroproduction experiments. 
This suggests an extra suppression of strange tensor mesons.
However OPAL~\cite{bib-OVTM} measures a larger rate
for K$_2^*$ and so the picture is not yet clear. 
Some evidence has also been reported~\cite{bib-Dlight} 
for a rise in the ratio f$_2/\rho^0$ with meson
momentum; in other words the fragmentation function of
the tensor meson may be harder than that
of the vector meson,
as generally expected of heavier hadrons~\cite{rbj}.
However, these are difficult resonances to
measure; they have large widths, large combinatorial
backgrounds and uncertainties in the resonance line
shapes. As usual, more results would help.  

Only one measurement of a scalar meson, the f$_0(980)$, has 
been reported~\cite{bib-Dlight}, with a ratio f$_0/\rho^0$ of
$0.14\pm0.03$. Thus if the f$_0$ is indeed a conventional 
$0^{++}$ meson
(see~\cite{rpdg} for a mini-review of the scalars) then
the scalar and tensor mesons are produced with similar
rates. But again the measurements are difficult, and other
studies~\cite{bib-AVM} of the inclusive $\pi^+\pi^-$ mass 
spectrum with higher statistics have failed to report
a measurement of the f$_0(980)$ because of systematic uncertainties. 
One should therefore be wary of too much interpretation 
of one measurement. 

The f$_0(980)$ and $a_0(980)$ have aroused interest~\cite{rfunny}
as potential probes of the Gribov confinement scenario~\cite{rgribov}. 
In this theory the QCD vacuum is likened to the intense QED
fields expected around super charged ions, $Z>180$ (or $>137$ for
point-like charges)~\cite{rpom}. This results in the production of 
spatially compact `novel
vacuum scalars', identified with the f$_0(980)$, which are expected to be
produced in relative isolation. Particular signatures include enhanced
production at central rapidities (with respect to the thrust axis) and
in low multiplicity events~\cite{rfunny}.

From a semiclassical point of view, the orbital angular momentum $l$
of a \qqbar\ pair from string fragmentation is given 
by $<p_T> \times d$ where $<p_T>$ is the
mean quark momentum transverse to the string and $d$ is
the size of the resulting hadron~\cite{rlund}. 
Typical values give $l \approx 0.05\hbar$ so that the rate for mesons 
with non-zero orbital angular momentum is expected
to be less than about 10\% of the rate for the 
corresonding L=0 mesons.
The large average quark-antiquark separation in radially excited
states also works to prevent production of these mesons in string
fragmentation. Although the $L=1$ mesons can be simulated in
JETSET at the correct rates by the adjustment of appropriate parameters, 
the large experimental rates are nevertheless a
problem for the basic model assumptions about production of hadrons
from strings. 

\subsubsection{Strangeness suppression}

Strangeness suppression is immediately apparent from both the meson and
baryon measurements given in the tables.
There is a large number of ways to determine, from the data,
values for $\gamma_{\rm s}$,
the quark-level strangeness suppression assumed in the string 
model.
Results are tabulated for example in~\cite{rlep2mc} where all
measurements agree on a value of 0.3, a result which accords 
with reasonable values for the strange quark mass~\cite{rpdg} and the
%
%
string energy density $\kappa$. Since the 
various methods in~\cite{rlep2mc} use both light and heavy 
quark states, this 
consistency suggests that the suppression occurs at the quark 
level, in agreement with the string-model assumption.
On the other hand the   
UCLA model also reproduces reasonably well the strange
particle rates, and previous versions of HERWIG have been
tuned to do so. And hadronization studies in ep collisions 
at HERA~\cite{bib-ssHERA} give a lower value of 
$\gamma_{\rm s} \approx 0.2$,
in apparent disagreement with the Z$^0$ decay measurements. 
Therefore it is fair to say that the data are not yet conclusive. 
A direct 
method to measure the strangeness suppression  
in \eetoZtoh\ has been proposed~\cite{bib-ssuppress}
which makes use of the electroweak forward-backward asymmetry
and which could possibly distinguish between quark-level and
hadron-level suppression. Recently SLD~\cite{SLDss} have
applied this method to their data on inclusive K$^{*0}$ 
and $\overline {{\rm K}^{*0}}$ production 
using 150k events, 
with the result $\gamma_{\rm s} = 0.26 \pm 0.12$.  

\subsubsection{Relative rates of vector and pseudoscalar mesons}
\label{sec-VtoP}

In the absence of mass effects, the ratio of direct
pseudoscalar to vector meson production may
depend simply on spin statistics, in which case the 
value P/(P+V) would be expected to be 3/4. However feed-down 
from decays is also important and 
may obscure the interpretation of the experimental results.
One approach to determine the underlying P/(P+V)
value is to tune the appropriate parameter in the
JETSET model, but this can only be
done within the limited knowledge available on 
production of the higher states. Alternatively, one
can use the measurements in the b and c sectors where 
vector to pseudoscalar mass differences are much smaller
and there are some hopes to measure the orbitally
excited states. The average ratio for 
primary B$^*$/(B+B$^*$) 
is found to be $0.75\pm0.04$~\cite{bib-DELB*,rbstar,rbstar1}, in
excellent agreement with simple spin counting. However
the picture is not so clear when charm is considered:
here the ratio D$^*$/(D+D$^*$) is measured~\cite{bib-AD} to be
$0.51\pm0.04$. The difficulty here arises from incomplete
knowledge of the production rate and decay modes of the 
orbitally excited D$^{**}$ states which are likely
to feed down into the D and D$^*$ production. The question
then arises as to why B$^{**}$ production does not 
similarly muddy the waters in the b sector; it may be
that production and decay rates of the four different
$J^P$ states of B$^{**}$ conspire to leave the value 
of P/(P+V) at 3/4. More measurements are needed before 
definite conclusions can be reached. 
 
\subsubsection{Baryon production}
\label{sbprod}

In the string model, baryon yields are
determined by many parameters. The overall baryon rates 
relative to mesons depend on the relative probability
to produce a diquark pair from the sea. Spin-1 diquarks
may be suppressed relative to spin-0 diquarks. The strangeness
suppression enters in a similar way as for meson 
production, but there is in addition the possibility of extra 
suppression of strangeness in a diquark. And the popcorn
mechanism may introduce one or more mesons locally in
phase space between a baryon and an antibaryon. 

Although JETSET does rather well, there are some
discrepancies with measured rates. Attempts by OPAL~\cite{bib-OPSB96}
to tune the parameters which control baryon production
have shown that it is not possible to reproduce
simultaneously all of the measured rates. 
The suggestion then is that the mechanisms for baryon production in 
the string model, and particularly for the strangeness suppression,
are deficient. 
However, as has been said, there is now clear experimental 
evidence also for orbitally excited states in baryon production. The
rate for the $J^P = {3\over 2}^-$ 
$\Lambda(1520)$, at 0.02 per hadronic Z$^0$ decay, is
around 5\% of that for the $J^P = {1\over 2}^+$ $\Lambda$.
And there are many similar baryon states which cannot be 
measured experimentally but which, it is fair to assume,
must be produced in the hadronization.
So since JETSET, like all of the other models, does not include
production of orbitally excited baryons, no clear 
conclusions can yet be reached.
 
Neither HERWIG nor UCLA,
both of which rely only on phase space, mass and spin,
are successful in the strange baryon sector, with
the former consistently overestimating the rates, and
the latter underestimating them. 
As presently implemented, baryon production in HERWIG does not take account
of the appropriate SU(6) Clebsch-Gordon coefficients and this will lead to
an overestimation of baryon production rates.
As with JETSET,
neither of these models includes the production of baryons other 
than the lowest lying $L=0$ states. In principle 
their inclusion would lower 
the predicted rates for primary low lying 
baryons, since some higher states would be produced in their
stead. However there would be a compensatory increase in the
rates due to feed down from decays. Since HERWIG consistently
overproduces and UCLA underproduces the baryons, it seems
reasonable to deduce that at least one of them is
incapable of being fixed up by this mechanism.

The conclusion then from the baryon yields is that the   
string model of JETSET, while certainly not perfect, is in 
reasonable accord with the measurements. Recent work on
the baryon production within the popcorn model has further 
improved the agreement~\cite{rpop2}.
Whether or not this is to be taken as a strong
endorsement of the model is however an arguable point, given
the large amount of freedom available to tune parameters 
in order to reproduce the observations. Both the HERWIG cluster
model and the UCLA string model are clearly in 
difficulty, and it remains to be seen whether they can
be rescued.

\subsubsection{Comparison to models of total yields}

A number of
models have been proposed to treat only the overall 
yields of identified particles. Such models are
necessarily of limited physical content, although they 
turn out to be reasonably successful in describing the 
inclusive rates. Why they do so is not at all clear. 

In the thermodynamic model~\cite{bib-Becattini} the source
of particles is assumed to be a hadron gas in thermal and
chemical equilibrium. The model has three parameters, a
temperature, a volume and a parameter to allow for 
incomplete strange chemical equilibrium (similar
to the strangeness suppression of the Lund string model). The
model gives a good fit to the LEP data, as well as to 
lower energy \ee\ annihilation data, with a temperature of
around 170~MeV, close to $\Lambda_{\rm QCD}$
and the temperature found in the earlier statistical model of
Hagedorn~\cite{rboot}.
The author
of the model speculates that the thermal equilibrium could
be a feature of the quark-hadron transition, brought 
about by strong interactions. 
This argument was also invoked by Fermi~\cite{rfermi} to justify his
phase space model for hadron production.
However it is difficult to 
reconcile this picture with the conventional view of
hadronization in \ee\ annihilation as occuring locally in the wake
of rapidly separating colour sources.

A ``striking regularity''~\cite{bib-Chliapnikov} has been noted
in the particle yields, and a simple formula proposed which 
reproduces well the observations (apart from the pions and
possibly the $\Omega^-$ baryon):
$N = (2J+1)/(I_m+1)\times a\exp{(-bM^2)}$. Here, 
$N$ is the yield for a particle of spin $J$ and mass $M$, and
$I_m$ is the isospin for baryons and a ``modified'' isospin
for K and K* mesons and isosinglet pseudoscalar mesons. 
The introduction of the ``modified''
isospin appears rather ad hoc, although there are 
plausible arguments to justify it. 
The model makes no attempt to explain the yield differences between
members of the same isomultiplet which, for example, are significant
for kaons.
The parameters 
$a$ and $b$ are fitted to the data,
and the slope parameter $b$, at about 3.9~GeV$^{-2}$, is found
to be the same for LEP as for lower energy measurements,
implying that the regularity may be universal.   
It is unclear as to the physical origin of the
expression, and it will be interesting to see how the formula 
copes with future measurements. 
This observed regularity has a less successful predecessor of the form
$N=(2J+1)\times a\exp(-bM)$~\cite{rhof}, which was applied
to lower energy data.

Another model~\cite{bib-YiPei} of the total yields is based
on string fragmentation, and proposes a simple formula with
only three parameters: a strangeness suppression, an effective
temperature and a relative normalization factor between mesons
and baryons. Following the string model, the rate $N$ of light 
meson and baryon production is taken as 
$N=(C/C_B)\times(2J+1)\times(\gamma_{\rm s})^{N_{\rm s}}
\times\exp{(-E_{\rm {bind}}/T)}$.
The normalization $C$ depends on the centre-of-mass energy, $C_B$ is
a relative suppression of baryons, $J$ is the particle spin,
$\gamma_{\rm s}^{N_{\rm s}}$ gives the suppression for a hadron containing
$N_{\rm s}$ strange quarks, $E_{\rm {bind}}$ is the hadron binding energy
and $T$ is the effective temperature.
The model gives a good simultaneous fit to LEP and lower energy 
data, with a temperature of $298\pm15$~MeV and a strangeness suppression 
$\gamma_{\rm s}$ of $0.29\pm0.02$. The model also gives a
good description of heavy flavour production. Its predictions for production
rates of excited charm states have recently been shown to agree
with OPAL measurements~\cite{OPALD**}.  

\subsection{Rates for heavy quarkonia}

Due to their narrow widths and the availability of clean leptonic decay
channels the principle heavy quarkonium 
states measured are the $J/\psi$ and $\Upsilon$(1S,
2S,3S), based upon which further excitations can be 
reconstructed~\cite{bib-AJpsi,bib-DJpsi,bib-LJpsi,bib-OJpsi, 
bib-Oupsilon,rAupsilon}. The
production of these heavy Q$\overline{\rm Q}$ bound states is 
thought to be rather
atypical of hadron production in general, especially for the $\Upsilon$,
due to the significant part
played by perturbative physics.

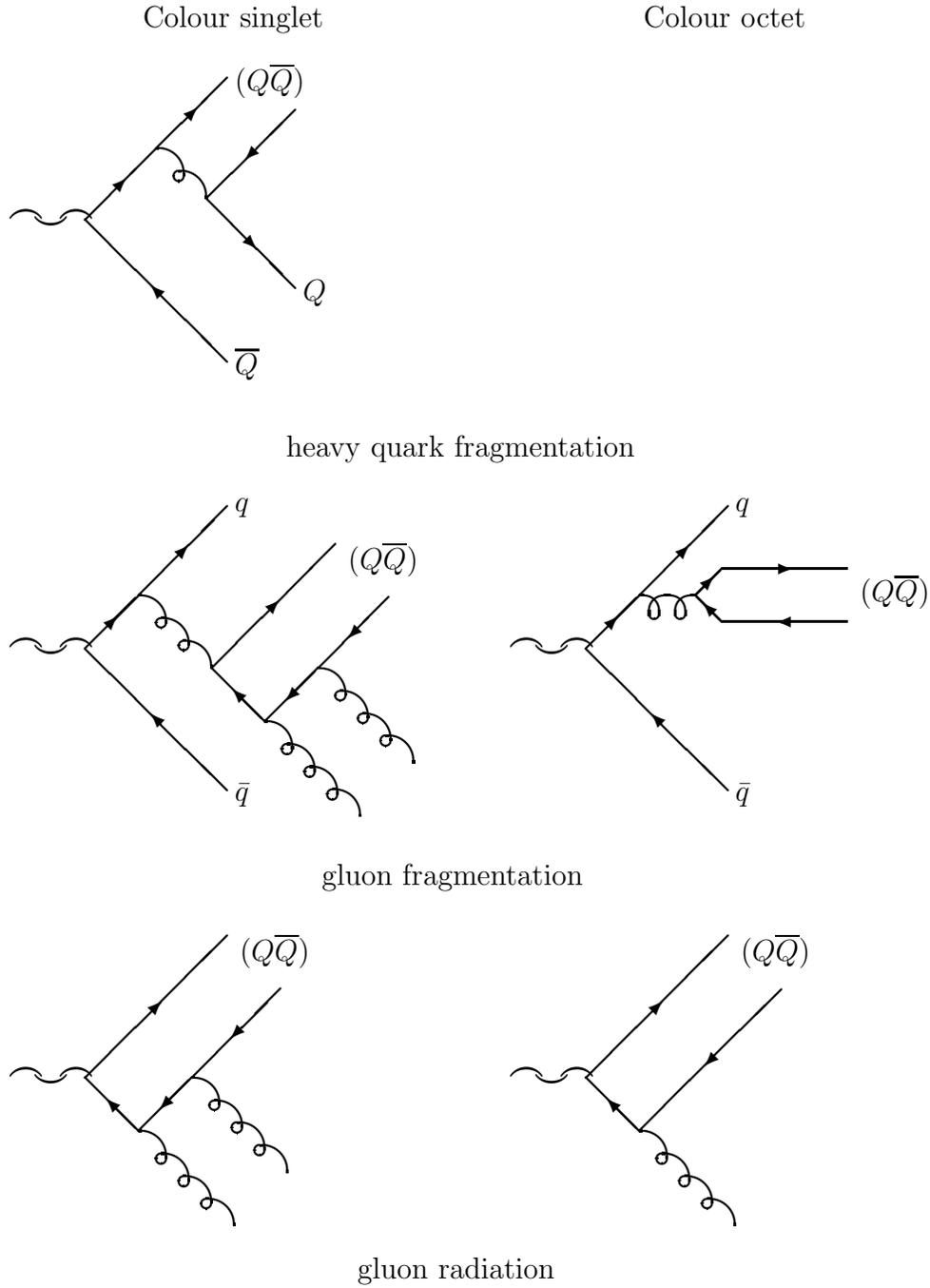
\begin{figure}
\unitlength 1mm
\begin{picture}(130,180)
\put(20,175){Colour singlet}
\put(90,175){Colour octet}
\put(0,120){
\Feynmanlength
\begin{picture}(16000,16000)
\THICKLINES
\bigphotons
\drawline\photon[\E\REG](0,8000)[3]
\drawline\fermion[\SE\REG](\pbackx,\pbacky)[8000]
\drawarrow[\NW\ATBASE](\pmidx,\pmidy)
\global\advance\pbackx by  300
\global\advance\pbacky by -500
\put(\pbackx,\pbacky){$\overline Q$}
\drawline\fermion[\NE\REG](\pfrontx,\pfronty)[4000]
\drawarrow[\NE\ATBASE](\pmidx,\pmidy)
\drawline\fermion[\NE\REG](\pbackx,\pbacky)[4000]
\drawarrow[\NE\ATBASE](\pmidx,\pmidy)
\global\advance\pbackx by  300
\global\advance\pbacky by -500
\put(\pbackx,\pbacky){$(Q\overline Q)$}
\drawline\gluon[\SE\REG](\pfrontx,\pfronty)[1]
\drawline\fermion[\NE\REG](\pbackx,\pbacky)[5000]
\drawarrow[\SW\ATBASE](\pmidx,\pmidy)
\drawline\fermion[\SE\REG](\pfrontx,\pfronty)[5000]
\drawarrow[\SE\ATBASE](\pmidx,\pmidy)
\global\advance\pbackx by  300
\global\advance\pbacky by -500
\put(\pbackx,\pbacky){$Q$}
\end{picture}}
\put(40,115){heavy quark fragmentation}
\put(70,60){
\Feynmanlength
\begin{picture}(16000,16000)
\THICKLINES
\bigphotons
\drawline\photon[\E\REG](0,8000)[3]
\drawline\fermion[\SE\REG](\pbackx,\pbacky)[8000]
\drawarrow[\NW\ATBASE](\pmidx,\pmidy)
\global\advance\pbackx by  300
\global\advance\pbacky by -400
\put(\pbackx,\pbacky){$\bar q$}
\drawline\fermion[\NE\REG](\pfrontx,\pfronty)[3000]
\drawarrow[\NE\ATBASE](\pmidx,\pmidy)
\drawline\fermion[\NE\REG](\pbackx,\pbacky)[5000]
\drawarrow[\NE\ATBASE](\pmidx,\pmidy)
\global\advance\pbackx by  300
\global\advance\pbacky by -200
\put(\pbackx,\pbacky){$q$}
\drawline\gluon[\E\REG](\pfrontx,\pfronty)[2]
\drawline\fermion[\NE\REG](\pbackx,\pbacky)[1500]
\drawarrow[\NE\ATBASE](\pmidx,\pmidy)
\drawline\fermion[\E\REG](\pbackx,\pbacky)[5000]
\drawarrow[\E\ATBASE](\pmidx,\pmidy)
\global\advance\pbackx by   500
\global\advance\pbacky by -1300
\put(\pbackx,\pbacky){$(Q\overline Q)$}
\drawline\fermion[\SE\REG](\gluonbackx,\gluonbacky)[1500]
\drawarrow[\NW\ATBASE](\pmidx,\pmidy)
\drawline\fermion[\E\REG](\pbackx,\pbacky)[5000]
\drawarrow[\W\ATBASE](\pmidx,\pmidy)
\end{picture}}
\put(0,60){
\Feynmanlength
\begin{picture}(16000,16000)
\THICKLINES
\bigphotons
\drawline\photon[\E\REG](0,8000)[3]
\drawline\fermion[\SE\REG](\pbackx,\pbacky)[8000]
\drawarrow[\NW\ATBASE](\pmidx,\pmidy)
\global\advance\pbackx by  300
\global\advance\pbacky by -400
\put(\pbackx,\pbacky){$\bar q$}
\drawline\fermion[\NE\REG](\pfrontx,\pfronty)[3000]
\drawarrow[\NE\ATBASE](\pmidx,\pmidy)
\drawline\fermion[\NE\REG](\pbackx,\pbacky)[5000]
\drawarrow[\NE\ATBASE](\pmidx,\pmidy)
\global\advance\pbackx by  300
\global\advance\pbacky by -200
\put(\pbackx,\pbacky){$q$}
\drawline\gluon[\SE\REG](\pfrontx,\pfronty)[2]
\drawline\fermion[\NE\REG](\pbackx,\pbacky)[7000]
\drawarrow[\NE\ATBASE](\pmidx,\pmidy)
\global\advance\pbackx by  500
\global\advance\pbacky by -1000
\put(\pbackx,\pbacky){$(Q\overline Q)$}
\drawline\fermion[\SE\REG](\pfrontx,\pfronty)[3000]
\drawarrow[\NW\ATBASE](\pmidx,\pmidy)
\drawline\gluon[\SE\REG](\pbackx,\pbacky)[3]
\drawline\fermion[\NE\REG](\pfrontx,\pfronty)[3000]
\drawarrow[\SW\ATBASE](\pmidx,\pmidy)
\drawline\gluon[\SE\REG](\pbackx,\pbacky)[3]
\drawline\fermion[\NE\REG](\pfrontx,\pfronty)[4000]
\drawarrow[\SW\ATBASE](\pmidx,\pmidy)
\end{picture}}
\put(45,55){gluon fragmentation}
\put(70,0){
\Feynmanlength
\begin{picture}(16000,16000)
\THICKLINES
\bigphotons
\drawline\photon[\E\REG](0,8000)[3]
\drawline\fermion[\NE\REG](\pbackx,\pbacky)[8000]
\drawarrow[\NE\ATBASE](\pmidx,\pmidy)
\global\advance\pbackx by  500
\global\advance\pbacky by -1000
\put(\pbackx,\pbacky){$(Q\overline Q)$}
\drawline\fermion[\SE\REG](\pfrontx,\pfronty)[3000]
\drawarrow[\NW\ATBASE](\pmidx,\pmidy)
\drawline\gluon[\SE\REG](\pbackx,\pbacky)[3]
\drawline\fermion[\NE\REG](\pfrontx,\pfronty)[8000]
\drawarrow[\SW\ATBASE](\pmidx,\pmidy)
\end{picture}}
\put(0,0){
\Feynmanlength
\begin{picture}(16000,16000)
\THICKLINES
\bigphotons
\drawline\photon[\E\REG](0,8000)[3]
\drawline\fermion[\NE\REG](\pbackx,\pbacky)[8000]
\drawarrow[\NE\ATBASE](\pmidx,\pmidy)
\global\advance\pbackx by  500
\global\advance\pbacky by -1000
\put(\pbackx,\pbacky){$(Q\overline Q)$}
\drawline\fermion[\SE\REG](\pfrontx,\pfronty)[3000]
\drawarrow[\NW\ATBASE](\pmidx,\pmidy)
\drawline\gluon[\SE\REG](\pbackx,\pbacky)[3]
\drawline\fermion[\NE\REG](\pfrontx,\pfronty)[3000]
\drawarrow[\SW\ATBASE](\pmidx,\pmidy)
\drawline\gluon[\SE\REG](\pbackx,\pbacky)[3]
\drawline\fermion[\NE\REG](\pfrontx,\pfronty)[5000]
\drawarrow[\SW\ATBASE](\pmidx,\pmidy)
\end{picture}}
\put(50,0){gluon radiation}
\end{picture}
\caption{The leading production mechanisms for heavy quarkonia assuming the
colour singlet and octet mechanisms. \label{fonia}}
\end{figure}

In the case of charmonium the dominant production mechanism is expected
to be weak b hadron decays: b$\to $c$+(\bar {\rm c}$s) plus subsequent
colour rearrangement. In the case of B$_{\rm u,d}$ mesons the J$/\psi$
branching ratios ($\approx$1.15\%) have been previously measured at the
$\Upsilon(4S)$~\cite{rjprev}, so allowing reliable predictions for
charmonium rates at the Z$^0$. (The presence of B$_{\rm s}$ and b baryons at
the Z$^0$ makes little difference to the inclusive b hadron branching
ratio~\cite{bib-DJpsi}). Perturbative fragmentation contributes only at
the few percent level to charmonium production~\cite{rqfrag,rgfrag}, as is 
indeed
observed~\cite{rjprom}, but 100\% to bottomonium production. Three basic
pQCD production mechanisms, illustrated in \fref{fonia}, 
are considered~\cite{rcho}:
\begin{itemize}
\item
Heavy quark fragmentation~\cite{rqfrag}, Z$^0\to (\Upsilon)$b$\bar{\rm b}$.
Here the production of a primary b$\bar{\rm b}$ pair is followed by the
radiation of a gluon which splits into a second b$\bar{\rm b}$; a b and
$\bar {\rm b}$ from these two pairs then bind in a colour singlet system.
It is noteworthy that in addition to the quarkonium two other heavy hadrons
occur in this process.
\item
Gluon fragmentation~\cite{rgfrag}, Z$^0\to$q$\bar{\rm q}$g$^\star,\;$g$^\star
\to($b$\bar{\rm b})$gg. The need to remove the colour from the b$\bar{\rm b}$
pair, whilst forming a positive $C$ parity $\Upsilon$ (or $J/\psi$ in the
case of c$\bar{\rm c}$), requires the emission of two perturbative gluons.
Thus this is an order-$\alpha^4_s$ process and typically the quarkonium state
has a relatively large transverse momentum.
\item
Gluon radiation~\cite{rgrad}, Z$^0\to ($b$\bar{\rm b})$gg. Here two gluons
are emitted from a primary heavy (anti)quark allowing it to recoil and form
a colour singlet with the other (anti)quark. This results in a very hard,
isolated, quarkonium state.
\end{itemize}
All of the above prompt production mechanisms give comparatively isolated
quarkonia originating from the interaction point, in contrast to charmonium
from b decays. Here the b$\bar{\rm b}$ system is produced in a colour singlet
state by the emission of perturbative gluons. The dominant process is quark
fragmentation~\cite{rcho}. However the theoretical predictions are
significantly low compared to the Z$^0$ data. Also no evidence for displaced
vertices, associated with additional heavy quarks, is observed in quarkonium
events. A similar situation has occurred at the TEVATRON where colour singlet
fragmentation mechanisms fail to account for the number of observed 
high-$p_T$ quarkonium states~\cite{rcdf,rcdf8}.

Fortunately recent theoretical developments suggest that colour octet
quarkonium production may play an important role~\cite{roctet}. Here the
Q$\overline{\rm Q}$ forms a colour octet system from which the colour is 
leached away (by exchange of soft gluons) in an, as yet unspecified,
non-perturbative mechanism. Two new contributions, also shown in
\fref{fonia}, arise: gluon fragmentation Z$^0\to$q$\bar{\rm q}$g$^\star,
\;$g$^\star\to($b$\bar{\rm b})$; and gluon radiation Z$^0\to$b$\bar{\rm
b}$g. These new diagrams are lower order in $\alpha_s$ but suppressed by
larger powers of $v$, the relative velocity of the b and $\bar{\rm b}$.
To calculate these processes requires knowledge of the octet matrix
elements; until recently these have been taken from fits to the CDF
data~\cite{rcdf8} though now a potential model based calculation is
available \cite{rocpot}. ({\it Ab initio} lattice calculations have so
far only been performed for octet decay matrix elements~\cite{rlatt} and
not the technically demanding production matrix elements). In the the
octet case the dominant fragmentation contribution to $\Upsilon$
production becomes gluon
fragmentation~\cite{rcho}. Taking into account all contributions, agreement
with the Z$^0$ data is possible. Although the data are insufficient
to be able to isolate components due to the individual processes, the lack
of an observed hard $\Upsilon$ spectrum does rule out a large contribution
from octet gluon radiation~\cite{rjprom}. It is also noteworthy that the
octet mechanism predicts a large transverse polarization for the vector
quarkonium states~\cite{rocpot,rjspn} particularly so at high momentum.

Looking at the J$/\psi$ data in table~4, one can conclude that
since b hadron decays dominate charmonium production, HERWIG is deficient in
this aspect of its b decay description. Possible remedies include adding
more explicit decay channels to the decay tables or refining the partonic b
decay model, particularly in its treatment of light clusters. Next, because
none of the models considered include the dominant octet production
mechanism the large underestimates of $\Upsilon$~\cite{bib-Oupsilon,rAupsilon}
production are not too surprising. Only ARIADNE, version 4.09 onwards, 
been extended to include an approximation to the perturbative colour octet
production mechanism~\cite{rjmc}. This program is therefore the only one
containing the necessary physics to attempt to describe the $\Upsilon$ data.

\subsection{Semi-inclusive momentum spectra}
\label{semi}

Having discussed the total production rates of
identified particles, we
now turn to their momentum distributions: $\sigma^{-1}\d\sigma/\d x$.
All particle spectra vanish as $x\to1$ and are expected to vanish as $x\to
0$, though measurements are rarely available at sufficiently low momentum
to see indications of this behaviour. However changing to the commonly used
variable $\xi=-\ln x$, a clear `hump-backed plateau' shape (see \sref{shump})
is seen. 

\subsubsection{Light hadrons}
A compilation of many of the measurements of the $x$ and $\xi$ spectra 
of identified hadrons at the Z$^0$
can be found in~\cite{bib-datarev} and~\cite{bib-Boehrer}.
References to the more recent measurements are listed
in table~6. In addition, a recent measurement of the charged particle
momentum spectrum is available in~\cite{rscvioex2}.

\fulltable{A summary of momentum spectrum measurements that 
have appeared since~[86]. The letters ADL indicate the
contributing experiments.}
\br
\centre{6}{Mesons} \\
\ns
\mr
Particle & \centre{2}{Reference} & Particle & \centre{2}{Reference} \\
\ns
\mr
$\pi^0$             & A    &~\cite{rAxpi0} &
f$'_2(1520)$        & D    &~\cite{bib-f2'} \\
$\eta$              & A    &~\cite{rAmega} &
K$^0$               & A    &~\cite{rAmega} \\
$\omega(782)$       & L    &~\cite{bib-L3ometa} &
K$^\star(892)^\pm$  & A    &~\cite{rAmega} \\
$\eta'(958)$        & AL   &~\cite{rAmega,bib-L3ometa} &
K$^\star(892)^0$    & D    &~\cite{bib-DEL96} \\
$\phi(1020)$        & D    &~\cite{bib-DEL96} & 
K$^\star_2(1430)^0$ & D    &~\cite{bib-DEL96} \\
\mr
\ns
\centre{6}{Baryons} \\
\ns
\mr
Particle & \centre{2}{Reference} & Particle & \centre{2}{Reference} \\
\ns
\mr
$\Lambda$           & A      &~\cite{rAmega} &
$\Xi^-$             & A      &~\cite{rAmega} \\
$\Sigma(1385)^\pm$  & A      &~\cite{rAmega} &
$\Xi(1530)^0$       & A      &~\cite{rAmega} \\
\br
\endtab

In \sref{sfull} the calculation of the charged particle momentum spectrum
was discussed; these results can also be applied to identified hadron
spectra~\cite{rbible}. Especially in the case of identified hadrons a
dependence on the primary quark flavour is anticipated: Monte Carlo
studies and recent SLD results~\cite{SLDss}
suggest that the main effects are at high momentum but that
residual effects could also occur at low momentum. Consequently investigations
have again concentrated on the peak position, $\xi^\star$, which occurs
at low momentum and for which the complete prediction is of 
the form~\cite{rxi*,rbrum}:
\begin{equation}
\label{exi*}
\xi^\star=F\left[\ln\left(\frac{Q_0}{\Lambda}\right)\right]
+\frac{1}{2}\ln\left(\frac{Q}{\Lambda}\right)+\cdots
\end{equation}
Observe that  the cut-off, $Q_0$, dependence only occurs through the first
term which has the property that $F(0)=0$ whilst the scale, $Q$, dependence
only arises in the second term: the constant $\Lambda$ is the usual
(effective) QCD scale. In order to create a particle of a given mass it is
reasonable to expect that the final-state partons will require virtualities
of the same magnitude, so that $Q_0$ is a simple function of the hadron mass
$m_{\rm h}$~\cite{rmasdep}. Invoking the LPHD concept one can hope that
these parton level calculations are then sufficient to predict an
identified hadron's energy spectrum.

Application of \eref{exi*} has caused some confusion. First, there are
small differences between experiments in the practical definition of
$\xi^\star$. Second, it is common to set $Q_0=\Lambda$ so that the first
term vanishes, the limiting spectrum case, but then somewhat
inconsistently to introduce a new variable $Q'_0=\Lambda$ in the second
term. That is \eref{exi*} is replaced by:
\begin{equation}
\label{exi'}
\xi^\star=\frac{1}{2}\ln\left(\frac{Q}{Q'}\right)
\end{equation}

Given the availability of low energy data it is possible to investigate the
$\ln Q$ dependence of $\xi^\star$ for the $\pi^+,\;\pi^0$, K$^+$, 
K$^0_{\rm s}$,
p and $\Lambda$ hadrons separately~\cite{rAxpi,rbrum}. The data for each
hadron species appear to lie on straight lines each of slope $\approx 1/2$
but with differing offsets, indicating that $F$ in \eref{exi*} decreases
with increasing hadron mass. ALEPH also finds that the linearity of these
fits can be improved by using JETSET to remove the effects of secondary
hadron decays~\cite{rAxpi}. For example kaons produced in the weak decays
of hard B mesons stiffen the kaon spectrum causing a decrease in their
$\xi^\star$. However this correction procedure is somewhat at odds with
the original idea of LPHD which was thought to account for such decays. At
the Z$^0$ it is also possible to study the hadron mass dependence of
$\xi^\star$ with $Q=M_{\rm Z}$ fixed~\cite{bib-Ocharged,bib-Dcharged,rbrum},
which
probes the relationship between $Q^{(\prime)}_0$ and $m_{\rm h}$. Adopting
\eref{exi'}
OPAL~\cite{bib-Ocharged} claim two linear relationships between $\xi^\star$
and $m_{\rm h}$ (an exponential dependence of $Q'$ on $m_{\rm h}$) one
describing the mesons, except pions, and the other the baryons. This
pattern is well reproduced by JETSET allowing a correction to remove
secondary decays to be applied after which all the points, including pions,
now fall on a single curve. However DELPHI~\cite{bib-Dcharged} report that
mesons
and baryons lie on two curves corresponding to a negative logarithmic
dependence of $\xi^\star$ on $m_{\rm h}$ (linear relationship between $Q'$
and $m_{\rm h}$) which again reduce to a single curve after correcting back
to primary hadrons using JETSET.

Whilst the full shape of the momentum spectrum can be calculated within
pQCD~\cite{rbible,rlphd,rhump,rgaus} comparison is more commonly made to
Monte Carlo predictions, particularly so in the case of identified 
particle momentum
spectra, reflecting the expected interplay between perturbative and
non-perturbative contributions. From a practical point of view the primary
concern is with a correct description of the pion momentum spectra
\cite{bib-datarev,bib-Ocharged,rAxpi0,bib-L3pi0,rAxpi} which dominate the
total event multiplicity. These fall by five orders of magnitude over the
measured momentum range. Here, and in general, the predictions of ARIADNE
and JETSET are rather similar and typically agree with the data a little
better than HERWIG, though all suffer problems at very low and very high $x$.
Meson spectra are described reasonably well on the whole, though the
K$^\pm$ spectrum is notably too soft~\cite{bib-datarev,bib-Ocharged,
bib-Dcharged,rAxpi} which might be attributable to an inadequate 
description of heavy hadron decays. However the description for baryons
is less satisfactory. In particular the number of fast, predominantly
primary, protons is consistently overestimated~\cite{bib-Ocharged,
bib-Dcharged,rAxpi}. In the case of ARIADNE and JETSET this may be
remedied~\cite{rAmega,rDtun2} by invoking a suppression of leading
uds-baryons~\cite{rjetset}, which may have an underlying physics
motivation \cite{rpop2}.

An intriguing possibility is that of an inequality in the fragmentation
function of strange quarks into protons and antiprotons, $D^{\rm s}_{\rm p}
(x,Q^2)\neq D^{\rm s}_{\bar{\rm p}}(x,Q^2)$~\cite{rassym}. This prediction
follows from a possible asymmetry, of non-perturbative origin, in the sea
quark structure functions of protons~\cite{rsea} and application of the
`reciprocity rule'~\cite{repro}. A test of this idea may be possible by SLD
using a strangeness tag~\cite{rlead} and exploiting their polarization
asymmetry to distinguish the quark from the antiquark jet.

\subsubsection{Heavy quark hadrons}
\label{sQfrag}

Hadrons containing heavy (c,b) quarks are special since the heavy quark is
expected to be principally (if not exclusively) of perturbative origin. In
practice non-perturbative physics also plays a role and the delineation of
the two contributions is not clear cut. If the fragmentation function can be
written as a convolution of perturbative ($PT$) and non-perturbative ($NP$)
parts, $D(x)=d_{PT}\otimes d_{NP}$, then it immediately follows that $\langle
x\rangle=\langle x\rangle_{PT}\cdot\langle x\rangle_{NP}$. 
In leading order pQCD predicts~\cite{rzqlo} 
$\langle x\rangle_{PT}=(\alpha_s(E_{\rm {jet}})/
\alpha_s(m_{\rm Q}))^{8/(9\pi\beta_0)}$; the NLO calculation 
of $d_{PT}(x)$~\cite{rzqnlo} is sharply forward-peaked, vanishing 
at $x=1$. The
perturbative result alone is too hard ($\langle x_b\rangle_{PT}|_{E=M_Z/2}
\approx0.8$) and a non-perturbative component is required, especially in
those rare cases (of order 1\%) where no gluon radiation occurs at all.

A simple argument~\cite{rbj} shows that the non-perturbative hadronization
should also be very hard for heavy quarks due to their inertia, $\langle x
\rangle_{NP}\approx 1-R^{-1}/m_{\rm Q}$, where $R$ is a typical hadron size,
with the remaining energy uniformly
distributed in rapidity between $\pm\ln(\sqrt s/2m_{\rm Q})$. A full
expression for $d_{NP}(x)$ depends on the details of the hadronization
mechanism assumed, and several are available.

The most commonly adopted standard is due to Peterson et al~\cite{rpet}. In
an independent fragmentation approach, the amplitude for the transition
Q$\to($Q$\bar {\rm q}')+$q$'$ in perturbation theory is proportional to the
inverse of the energy transfer, assuming a constant matrix element. The
fragmentation function is then given by the amplitude squared and the
appropriate flux factor as:
\begin{equation}
\label{epeter}
\fl D^{\rm Q}_{\rm h}(x)=\frac{N}{x}\left(1-\frac{1}{x}
-\frac{\epsilon_{\rm Q}}{1-x}\right)^{-2}\stackrel{x\to1}{\longrightarrow}N
\frac{(1-x)^2}{\epsilon^2_{\rm Q}}\hspace{10mm}N\approx\frac{4
\sqrt{\epsilon_{\rm Q}}}{\pi}
\end{equation}
Theoretically one predicts $\epsilon_{\rm Q}=R^{-2}/m_{\rm Q}^2$; experimental
fits yield small $\epsilon$ values with $\epsilon_{\rm c}/\epsilon_{\rm b}
\sim10$, consistent
with $(m_{\rm b}/m_{\rm c})^2$~\cite{rQfComp,rQfA,rQfD,rQfL,rQfO,rQfS}. 
On average, $\langle x\rangle=1-
\sqrt{\epsilon_{\rm Q}}\approx 1-R^{-1}/m_{\rm Q}$ as anticipated above.

A perceived problem with the Peterson form is its $\sim(1-x)^2$ behaviour (it
is sometimes argued that including gluon radiation is equivalent to an
additional $(1-x)$ factor) as $x\to 1$ which is in conflict with the
`reciprocity' rule~\cite{repro}. This posits that the $x\to1$ behaviour of
the Q$\to$h fragmentation function should equal the $x\to1$ behaviour of the
h$\to$Q structure function $F^{\rm Q}_{\rm h}(x,Q^2)$. Furthermore it is
expected from dimensional counting arguments~\cite{rdimc} that as $x\to1$,
$F^{\rm q}_{\rm h}(x,Q^2)\sim(1-x)^{2n_s-1}$ where $n_s$ is the number of
spectators in the hadron H. In a meson $n_s=1$ and therefore the $x\to1$
behaviour of the fragmentation function should be $(1-x)$. Two alternatives
are available. By adopting an explicit meson wavefunction~\cite{rcol} and
thereby introducing a non-trivial matrix element, a `refined' Peterson form
may be derived:
\begin{equation}
\label{ecoli}
\fl D^{\rm Q}_{\rm h}(x)=N(1+x^2)\!\left(\frac{1-x}{x}+\frac{\epsilon(2-x)}
{1-x}\right)\!\left(1-\frac{1}{x}-\frac{\epsilon}{1-x}\right)^{-2}
\stackrel{x\to1}{\longrightarrow}N\frac{2(1-x)}{\epsilon}
\end{equation}
An earlier approach~\cite{rkart}, based directly on the structure
function analogy, leads to:
\begin{equation}
\label{ekart}
D^{\rm Q}_{\rm h}(x)=(\alpha+1)(\alpha+2)x^\alpha(1-x)\stackrel{x\to1}
{\longrightarrow}(\alpha+1)(\alpha+2)(1-x)
\end{equation}
Here $\alpha\sim m_{\rm Q}$ so that $\langle x\rangle=1-2/(\alpha+3)$ in
accord with expectation.

In string models two approaches have been elaborated according to whether
the momentum or space-time aspects are emphasized. In the presence of
massless quarks, which move along linear light-cone trajectories, these
pictures are simply related, $\Delta p=\kappa\Delta t,\;\Delta E=\Delta p_z$;
however massive
quarks move along displaced hyperbolae. In the Lund momentum-space
approach the relevant string area is given, as for light quarks, by
$m^2_\perp/(x\kappa^2)$ and the same reasoning then leads to the LSFF
\eref{elsff}. However this gives $\langle x\rangle\approx1-(1+a)/b
m^2_\perp$ which is too hard compared to both theoretical expectation and
experiment~\cite{rQfComp,rQfA,rQfD,rQfL,rQfO,rQfS}. Adopting a space-time
based
approach~\cite{rbowl}, {\it \`a la} Artru Mennessier, the string fragments
randomly into clusters and the relevant string area becomes:
\begin{equation}
\frac{m^2_{\rm Q}}{2\kappa^2}\left[\frac{m^2_{\rm str}}{m^2_{\rm Q}}
\frac{1}{x}-1-\ln\left(\frac{m^2_{\rm str}}{m^2_{\rm Q}}\frac{1}{x}
\right)\right]
\end{equation}
where $m_{\rm str}$ is the mass of the string segment containing the
heavy quark. In the $Q/M_{\rm Q}\to\infty$ limit this results in a
generalization of the LSFF but with no $a$-term present. Elaborating this
scheme~\cite{rmor} to multiple string breaks along a finite size string
leads to the following effective fragmentation function:
\begin{equation}
\label{ebowl}
D^{\rm Q}_{\rm h}(x)\frac{1}{x^{1+bm^2_{\rm H}}}x^{a_\alpha}\left(\frac{1-x}{x}
\right)^{a_\beta}\exp\left(\frac{-bm^2_{\rm H}}{x}\right)\stackrel{x\to1}
{\longrightarrow}(1-x)^{a_\beta}
\end{equation}
Now $m_{\rm str}\approx m_{\rm H}$ should be identified with the mass of
the lightest Q hadron, where in fact the cluster mass spectrum peaks. This
has a softer spectrum, $\langle x\rangle\approx 1-R^{-1}/m_{\rm Q}$, and is
very similar to the Peterson form in practice.

It should be noted that the non-perturbative fragmentation functions can
be expected to describe effects due to the perturbative emission of soft
gluons, as found in a parton shower. However effects due to the emission
of hard gluons cannot be accounted for. In particular when defining $x$ as
a light-cone momentum fraction this may lead to a difference between the
reconstructed value $x_{\rm rec}$, measured along a jet axis, and the
primary value $x_{\rm pri}$, generated with respect to the `string' 
axis: the Lund model indicates~\cite{rshift} that $\langle x_{\rm rec}
-x_{\rm pri}\rangle_{\rm b}\approx0.08$.

The experimental measurement of the b fragmentation function is difficult; to
date it has been based on reconstructed B$\to$D$^{(\star)}\ell\nu_\ell(X)$
decays or the rapidity method~\cite{rpod}. Also, the interpretation of the
results is delicate, and ambiguous conclusions have been drawn. Two sources
of confusion are the delineation of the perturbative and non-perturbative
contributions and whether a distinction is made between primary and secondary
b hadrons cascading down from excited states. At the Z$^0$ there is substantial
production of B$^\star$~\cite{bib-DELB*,rbstar,rbstar1} and the four B$^{\star
\star}$~\cite{bib-DELB**,rbstar,rbstar2} mesons: the primary rate is
approximately 1:3:2 for B:B$^\star$:B$^{\star\star}$. OPAL~\cite{rQfO} finds
reasonable agreement with the Peterson \etal \eref{epeter}, Collins and
Spiller \eref{ecoli}, Kartvelishvili \etal\ \eref{ekart} and Lund-Bowler
\eref{ebowl} fragmentation functions. On the other hand, 
the ALEPH data~\cite{rQfA}
would rule out the Collins and Spiller form whilst favouring the 
Kartvelishvili
\etal\ form; the data also disfavour an untuned version of HERWIG, which
is too soft.
Preliminary studies from DELPHI~\cite{rQfD} indicate that JETSET with
parton showers (PS) and the 
Peterson form, \eref{epeter}, gives the best description of the data 
but with a
peak that is appreciably too wide. Using JETSET PS with Lund-Bowler
functions,
\eref{ebowl}, gives a very poor fit as does using JETSET with matrix
elements and the Peterson function. This last combination 
gives too narrow a peak, indicating the need for a contribution from gluon
radiation. HERWIG gives a reasonable description of the data but must be
tuned to avoid too many heavy b clusters. DELPHI~\cite{rQfD} also 
give measurements of
the primary B$^\star$ and B$^{\star\star}$ fragmenation functions. 
L3~\cite{rQfL} find
agreement within errors with the Peterson \etal\ fragmentation function.

\section{Spin phenomena}

As discussed in~\ref{selwk}, the primary quarks
produced via \eetoZtoqqbar\ are highly polarized. 
Whether, and in what circumstances, this polarization
survives the hadronization or whether
spin-spin forces wash out any memory of the
initial polarization is an open question. 
The primary quarks may become constituents of unstable 
baryons or of vector or tensor mesons, the angular   
distribution of whose decay products may be used to 
extract information about spin states.
 
In the particular case of
leading (large $x$) spin-1/2 $\Lambda$-type baryons,  
considerable polarization is expected~\cite{bib-poltheory}, though
also see~\cite{rpoldis}. 
In the constituent quark model, the spin of such baryons 
is carried by the heavy quark, with the light diquark system
in a spin zero, isospin zero state (though see the discussion
in \sref{spop}). Thus fast \Lambdab\ particles could
carry a substantial fraction of the polarization of the initial 
b quark, with the light diquark system carried along as
a spectator.
Similarly, high-$x$ \Lambdac, and $\Lambda$ 
baryons formed in fragmentation of s quarks, are expected 
to be polarized. Since $\Lambda$ particles can also arise 
from hadronization of initial u and d
quarks, the polarization in this case is considerably reduced.

\subsection{\Lambdab\ polarization}

ALEPH have measured the polarization of 
\Lambdab\ baryons~\cite{bib-ApolLb} using semileptonic decays,
\Lambdab~$\rightarrow l^-{\overline \nu}_l$+charmed hadrons. 
The method~\cite{bib-LBmethod} is based on measurement of the 
ratio of the average lepton to average neutrino energy.
The measured longitudinal polarization, 
$P_{\Lambda_{\rm b}} = -0.23^{+0.25}_{-0.21}$, 
is well below the theoretical expectation of $-0.69\pm0.06$
although the error is large.
This is a surprising result and, if confirmed, would indicate
the likely existence 
of depolarizing mechanisms in the b quark hadronization.

\subsection{$\Lambda$ polarization}

Because of the parity-violating nature of the 
decay, $\Lambda \rightarrow {\rm p}\pi^-$, the 
distribution of the polar angle, $\theta^*$, of the 
proton direction in the $\Lambda$ rest frame,
relative to the $\Lambda$ direction in the laboratory,
is proportional to $1+\alpha P_L \cos \theta^*$, where $P_L$ 
is the degree of longitudinal polarization.
The value of the weak decay parameter $\alpha$ is well 
measured~\cite{rpdg}. ALEPH~\cite{bib-ApolL} have
exploited this distribution to measure a value
$P_L = -0.32\pm0.07$ for leading ($x_p > 0.3$) 
$\Lambda$ baryons. When all sources of $\Lambda$ baryons
are taken into account, a result of $-0.39\pm0.08$
is expected if $\Lambda$'s containing a primary s quark
carry all of its initial polarization. Thus the ALEPH
measurement is in agreement with standard electroweak
theory together with  
the assumption that the initial strange quark polarization 
survives hadronization to become a leading $\Lambda$ baryon.  

This result is clearly
at odds with the conclusion from the \Lambdab\
study which indicated significant depolarization in
the hadronization. Indeed, the polarization of the 
heavier b quark may be expected  to survive more easily
than that of the s quark. 

\subsection{Vector meson spin alignment}

Study of spin alignment of vector mesons, particularly at
large $x$, where the meson may be expected to contain one
of the primary quarks, may provide information on the
nature of the quark to meson transition. Such analyses are
normally done in terms of the vector meson helicity density 
matrix, $\rho_{\lambda \lambda'}$, some of whose elements
can be determined by measuring the distribution of 
the vector meson decay products~\cite{bib-sdmatrix}. 
The element \r00\ is the fraction of mesons
which are in the helicity zero state. 
In strong vector meson decays it is not possible to infer 
separately the
values of elements $\rho_{11}$ and $\rho_{-1-1}$, since the
decay angular distributions are the same for helicity
$+1$ and helicity $-1$ vector mesons.  

In statistical
models~\cite{bib-statmodels}, the fragmentation is assumed
to produce extra quarks with both helicities equally
likely. Parallel alignment of primary and secondary quark
spins will produce
a vector meson with helicity $\lambda = \pm 1$.
If the spins are initially antiparallel,
the value of \r00\ will depend on the relative probability
to produce a vector or a pseudoscalar meson; in this case
\r00$=(1-P/V)/2$, with a maximum value of 1/2 when
$P/V = 0$ and a value of 1/3 if there is no suppression of
vector mesons. 
In the model of~\cite{bib-donoghue} vector
mesons are produced via vector currents
q$\rightarrow$qV which conserve the quark helicity. 
The vector meson then has helicity zero. 
Another model~\cite{bib-augustin}, for production of leading
mesons, assumes multiple emission of soft gluons by the
fragmenting quark, a process which conserves the quark
helicity. The leading vector meson may be formed when the 
leading quark combines with a soft antiquark in a process which
results in meson helicity $\pm 1$. The basic string and cluster
models have little to say about vector meson spin 
alignment, although~\cite{rlund} does point out that
no alignment may be expected in the simplest string picture.

ALEPH~\cite{rbstar}, DELPHI~\cite{bib-DELB*} and 
OPAL~\cite{bib-OPALB*} have 
measured the alignment of B$^*$ mesons using the
angular distribution of $\gamma$ rays from the decay
B$^* \rightarrow$B$\gamma$. The measurements agree, and 
produce a weighted average of 
\r00$=0.33\pm0.04$ (although they all express the
result in terms of a relative contribution of longitudinal
polarization states). These results, which 
imply no spin alignment, are consistent with simple spin counting 
and with heavy quark effective theory (HQET). They are
also in accord with the measurements of the ratio B$^*$ to B 
meson production~\cite{bib-DELB*,rbstar,rbstar1} which imply no 
suppression of the vector state (see~\ref{sec-VtoP}).
 
Results from OPAL~\cite{bib-OPALspin} on lighter vector mesons
show deviations from \r00$=1/3$. For D$^{*\pm}$ 
mesons, a 
value \r00~$ = 0.40\pm0.02$ has been measured, consistent
with lower energy results~\cite{D*CLEOHRS}. And for
primary $\phi(1020)$ mesons at $x > 0.7$, OPAL report an even
larger value, \r00$=0.54\pm0.08$. 
These mesons therefore appear to be preferentially in the
helicity zero state. Measurements of some off-diagonal elements
of the helicity density matrix show small deviations from zero for
both mesons. Such non-zero off-diagonal elements are a natural
consequence of coherence in the hadronization, and are a 
firm prediction~\cite{bib-Anselmino1} 
of any general model other than independent fragmentation.
There is clearly more to be learned about hadronization from 
such measurements. 

\section{Correlation phenomena}
\label{scorr}

Correlations in hadronic systems may be defined as departures 
from phase 
space in distributions for groups of two or more particles.
Such correlations may be associated
with 
\begin{itemize}
\item quantum mechanics --- Bose-Einstein and Fermi-Dirac effects
\item hadron dynamics --- resonances, reflections and final-state interactions
\item local baryon number conservation --- baryon-antibaryon phase space
correlations
\item local strangeness conservation --- strange particle rapidity correlations
\item soft gluon coherence in QCD showers --- two-particle momentum 
correlations and possibly intermittency
\item fragmentation dynamics --- transverse-momentum limitation of phase space
\end{itemize}

In so far as one wishes to understand the underlying dynamics
of the hadronization phase, it is important to take into account all
of these effects. 

\subsection{Bose-Einstein correlations}
\label{sbe}

Bose-Einstein correlations (BECs) have been extensively
studied in hadronic systems from Z$^0$ decay. A review
may be found in~\cite{bib-Haywood}. The correlations are interesting
in their own right as a quantum mechanical phenomenon whose
experimental details can give information on the space-time
structure of the source of hadrons. The correlations arise from
the necessity to symmetrize the wavefunction
for systems of two or more identical bosons. In most 
experimental analyses, a simple model is assumed where 
the source of particles is spherical with a Gaussian density.
Then the two-particle phase space is enhanced by a factor
$C(Q) = 1+\lambda \exp(-Q^2 R^2)$
relative to its density in the absence of the correlations.
Here $Q$, the square of the 4-momentum difference between
two bosons, is the measure of the separation in phase space, 
$\lambda$ measures the degree of coherence in
the particle emission ($\lambda=0$ corresponds to full
coherence) and $R$ is the radius of the Gaussian source.

\fulltable{Measurements of Bose-Einstein correlations in Z$^0$ decay.}
\br
Particle System   & R (fm)        & $\lambda$      
& References           & Comments \\
\mr
$\pi^\pm\pi^\pm$  & $0.65\pm0.16$ & $0.51\pm0.12$ & A~\cite{bib-ABE}&      \\
                  & $0.49\pm0.05$ & $1.06\pm0.17$ & D~\cite{bib-DBE}&
                                                              Direct $\pi$ \\
                  & $0.93\pm0.15$ & $0.87\pm0.14$ & O~\cite{bib-OBE} &     \\
$\pi^\pm\pi^\pm\pi^\pm$
                  & $0.62\pm0.05$ & $0.28\pm0.09$ & D~\cite{bib-D3BE}&     \\
K$^\pm$K$^\pm$    & $0.48\pm0.08$ & $0.82\pm0.27$ & D~\cite{bib-DKBE}&      \\
K$^0_{\mathrm S}$K$^0_{\mathrm S}$
                  & $0.71\pm0.07$ & $1.4\0\pm0.3$ & A~\cite{bib-AK0}&
                                                not corrected for f$_0$    \\
                  & $0.55\pm0.14$ & $0.61\pm0.23$ & D~\cite{bib-DKBE}&     \\
                  & $0.76\pm0.15$ & $1.14\pm0.39$ & O~\cite{bib-OK0}&     \\
\br
\endtab

BECs have been studied at LEP
for systems of $\pi^\pm\pi^\pm$, $\pi^\pm\pi^\pm\pi^\pm$,
$\pi^+\pi^-\pi^\pm$,
K$^\pm$K$^\pm$ and  K$^0_{\rm S}$K$^0_{\rm S}$. A summary 
of measured values of $\lambda$ and $R$ is
given in table~7. 
It is not straightforward to compare the various results, nor
to interpret the findings. For example,
in order to measure the enhancement $C(Q)$ due to BECs it is
necessary to know the phase space density in their absence, and
there are several ways to tackle this. In addition
particles which arise from resonance or weak
decays may be removed (if the detector has the 
capability).
The BEC effect will be diluted in data samples containing
mixtures of different particle types, and the purity of pion 
or kaon samples may be increased, 
depending again on the capabilities of the detector. 
In the LEP analyses, various different approaches to these 
problems have been used. 
 
Nevertheless, in all cases it is
clear that the model based on a spherical source with Gaussian
density gives a reasonable fit to the observations, although
there is no real evidence that other models would not fit
equally well. The coherence
parameter $\lambda$ varies from close to zero to 
greater than one; the latter is unphysical and possibly indicates
deviations from the model. The source has a typical size of 1~fm, 
which at first sight seems inconsistent with a picture of hadrons arising
from a rapidly expanding linear colour string. However
the length scale measured by the BECs is not the longitudinal size of the 
string, but the distance in production points
for which particles are close together in momentum space~\cite{rstbe}. 
Recently BEC studies at LEP have been extended to three-particle 
systems~\cite{bib-D3BE}, and the multiplicity dependence of 
the correlations has been investigated~\cite{bib-OBEmult}.    

Even if all of the measurements had been made in
a consistent way, the interpretation of the results would not be
straightforward: the correlations depend in a poorly understood
way on final-state interactions, resonance production 
and rescattering of resonance decay
products. According to~\cite{rberes1} the
situation may be ``impossibly complicated''. And~\cite{bib-Haywood}
says ``it seems very difficult to make progress in studying the
Bose-Einstein effect in the context of \ee\ physics, and it
is not clear to what extent it can be considered a useful and
interesting activity.'' 

It is nevertheless important to understand
at least the phenomenology of the Bose Einstein correlations since they
impact on studies of other features of hadronic systems. 
For example, although the effect primarily influences systems of 
identical bosons, in the relatively high-multiplicity, 
jet-like environment of Z$^0$ decay, 
``residual'' correlations~\cite{bib-residBE} 
arise between pairs of unlike particles. The
BECs produce a general collimation of the jets and a 
tendency to reduce the mean transverse momentum; 
this brings all particles closer together in momentum space.
The effect is to produce a distortion of $\pi^+\pi^-$ mass spectra,
especially at low momentum where the
multiplicity is highest. This means for example that the use
of opposite charge particle pairs to determine
the phase space in the absence of BECs can result in
biased values of $\lambda$ and $R$. An important practical 
result of the residual correlations is the considerable 
difficulty in inclusive measurements of $\pi^+\pi^-$ 
resonances~\cite{bib-AVM,bib-Dlight,bib-OPALVM1} 
such as the $\rho(770)^0$ and the f$_0(982)$ whose line shapes 
are distorted by the correlations and possibly also by other 
mechanisms~\cite{bib-residBE}.

In the process \eetoWWtohadrons\ at LEP 2, 
correlations may arise between the hadrons from one W and 
those from the other, an effect recently investigated
by DELPHI~\cite{bib-DELWW}.  
Such an effect could result in a shift of the   
reconstructed W mass in multihadronic W decays~\cite{WWBEC}.
A good understanding of the role of BECs in hadronization at
the Z$^0$ may help to reduce uncertainties introduced by this
effect.

\subsection{Fermi-Dirac anticorrelations}
While BECs between identical bosons are firmly established 
experimentally, 
anticorrelations between identical fermions are more
difficult to observe. In Z$^0$ decay, baryons are
produced at much lower rates than mesons, and local 
conservation of baryon number suppresses production of 
identical baryons close together in phase space. 
The only evidence for anticorrelations comes from a 
study by OPAL~\cite{bib-OPALFD} of
$\Lambda \Lambda$, 
$\overline {\Lambda}$\ $\overline {\Lambda}$ and 
$\Lambda \overline {\Lambda}$ pair 
production. Close to threshold it is found the the spin states of
the di-hyperon systems agree with expectations of a simple 
statistical mixture, with no indication of any resonance
in the $\Lambda \overline {\Lambda}$ threshold enhancement.
In the case of the identical baryon pairs, there is
a tentative indication of suppression close to threshold,
as would be expected by the Pauli exclusion principle.
However, further measurements are needed to confirm this
result.

\subsection{Baryon-antibaryon phase space correlations}
\label{sbcor}

Since baryons, and strange hadrons, are both heavier and less frequently
of secondary origin than ordinary hadrons investigating their production
and pairwise correlations appears to offer a more direct probe of the
momentum and quantum number flow during hadronization. However, allowance
must always be made for secondary hadrons coming from decays, such as
$\Sigma^0\to\Lambda\to{\rm p}$ (here further data~\cite{rpopex,rbylow} on
$\Lambda^{\tiny (}\bar{\rm p}^{\tiny )}$ correlations would be welcome).
Baryon number conservation implies that a baryon is always accompanied by
an antibaryon; flavour conservation, via diquark pairs, suggests that the
baryons may be preferentially particle-antiparticle pairs; and LPHD argues
that these baryons occur close by in phase space. So far, studies have
concentrated on measuring the proximity in phase space of 
\ppair~\cite{rAmega,rSbbcor}, \Lpair~\cite{bib-AK0,bib-Dlam,rOlam2} 
and $\Lambda^{\tiny (}\bar{\rm p}^{\tiny )}$~\cite{rpopex}
baryon
pairs. By introducing an event axis, typically the thrust or sphericity
axis, correlations can be studied in rapidity, azimuthal angle 
or polar angle with respect to the axis. 

Several models of baryon production are available~\cite{rbrev}. \Sref{shad}
discusses
the independent fragmentation~\cite{rff,rfif}, the diquark~\cite{rstdiq}
and popcorn~\cite{rpop,rpop2} (triplet) string options and 
the cluster~\cite{rclus} models. In addition a 
possible contibution from direct,
$\gamma^\star\to({\rm qq}')(\bar{\rm q}\bar{\rm q}')$, diquark production
has been proposed~\cite{rdiq,rdirdiq}. Also there is the recombination
model of hadron production~\cite{reco}. In this approach a spectrum of
partons occuring after a shower at the fixed scale $Q^2_0$ is convoluted
with an explicit wavefunction for a particular hadron. In the case of a
baryon~\cite{recobar} this gives (c.f. \eref{econv}): 
\eject
\begin{eqnarray}
D^a_{\rm B}(x,Q^2)=\sum_{a_1,a_2,a_3}\int\d x_1\d x_2&\d x_3
D_{a_1,a_2,a_3}^a(x_1,x_2,x_3,Q^2;Q^2_0) \nonumber \\ 
    & \times\;R_{a_1,a_2,a_3}^{\rm B}(x_1,x_2,x_3,x;Q^2_0)
\end{eqnarray}
Here a typical wavefunction is $R\propto(x_1x_2x_3)^2\delta(x-x_1-x_2-x_3)$,
consistent with reciprocity~\cite{repro} and quark counting rules~\cite{rdimc},
and on the assumption of uncorrelated partons $D_{a_1,a_2,a_3}^a(x_1,x_2,x_3,
Q^2;Q^2_0)=\prod_iD_{a_i}^a(x_i,Q^2;Q^2_0)$. It may be mentioned in passing
that the recombination model makes interesting predictions for baryon
polarizations~\cite{rtom}.

\subsubsection{Rapidity correlations} 

At the Z$^0$ strong, short-range correlations in rapidity are seen between
baryon-antibaryon B$\overline {\rm B}$ pairs. (Unlike-baryon distributions
have the like-baryon (BB) distributions subtracted to remove secondary
correlations due to more than one B$\overline{\rm B}$ pair in an event.) For
example, the distribution of baryon rapidity ($y$) in such events,
$\d n/\d y(y_{\rm B}|y_{\overline{\rm B}})$, is measured
to be compact and centred on 
$y_{\overline{\rm B}}$~\cite{rAmega,rpopex,bib-AK0,rOlam2}. In a 
B$\overline{\rm B}$ pair, given the rapidity of the $\bar{\rm p}\;
(\overline\Lambda)$ there is $\approx70\;(50)$\% probability 
that the p ($\Lambda$)
will be found within $|y_{\rm B}-y_{\overline B}|\leq 1(0.6)$ and 
vice versa~\cite{rAmega,rOlam2}; a 
marginally weaker short-range $\Lambda\bar{\rm p}$
correlation is found~\cite{rpopex}. A much weaker long-range anticorrelation
is also seen for far forward/backward B$\overline{\rm B}$ pairs, as anticipated
from a leading particle effect. In contrast, when BB pairs occur in an event,
the $\d n/\d y(y_{\rm B_1}|y_{\rm B_2})$ distribution 
is nearly flat
with just a weak short-range anticorrelation, particularly so away from the
central region~\cite{rOlam2} where phase space constraints become important.
Similar, though statistically limited, results have been seen 
at lower energy~\cite{rmark2,rbylow}.

Qualitatively these features are reproduced by both the HERWIG and JETSET
models, though quantitatively the strength of the correlations is
overestimated by HERWIG and by JETSET without popcorn~\cite{rAmega,rOlam2}.
The recombination model predicts rapidity correlations which are much too
weak whilst direct diquark production predicts long-range correlations
that are too strong; both are disfavoured by the data~\cite{bib-Dlam}.

\subsubsection{Azimuthal angle correlations} 

Somewhat weaker correlations are seen between the azimuthal angles in baryon
pairs at the Z$^0$~\cite{rAmega,bib-AK0}. B$\overline{\rm B}$ pairs show a
tendency towards $\Delta\phi=0$, though in those pairs which lie out of the
event plane there is a tendency towards $\Delta\phi=\pi$. BB pairs show a
weaker tendency towards $\Delta\phi=\pi$. Two competing mechanisms might be
envisaged to explain the azimuthal angle correlations. First is a local
compensation of transverse momentum which leads to an enhancement for
$\Delta\phi=\pi$. Second is a tendency for any off-axis boost to be shared
by neighbouring baryons, leading to an enhancement at $\Delta\phi=0$. The
latter effect might be expected to become more important at higher energies
where three-jet effects come into play (these by definition involve higher
transverse momentum scales than hadronization), more so for baryons
lying in the event plane. At low CoM energy (10~GeV) the first, back-to-back,
effect appears to dominate~\cite{rb2b10} but this is seen to weaken at
higher energy (30~GeV)~\cite{rmark2,rb2b30,rtpc} and at the Z$^0$ the second,
side-by-side, effect is more important.

\subsubsection{Polar angle correlations} 

Historically a very powerful way to discriminate between hadronization
models is the orientation of the B$\overline{\rm B}$ pair with 
respect to the event axis~\cite{rtpc}. If $\theta^\star$ is the angle between 
the axis of the B$\overline{\rm B}$ pair and the event axis 
as seen in the B$\overline{\rm B}$ pair rest
frame then $\cos\theta^\star$ is measured to be highly forward-backward
peaked~\cite{rAmega,bib-Dlam}. By utilizing the lepton beam polarization
information (see section \ref{selwk}) SLD are capable of determining the
primary quark direction and have recently been able to demonstrate that
the baryon preferentially follows the quark direction~\cite{rSbbcor}.

In string models the colour field typically aligns along the event axis, so
that provided that the transverse momenta acquired by the baryons are small
compared to the longitudinal momenta transferred from the string, the baryons
will retain a strong memory of the string/event axis direction. In contrast,
because clusters are deemed to be structureless, they have no means of
retaining any information about an original event axis (they are unpolarized)
and so decay isotropically in their rest frame. Not surprisingly then the
highly peaked $\cos\theta^\star$ strongly disfavours the essentially flat
prediction from HERWIG whilst JETSET offers a good description of the data.
In HERWIG version 5.7 an option was introduced to allow non-isotropic
decays of clusters containing primary quarks; this was implemented to
stiffen the momentum spectrum of c and b hadrons and has little influence on
B$\overline{\rm B}$ pair polar angle distributions.

\subsection{Strangeness correlations}

In the previous section, \ref{sbcor}, phase space correlations were discussed
for baryons; we now turn to similar measurements made on strange hadrons.
The conservation of strangeness in strong interactions implies that strange
hadrons are pair-produced during hadronization. The probability that hadron
$h_1$ is accompanied by $h_2$ is defined as $P(h_1,h_2)=2\times\langle
n_{h_1,h_2}\rangle/\langle n_{h_1}+n_{h_2}\rangle$ where the 2 is included
because of double counting. Measurements give $P(\Lambda,\overline\Lambda)=49
\pm6$\% whilst $P(\Lambda,\Lambda)=13\pm1$\%~\cite{bib-AK0,bib-Dlam,rOlam2},
$P(\Lambda,{\rm K}^0_{\rm s})=17\pm2$\% 
and $P({\rm K}^0_{\rm s},{\rm K}^0_{\rm s})=29\pm4$\%~\cite{bib-AK0},
$P(\Xi^-\overline\Lambda)=40\pm7$\%~\cite{bib-DELB95,rOlam2} 
and $P(\Xi^-\overline
\Xi^+)=4\pm6$\%~\cite{rOlam2}.

Two sources of strange hadrons can be anticipated: leading hadrons associated
with initial s$\bar {\rm s}$ quarks and those pair-produced locally in the event's
colour field. To help distinguish these possibilities requires information on
the phase-space correlations between pairs of strange hadrons. A short-range
rapidity correlation occurs between ${\rm K}^0_{\rm s}{\rm K}^0_{\rm s}$ pairs 
and a slightly weaker
one between ${\rm K}^0_{\rm s}\Lambda$ pairs~\cite{bib-AK0} indicating a 
local mechanism
for strangeness compensation. Also visible are: weaker long-range
correlations, as expected from leading quarks; and evidence of phase-space
suppression, particularly when both hadrons are leading. Weak correlations
are also seen at $\Delta\phi=0,\pi$ for 
centrally produced ${\rm K}^0_{\rm s}{\rm K}^0_{\rm s}$ and
${\rm K}^0_{\rm s}\Lambda$ pairs.

The possible influence of introducing the popcorn baryon production mechanism,
see \sref{spop}, is of particular interest for string models. The presence
of an intermediate meson, BM$\overline{\rm B}$ tends to soften all 
correlations.
At present, taking account of systematic errors, measurements of baryon and
strangeness correlations are insufficient to place any significant constraint
on the level of popcorn production required. The B$\overline{\rm B}$ pair
rapidity
difference is mainly sensitive to the amount of popcorn production, with data
favouring a substantial component~\cite{bib-Dlam,rOlam2}. The substantial
rate of $\Xi^-\overline\Lambda$ pair production also favours a high level of
popcorn production in order to supply an intermediate kaon~\cite{bib-DEL96,
rOlam2}. Interestingly the number of B$^{\scriptsize (}\overline{\rm
B}^{\tiny )}$
pairs
decreases linearly with the amount of popcorn introduced so that their
measured multiplicity can be (rather simplistically) used to constrain the
amount of popcorn to around 50\%~\cite{bib-AK0}.

A more direct test of the popcorn mechanism is to look at rapidity ordered
BM$\overline{\rm B}$ triples~\cite{rpopex}. Using nearby baryon pairs,
$|y_{\rm B}
-y_{\overline{\rm B}}|<1$, a probability of 7\% (25\%) is obtained for finding an
intervening kaon (pion) in a \ppair\, \Lpair\ or $\Lambda\bar{\rm p}$ pair.
In addition to there being no enhancement of popcorn-favoured $\Lambda$K$^+
\bar{\rm p}$ triples, unfavoured $\Lambda$K$^-\bar{\rm p}$ triples are found
equally likely. However, in events containing a kaon and a close $\Lambda$p
pair, the kaon is found very close in rapidity to the baryon pair.

\subsection{Intermittency}

Intermittency~\cite{bib-intermittency}, the non-random clustering 
of particles in phase space, is a somewhat obscure phenomenon  
of uncertain dynamical origin. In essence, 
intermittency corresponds with large, non-statistical, fluctuations
in the numbers of particles in particular events which are
found in narrow rapidity bins. 
It is normally studied by measuring factorial moments of
multiplicity distributions in rapidity bins. 
Intermittent behaviour has
been observed in hadronic systems from Z$^0$ decay at both 
LEP~\cite{bib-LEPinter} and SLC~\cite{bib-SLCinter} as well
as in lower energy \ee\ collisions.

The intermittency observed in hadronic Z$^0$ decays is in fact 
reproduced by the Lund parton shower Monte Carlo model
with string fragmentation. It is a possibility that the
self-similarity inherent in the QCD parton shower,
with its successive q$\rightarrow$qg and g$\rightarrow$gg
branchings, is
the source of the intermittency seen in the distribution
of the final-state hadrons. Certainly the JETSET model 
contains no feature explicitly introduced to simulate 
the dynamics of intermittency. 

The appearance of intermittency in the Z$^0$ decay data would
seem to be an ideal opportunity to gain a good understanding
of the mechanisms which lie behind it. Previously its
interpretation in hadroproduction experiments has been 
obscured by the complicated nature of the final states and
the effects of beam and target fragments. 

\section{Quark-gluon jet differences}

The determining property of quark and gluon jets is the colour charge of
the initiator partons, $C_F=(N^2_c-1)/(2N_c)$ and $C_A=N_c$ respectively.
Due to the gluon's larger charge it should radiate more subsequent gluons
in a parton shower. This leads to the anticipation that gluon jets, as
compared to quark jets, will have: a higher multiplicity, softer momentum
spectrum and wider angular distribution~\cite{rqgrat,rqgprop}. These global
features are largely borne out by experimental 
results~\cite{rqgex1,rstrng4,rqgex4,rqgex2,rqgex3,rtopol}. However 
whilst clear differences
are now established between gluon and light-quark jets, b-jets appear to be
rather like gluon jets at Z$^0$ energies~\cite{rqgex4,rqgex2}; the ratios
of measured properties typically fall short of the naive asymptotic
predictions. For example the multiplicity ratio is, in leading order,
predicted to be $C_A/C_F=9/4$~\cite{rqgrat}; at NNLO this 
becomes $\approx2$~\cite{rmmom,rqgcorr} and after imposing energy 
(but not momentum) conservation on
the shower this drops to $\approx 1.6-1.8$~\cite{rqgengy} (the exact
predictions depend on jet energy and the scale used for $\alpha_s$). The
measured ratio, which is seen to be sensitive to the precise jet definition,
is typically in the range $1.1-1.3$, though OPAL has obtained 1.55 in an
event hemisphere-based analysis~\cite{rqgex3}. It should be borne in mind
that the these basic predictions are made for back-to-back pairs of quark or
gluon jets whilst growing evidence suggests that the relative topology of a
jet is important in determining the appropriate scale~\cite{rtopol}.

The above discussion relates to the perturbative properties of quark and
gluon jets. Current cluster and string models of hadronization make no
distinction between whether a set of final-state partons arose from a
fragmenting primary quark or gluon; they are treated the same. This does
allow the possibility of a `leading particle' effect (for example one
should expect more leading kaons in an s-quark jet than a gluon jet)
but no other `anomalous' effects~\cite{rlead}. An interesting possibility
is that some isoscalar states, $\eta,\;\eta',\;\phi,\;\omega,\ldots$, may
contain a significant gg component and hence might appear more frequently
as leading particles in gluon jets. These particles are very often
primary hadrons coming directly from a cluster or string. At present the
possibility of hadrons containing gluons is not allowed for and only quark
constituents are considered. Indeed in the cluster framework gluons are
split into light \qqbar\ pairs whilst in the string approach they
represent energy-momentum `kinks' on a string. However modifying the
models to accommodate gluonium would not, {\it a priori}, appear to pose
significant problems of principle or practice.

A more radical scenario is offered by the independent fragmentation model of
Peterson and Walsh~\cite{rgeta}. This is based on the suggestion that a gluon
is attached to an octet colour flux tube and quarks to triplet flux tubes
(see \sref{string}). A gluon now fragments into a sequence of isoscalar gg or
gq$\bar{\rm q}$g clusters leading to a prediction of greatly enhanced $\eta,
\;\eta'$ etc production and harder momentum spectra as compared to quark
jets.

A third alternative scenario for gluon jet hadronization is provided within
the recombination model~\cite{reco} discussed in \sref{sbcor}. Calculations
predict: a softer pion spectrum~\cite{recopi}, an enhanced multiplicity
ratio $\langle n_{\eta'}\rangle/\langle n_{\pi^0}\rangle$~\cite{recoeta} and
enhanced baryon production~\cite{recobar} in gluon jets compared to quark
jets.

The L3 Collaboration have reported tentative indications that $\eta$
production is enhanced in gluon jets~\cite{bib-L3eta}. Studying the lowest
energy --- gluon --- jet in three-jet events they see a a harder $\eta$
momentum spectrum than predicted by both the HERWIG and JETSET Monte Carlos
with an enhancement in the ratio $\langle n_\eta\rangle/\langle n_{\pi^0}
\rangle$. The use of a ratio takes into account the established increase
in multiplicity found in gluon compared to quark jets and is designed to
make the measurement sensitive to any additional enhancement or suppression.
The Monte Carlos provide a satisfactory description of the spectrum and
$\langle n_\eta\rangle/\langle n_{\pi^0}\rangle$ ratio in quark jets.
DELPHI have studied the production rates of kaons, $\Lambda^0$ and $\Xi^\pm$
in multi-jet events~\cite{bib-DELB95}. They find that the relative yields
of strange hadrons in multi-jet, normalized to two-jet, events is constant
in events with widely separated jets but favours increased production 
in multi-jet events at small resolutions, particularly so for kaons. 
In a more direct study
DELPHI have looked at identified particle production rates in actual quark
(natural flavour mix) and gluon jets, normalized to the charged multiplicity
in the jet: the double ratios
$(\langle n_{\rm H}\rangle/\langle n_{\rm ch}\rangle)_g/(
\langle n_{\rm H}\rangle/\langle n_{\rm ch}\rangle)_q$~\cite{rDidqg}. The
evidence suggests values for the double ratios of  
approximately $1.1$ for the K$^0$, $0.9$ for K$^+$, $1.2$ for p and 
$1.4$ for $\Lambda$. In general these
results are in qualitative agreement with the Monte Carlos but quantitatively
the deviations from unity
are larger than typically predicted, apart for the K$^0$ result
where a small suppression was expected. However the errors are
relatively large. OPAL have also reported
preliminary studies of the double ratios~\cite{rOidqg}. For K$^0_{\rm s}$ and
$\phi$ mesons they report a slight, $\leq10$\%, increase in the relative
production rates in gluon jets, whilst for p and $\Lambda$ baryons they
measure a significant 30--40\% increase. These enhanced production rates
are presumably at the expense of pion production. At present more 
measurements are required before hard conclusions can be drawn.

The possibility of enhanced identified hadron production in gluon as compared
to quark jets has received previous attention. The results for mesons, in
particular hard $\eta$ relative to $\pi^0$, production are not conclusive.
At $\sqrt s=10$~GeV ARGUS~\cite{rargeta} saw no evidence for enhanced $\eta$
or $\phi$ production in the continuum, $\gamma^\star\to $q$\bar{\rm q}$. 
Also comparison has been made between continuum and $\Upsilon(1,
2S)\to $ggg events; here Crystal Ball~\cite{rcrybeta} see no enhancement
whilst DASP-II, CLEO and ARGUS~\cite{rbary} see a slight enhancement for a
number of mesons. At $\sqrt s=30$ GeV JADE~\cite{rjadeta} reported very weak
(statistically insignificant) evidence for a small enhancement in the
$\langle n_\eta\rangle/\langle n_{\pi^0}\rangle$ ratio in acollinear, gluon
rich, events. The situation is clearer for baryons. The production rate of
baryons in $\Upsilon(1,2S)$ decay (gluon dominated) to continuum (quark
dominated) events shows an excess of 200--300\%~\cite{rbary}.

In the context of conventional Monte Carlos a number of partial explanations
have been offered. A study using JETSET indicates that the relative production
rates of mesons in quark and gluon jets is energy independent and just less
than one, whilst for baryons it is 20--25\% larger and shows a slight increase
with jet energy~\cite{rOidqg}. This latter effect may be attributed to an edge
effect associated with the suppression of leading baryons. In cluster models
it has been argued that in gluon rich environments, the topology, rather than
any intrinsic properties of the jets, leads to heavier clusters and hence
larger baryon production rates~\cite{rbarcl}. At the $\Upsilon$ it has also 
been emphasized~\cite{rbarexp} that secondary decays are important, and 
that 40\% of the events are in the continuum to b and c quarks 
(see \fref{fbrs}) which, being heavy,  `eat up' the
available phase space for baryon production.

\section{Outlook}

That Monte Carlo event generators, solidly based on sound physics, are 
essential in modern high energy experiments, from detector conception 
to data analysis, ought not to be forgotten. 
It therefore almost goes without saying that the
reliability of Monte Carlo predictions should be a prime concern if only
for mere practical reasons. To this end the large event rates and pristine
conditions available in hadronic Z$^0$ decays at \ee\ colliders
play a particularly important role. Here, as nowhere else, 
the physics assumption built into the Monte Carlo models can be 
confronted with ever more exacting tests. The relatively complex 
conditions associated
with initial-state hadrons have largely precluded this activity using ep
and p$\bar {\rm p}$ data, thereby making physicists reliant on the quality 
of the Z$^0$ data. Two caveats to the wider application of the models 
are the questions of the reliability of the factorization theorem in 
perturbative QCD and the presumption of universality in the hadronization
process.

Focusing on the models for the hadronization processes one may ask what is
left that should be  done with the Z$^0$ data and what might be learned
from it. We list a number of topics, perhaps not all of which can be done
with available data (but one can always hope for more Z$^0$ data):

\begin{itemize}
\item
A study of the transverse momentum distributions of identified particles,
especially pions where low mass effects and correlations may 
occur~\cite{rpicor} --- In string models the fact that the predicted width
of the Gaussian $p_\perp$-distribution, $\sqrt{\kappa/\pi}$, proves too
narrow has raised questions about the adequacy of a tunnelling mechanism
explanation: is unresolved (non-perturbative) gluon emission~\cite{rbo}
the real explanation?

\item
An attempt to establish properties of the relatively rare, directly produced
pions --- Pions are special particles by virtue of their nature as Goldstone
bosons (of the chiral symmetry); this mandates them to have small masses
which in a string model implies a very small size, in fact uniquely less
than a string's width.

\item
Further measurements of orbitally excited mesons and baryons --- Are the
production rates for these states too high to be compatible with the
string model?  

\item
A search for D-wave states, such as the K$^\star_3(1780)$ --- These could
help elucidate the role of the wavefunction in determining a hadron's
production rate.

\item
A study of f$_0(980)$ production as a function of rapidity and of total
event multiplicity --- This would test the Gribov confinement
scenario~\cite{rfunny}.

\item
A search for deuterium production --- This has been reported previously in
studies at the $\Upsilon$~\cite{rdeutexp} and is expected, on the basis 
of a string model calculation~\cite{rdeut}, at the level
of $5\times10^{-5}$ per hadronic Z$^0$ event.

\item
Further measurements of strangeness suppression ($\gamma_{\rm s}$) which
ideally can be directly compared to those available in ep collisions,
particularly those associated with the current region of the Breit
frame~\cite{rbreit} --- If it turns out that $\gamma^{\rm e^+e^-}_{\rm s}
\neq\gamma^{\rm ep}_{\rm s}$ then effort should go to establishing any
other differences.

\item
A measurement of the s-quark to p and $\bar{\rm p}$ fragmentation functions
--- Is there a measurable difference as suggested~\cite{rassym} by the
possible asymmetry in sea quark structure functions?

\item
Measurements of identified particle production rates in quark and gluon jets
--- Are there measurable differences and, if so, what is the mechanism?

\item
A search for glueball candidates, particularly in the gluon-rich
environment provided by gluon jets.

\item 
Further measurements of leading baryon polarization ---  Is there a
depolarizing mechanism for the b baryons which does not apply to those
arising from s quark fragmentation?

\item 
Measurements of the helicity density matrix for the light and heavy vector
mesons --- Does hadronization produce spin-aligned vector mesons, and does
any alignment depend on $x$? Are some off-diagonal elements non-zero as
predicted by coherent hadronization? Spin physics has a long history of
producing surprises. 

\item
More detailed studies of intermittency --- Can it be firmly established
that the phenomenon is due to the self-similar evolution of the initial
\qqbar\ state via a QCD parton shower?

\item
A search for direct evidence of popcorn-type baryon production via, for
example, further study of $\Lambda$K$^+\bar{\rm p}$ type correlations
--- Three-body cluster decays would also induce such correlations but
perhaps with different intensity.

\item
A search for $\Omega^-\bar{\rm p}$ type correlations --- These are possible
in a generalized popcorn mechanism although they will be hard to find 
experimentally because of low rates.

\item
An attempt to establish directly the existence of the `dead cone'
\cite{rdead} in non-leading particles, by removing the leading particles
using fully reconstructed b-hadron decays.

\item
Direct measurement of the rates and momentum spectra of hadrons produced in
b-quark events to ensure adequate descriptions of the weak decays in the
Monte Carlos --- b hadrons contribute a tenth of all particles 
in hadronic Z$^0$ decays, and significantly more at large $x$.

\item
A tuning of the Monte Carlo models to Z$^0$ data, incorporating colour 
rearrangement --- This is important in order to determine the effects
of colour rearrangement and to provide constraints which may prove
significant for later W mass measurements.

\item
A simultaneous tuning of the Monte Carlo programs to the Z$^0$ data and lower
energy data, particularly from PEP and PETRA --- At lower energies the
relative contribution to event properties made by hadronization is
more important, whilst the influence of the perturbative shower can be
tested using the CoM energy dependence of observables.

\item
An `ultimate' tuning of the Monte Carlo models to the final Z$^0$ data
--- This will be an invaluable service to experimentalists and theorists
allowing the experience gained from LEP1/SLC to be applied at future
machines.

\item
A continuation of the search for tests which discriminate between the competing
models of the non-perturbative physics underlying the hadronization, and
continuing development of these models.
\end{itemize}

In summary, the \eetoZtoh\ data have already provided a wealth of
information on the phenomena of parton hadronization. Of the available
models of the non-perturbative physics involved, the Lund string model,
as implemented in JETSET, has met with most success, particularly in 
the baryon sector, and most notably in its prediction of the angular
distribution of correlated baryon-antibaryon pairs in their rest frame. 
However, the model has many free parameters and consequently has little 
predictive power. But the parameters are not arbitrary --- most are based on 
incomplete knowledge of physics. Therefore it could be argued that their
values, when fully tuned to reproduce observations, provide important 
information about hadronic physics. On the other hand, while the weight 
of evidence tends to favour the string picture, the other models,
particularly the cluster model of HERWIG, are not dead, and more
analyses of the existing and future data are essential to provide
further discrimination between the models and to help elucidate the 
physics of hadronization.   

\section*{Acknowledgements}
IGK wishes to thank N. Alguacil Conde for encouragement.



\section*{References}
\begin{thebibliography}{999}

\bibitem{r1qcd}
Kunszt Z, Nason P, Marchesini G and Webber B R 1989 {\it Z Physics at LEP 1}
vol~1 yellow report CERN 89-08 p~373

\bibitem{rlep1}
Altarelli G \etal\ 1989 {\it Z Physics at LEP 1} vol~1
yellow report CERN 89-08

\bibitem{rlepew}
Bardin D \etal\ 1995 {\it Reports of the Working Groups on Precision
Calculations for the Z Resonance} yellow report CERN 95-03

\bibitem{r2isr}
Altarelli G \etal\ 1996 {\it Physics at LEP 2} vol~1
yellow report CERN 96-01

\bibitem{r3loop}
Gorishny S G, Kataev A L and Larin S A 1991 \PL {\bf B259} 114 \\
Surguladze L R and Samuel M A 1991 \PRL {\bf 66} 560 and {\it erratum} 2416 \\
Chetyrkin K G, K\"uhn J H and Kwiatkowski A 1996 {\it Phys. Rep.} {\bf 277}
189  

\bibitem{rlep1mc}
Sj\"ostrand T \etal\ 1989 {\it Z Physics at LEP 1} vol~3 
yellow report CERN 89-08 p~143

\bibitem{rpdg}
Particle Data Group, Barnett R M \etal\ 1996 \PR {\bf D54} 1

\bibitem{rbible}
Dokshitser Yu L, Khoze V A and Troyan S I 1989 {\it Perturbative QCD}
ed A H Mueller (World Scientific: Singapore) p~241 \\
Dokshitser Yu L, Khoze V A, Mueller A H and Troyan S I 1991 {\it Basics
of Perturbative QCD} (Gif-sur-Yvette: Editions Fronti\`eres) 

\bibitem{rinout}
Bjorken J D 1973 {\it Proc. SLAC Summer Inst. on part. Phys.} vol~1 
report SLAC-167 p~1

\bibitem{rplumb}
Bjorken J D 1992 \PR {\bf D45} 4077

\bibitem{rlep2mc}
Knowles I G \etal\ 1996 {\it Physics at LEP 2} vol~2
yellow report CERN 96-01 p~103

\bibitem{rariadne}
L\"onnblad L 1992 {\it Comp. Phys. Comm.} {\bf 71} 15 \\
URL: http://surya11.cern.ch/users/lonnblad/ariadne/

\bibitem{rdcm}
Gustafson G 1986 \PL {\bf B175} 453 \\
Gustafson G and Pettersson U 1988 \NP {\bf B306} 746 \\
Andersson B, Gustafson G and L\"onnblad L 1990 \NP {\bf B339} 393

\bibitem{rlund}
Andersson B, Gustafson G, Ingelman G and Sj\"ostrand T 1983 {\it Phys. Rep.}
{\bf 97} 31

\bibitem{rcojets}
Odorico R 1992 {\it Comp. Phys. Comm.} {\bf 72} 238 \\
URL: http://www.bo.infn.it/preprint/odorico.html

\bibitem{rff}
Field R D and Feynman R P 1978 \NP {\bf B136} 1

\bibitem{rherwig}
Marchesini G \etal\ 
1992 {\it Comp. Phys. Comm.} {\bf 67} 451 \\
URL: http://surya11.cern.ch/users/seymour/herwig/

\bibitem{rhersh}
Marchesini G and Webber B R 1984 \NP {\bf B238} 1 \\
Knowles I G 1988 \NP {\bf B310} 571 and 1990 {\it Comp. Phys. Comm.}
{\bf 58} 271 \\
Catini S, Webber B R and Marchesini G 1991 \NP {\bf B349} 635 \\
Seymour M H 1995 {\it Comp. Phys. Comm.} {\bf 90} 95

\bibitem{rdepl}
Marchesini G and Webber B R 1990 \NP {\bf B330} 261

\bibitem{rclus}
Webber B R 1984 \NP {\bf B238} 492

\bibitem{rjetset}
Sj\"ostrand T 1994 {\it Comp. Phys. Comm.} {\bf 82} 74 and Lund University
report LU TP 95-20 \\
URL: http://thep.lu.se/tf2/staff/torbjorn/

\bibitem{run}
Gross D J and Wilczek F 1973 \PRL {\bf 30} 1343 and 1973 \PR {\bf D8} 3633 \\
Politzer H D 1973 \PRL {\bf 30} 1346 \\
Tarasov O V, Vladimirov A A and Zharkov A Yu 1980 \PL {\bf B93} 429 \\
van Ritbergen T, Vermaseren J A M and Larin S A 1997 preprint archive
hep-ph/9701390

\bibitem{rert}
Ellis R K, Ross D A and Terrano A E 1981 \NP {\bf B178} 421 \\
Fabricus K, Kramer G, Schierholtz G and Schmitt I 1981 \ZP {\bf C11} 315

\bibitem{rertor}
K\"orner J G, Sch\"uler G A and Barreiro 1980 \PR {\bf D1} 1416 \\
Kramer G and Lampe B 1985 {\it Comm. Math. Phys.} {\bf 97} 257 \\
Zijlstra E B and van Neerven W L 1992 \NP {\bf B383} 525

\bibitem{rertb}
Rodrigo G V 1997 preprint archive hep-ph/9703359 \\
Bernreuther W, Brandenberg A and Uwer P 1997 preprint archive hep-ph/9703305

\bibitem{r4jet}
Signer A and Dixon L 1997 \PRL {\bf 78} 811 \\
Glover E W N and Miller D J 1997 \PL {\bf B396} 257 \\ 
Bern Z, Dixon L, Kosower D A and Weinzierl S 1997 \NP {\bf B489} 3

\bibitem{r5jet}
Ali A \etal\ 1979
\PL {\bf B82} 285 and 1980 \NP {\bf B167} 454 \\
Berends F A, Giele W T and Kuijf H 1988 \NP {\bf B321} 39 \\
Falck N K, Graudenz D and Kramer G 1989 \PL {\bf B220} 299

\bibitem{rkln}
Lee T D and Nauenberg M 1964 \PR {\bf 133} 1549 \\
Kinoshita T 1965 {\it J. Math. Phys.} {\bf 3} 56 \\
Sterman G 1978 \PR {\bf D17} 2789 and 2773

\bibitem{rlogs}
Gribov V N and Lipatov L N 1972 {\it Sov. J. Nucl. Phys.} {\bf 15} 438 \\
Dokshitzer Yu L 1977 {\it Sov. Phys. JETP} {\bf 46} 641

\bibitem{rap}
Altarelli G and Parisi G 1977 \NP {\bf B126} 298 \\
Owens J F 1978 \PL {\bf B76} 85 \\
Uematsu T 1978 \PL {\bf B79} 97

\bibitem{rmcsh}
Konishi K, Ukawa A and Veneziano G 1979 \NP {\bf B157} 45 \\
Bassetto A, Ciafaloni M and Marchesini G 1983 {\it Phys. Rep.} {\bf 100} 201 \\
Webber B R, {\it Ann. Rev Nucl. Part. Sci} 1986 {\bf 36} 253

\bibitem{resum}
Catani S, Turnock G and Webber B R 1991 \PL {\bf B272} 368 and 
1992 \PL {\bf B295} 368 \\
Catani S, Trentadue L, Turnock G and Webber B R 1991 \PL {\bf B263} 491 and
1993 \NP {\bf B407} 3 \\
Dissertori G and Schmelling M 1995 \PL {\bf B361} 167

\bibitem{rmatch}
Catani S \etal\ 1991 \PL
{\bf B269} 432

\bibitem{rdur}
Stirling W J 1991 \jpg {\bf 17} 1537 \\
Bethke S, Kunszt Z, Soper D E and Stirling W J 1992 \NP {\bf B370} 310 \\
Brown N and Stirling WJ 1992 \ZP {\bf C53} 629

\bibitem{rlows}
Ceradini F \etal\ 1972 \PL {\bf B42} 501

\bibitem{rmeds}
SLAC-LBL Collaboration, Hanson G \etal\ 1975 \PRL {\bf 35} 1609 \\
TASSO Collaboration, Althoff M \etal\ 1984 \ZP {\bf C22} 307

\bibitem{r30s}
S\"oding P and Wolf G 1981 {\it Ann. Rev. Nucl. Part. Sci.} {\bf 31} 231 \\
Duinker P 1982 {\it Rev. Mod. Phys.} {\bf 54} 325 \\
Wu S L 1984 {\it Phys. Rep} {\bf 107} 59 \\
MARK-J Collaboration, Adeva B \etal\ 1984 {\it Phys. Rep.} {\bf 109}
131 \\
Narosoka B 1987 {\it Phys. Rep.} {\bf 148} 67

\bibitem{rjets}
DELPHI Collaboration, Abreu P \etal\ 1990 \PL {\bf B247} 167 \\
L3 Collaboration, Adeva B \etal\ 1990 \PL {\bf B248} 464 \\
OPAL Collaboration, Akrawy M Z \etal\ 1991 \ZP {\bf C49} 375 and
Akers R \etal\ 1994 \ZP {\bf C63} 197 \\
SLD Collaboration, Abe K \etal\ 1993 \PRL {\bf 71} 2528 

\bibitem{rertmc}
Kunszt Z and Nason P in~\cite{r1qcd}\\
Giele W T and Glover E W N 1992 \PR {\bf D46} 1980 \\
Frixione S, Kunszt, Z and Signer A 1996 \NP {\bf B467} 399 \\
Catani S and Seymour M H 1996 \PL {\bf B278} 287

\bibitem{revsh}
ALEPH Collaboration, Decamp D \etal\ 1991 \PL {\bf B255} 623 and
1991 \PL {\bf B257} 479 \\
DELPHI Collaboration, Abreu P \etal\ 1992 \ZP {\bf C54} 55 \\
L3 Collaboration, Adriani O \etal\ 1992 \PL {\bf B284} 471 \\
OPAL Collaboration, Acton P D \etal\ 1992 \ZP {\bf C55} 1 \\
SLD Collaboration, Abe K \etal\ 1994 \PR {\bf D50} 5580 and
1995 \PR {\bf D52} 4240 

\bibitem{rsfit}
ALEPH Collaboration, Decamp D \etal\ 1992 \PL {\bf B284} 163 \\
DELPHI Collaboration, Abreu P \etal\ 1993 \ZP {\bf C59} 21 \\
OPAL Collaboration, Acton P D \etal\ 1993 \ZP {\bf C59} 1 \\
SLD Collaboration, Abe F \etal\ 1995 \PR {\bf D51} 962

\bibitem{rbw}
Webber B R 1995 {\it Proc. Int. Conf. on H.E.P. (Glasgow 1994)}
vol~1 eds P J Bussey and I G Knowles (Bristol: Inst. of Physics) p~213 
 
\bibitem{rAmega}
ALEPH Collaboration, Barate R \etal\ 1996 preprint CERN-PPE/96-186,
submitted to {\it Phys. Rep.}

\bibitem{rDtun2}
DELPHI Collaboration, Abreu P \etal\ 1996 \ZP {\bf C73} 11

\bibitem{rqfb}
ALEPH Collaboration, Buskulic D \etal\ 1996 \ZP {\bf C71} 357

\bibitem{rtunxi}
ALEPH Collaboration, Decamp D \etal\ 1992 \ZP {\bf C55} 209

\bibitem{rtun1}
OPAL Collaboration, Akrawy M Z \etal\ 1990 \ZP {\bf C47} 505 \\
L3 Collaboration, Adeva B \etal\ 1992 {\bf C55} 39 


\bibitem{rren}
Mueller A H 1993 {\it QCD 20 years later} vol~1 ed P M Zerwas and H A Kastrup
(Singapore: World Scientific) p~162

\bibitem{rope}
Wilson K 1969 \PR {\bf 179} 1499 \\
Wilson K and Kogut J 1974 {\it Phys. Rep.} {\bf 12} 75 

\bibitem{rpowid}
Bigi I I, Shifman M A, Uraltsev N G and Vainshtein A I 1994 \PR {\bf D50} 2234 

\bibitem{rpow}
Webber B R 1994 \PL {\bf B339} 148: see also \\
Manohar A V and Wise M B 1995 \PL {\bf B344} 407

\bibitem{rfine}
Dokshitzer Yu L and Webber B R 1995 \PL {\bf B352} 451 \\
Dokshitzer Yu L, Marchesini G and Webber B R 1995 preprint archive
hep-ph/9512336

\bibitem{runiv}
Akhoury R and Zakharov V I 1995 \PL {\bf B357} 646

\bibitem{rtube}
Feynman R P 1972 {\it Photon Hadron Interactions} (New York: W A Benjamin)

\bibitem{rphoton}
L3 Collaboration, Adriani O \etal\ 1992 \PL {\bf B292} 472 \\
ALEPH Collaboration, Buskulic D \etal\ 1993 \ZP {\bf C57} 17 \\
OPAL Collaboration, Acton P D \etal\ 1993 \ZP {\bf C58} 405 \\
DELPHI Collaboration, Abreu P \etal\ 1995 \ZP {\bf C69} 1 

\bibitem{rmike}
Seymour M H, 1994 \ZP {\bf C64} 445

\bibitem{rcart}
Cartwright S \etal\ {\it Photon radiaion from quarks} 1992
yellow report CERN 92-04

\bibitem{r1jgam}
OPAL Collaboration, Akers R \etal\ 1993 \ZP {\bf C67} 15 \\
ALEPH Collaboration, Buskulic D \etal\ 1996 \ZP {\bf C69} 365 

\bibitem{rwide}
OPAL Collaboration, 1996 submission to {\it ICHEP96 (Warsaw)} PA04-026

\bibitem{rstexp}
Azimov Ya I, Dokshitser Yu L, Khoze V A and Troyan S I 1985 \PL {\bf B165} 147

\bibitem{rdom}
Khoze V A and L\"onnblad L 1990 \PL {\bf B241} 123

\bibitem{rao}
Dokshitser Yu L, Khoze V A, Troyan S I and Mueller A H 1988
{\it Rev. Mod. Phys.} {\bf 60} 373

\bibitem{rlphd}
Azimov Ya A, Dokshitzer Yu L, Khoze V A and Troyan S I 1985 \ZP {\bf C27} 65

\bibitem{rlnc}
't Hooft G 1974 \NP {\bf B72} 461

\bibitem{rgood}
Andersson B, Gustafson G and Sj\"ogren C 1992 \NP {\bf B380} 391

\bibitem{rstrng1}
JADE Collaboration, Bartel W \etal 1981 \PL {\bf B101} 129 and
1985 \PL {\bf B157} 340 \\
TASSO Collaboration, Althoff M \etal\ 1985 \ZP {\bf C29} 29 \\
TPC/2$\gamma$ Collaboration, Aihara H \etal\ 1985 \ZP {\bf C28} 945

\bibitem{rsteff}
Andersson B, Gustafson G and Sj\"ostrand T 1980 \PL {\bf B94} 211

\bibitem{rstrng6}
OPAL Collaboration, Akrawy M Z \etal\ 1991 \PL {\bf B261} 334

\bibitem{rstrng2}
L3 Collaboration, Acciarri M \etal\ 1995 \PL {\bf B345} 74 

\bibitem{rstrng3}
MARK II Collaboration, Sheldon P D \etal\ 1986 \PRL {\bf 57} 1398 \\
TPC/2$\gamma$ Collaboration, Aihara H \etal\ 1986 \PRL {\bf 57} 945 \\
JADE Collaboration, Ould Saada F \etal\ 1988 \ZP {\bf C39} 1 

\bibitem{rstgam}
Azimov Ya I, Dokshitser Yu L, Troyan S I and Khoze V A 1986
{\it Sov. J. Phys.} {\bf 43} 95

\bibitem{rstrng4}
DELPHI Collaboration, Abreu P \etal\ 1996 \ZP {\bf C70} 179

\bibitem{rstrng8}
OPAL Collaboration, Akers R \etal\ 1995 \ZP {\bf C68} 531

\bibitem{rstrng5}
OPAL Collaboration, Acton P D \etal\ 1993 \ZP {\bf C58} 207

\bibitem{remmc}
Dokshitzer Yu L, Marchesini G and Oriani G 1992 \NP {\bf B387} 675

\bibitem{rappc}
Chmeisanni M 1992 {\it Annual Report} I.F.A.E. Universitat Aut\`onima de
Barccelona p~49

\bibitem{rstrng7}
L3 Collaboration, Acciarri M \etal\ 1995 \PL {\bf B353} 145

\bibitem{rhump}
Mueller A H 1981 {\it Proc. 1981 Int. Symp. on Lepton and Photon Ints. at
High Energies (Bonn 1981)} ed W Pfeil (Bonn: Phys. Inst. U. Bonn) p~689 \\
Dokshitzer, Yu L, Fadin V S and Khoze V A 1982 \PL {\bf B115} 242

\bibitem{rgaus}
Fong C P and Webber B R 1989 \PL {\bf B229} 289 and 1991 \NP {\bf B355} 210

\bibitem{rxiexp}
DELPHI Collaboration, Aarnio D \etal\ 1990 \PL {\bf B240} 271 \\
OPAL Collaboration, Akrawy M Z \etal\ 1990 \PL {\bf B247} 617

\bibitem{rmult2}
L3 Collaboration, Adeva B \etal\ 1991 \PL {\bf B259} 199

\bibitem{rxiexp2}
MARK-II Collaboration, Abrams G \etal\ 1990 \PRL {\bf 64} 1334

\bibitem{rscvioex1}
ALEPH Collaboration, Buskulic D \etal\ 1995 \PL {\bf B357} 487

\bibitem{rscvioex2}
DELPHI Collaboration, Abreu P \etal\ 1996 preprint CERN-PPE/96-185,
submitted to \PL {\bf B}

\bibitem{rcohq}
Boudinov E R, Chliapnikov P V and Uvarov V A 1993 \PL {\bf B309} 210

\bibitem{bib-datarev}
Lafferty G D, Reeves P I and Whalley M R 1995 \jpg {\bf 21} A1

\bibitem{rxi*cor}
Mueller A H 1983 \NP {\bf B213} 85 and {\it erratum} 1984 \NP {\bf B241} 141

\bibitem{rbreit}
ZEUS Collaboration, Derrick M \etal\ 1995 \ZP {\bf C67} 93 \\
H1 Collaboration, Aid S \etal\ 1996 \ZP {\bf C72} 573 

\bibitem{r2pcmea}
OPAL Collaboration, Acton P D \etal\ 1992 \PL {\bf B287} 401

\bibitem{r2pcpre}
Fong C P and Webber B R 1991 \NP {\bf B355} 54

\bibitem{r2pcorr}
Webber B R 1993 {\it Proc. XXVI Int. Conf. on H.E.P.} ed J R Sanford
(AIP Conference Proceedings: New York) p~878

\bibitem{rscvioth}
Nason P and Webber B R 1994 \NP {\bf B421} 473

\bibitem{rmmom}
Malaza E D and Webber B R 1986 \NP {\bf B267} 702

\bibitem{rmmlla}
Mueller A H 1981 \PL {\bf B104} 161 \\
Bassetto A, Ciafaloni M, Marchesini M and Mueller A H 1982 \NP {\bf B207} 189

\bibitem{rmlla}
Furmanski W, Petronzio R and Pokorski S 1979 \NP {\bf B155} 253

\bibitem{rmult}
Webber B R 1984 \PL {\bf B143} 501

\bibitem{bib-mult1}
ALEPH Collaboration, Decamp D \etal\ 1991 \PL {\bf B273} 181

\bibitem{bib-mult2}
DELPHI Collaboration, Abreu P \etal\ 1991 \ZP {\bf C52} 271

\bibitem{bib-mult3}
OPAL Collaboration, Acton P D \etal\ 1992 \ZP {\bf C53} 539 

\bibitem{rigk}
Knowles I G 1996 {\it Proc. of Int. Europhysics Conf. on H.E.P. (Brussels
1995)} eds J Lemonne \etal\ (World Scientific: Singapore) p~349

\bibitem{rfermi}
Fermi E 1950 {\it Prog. Theor. Phys.} {\bf 5} 550

\bibitem{rmdist}
DELPHI Collaboration, Abreu P \etal\ 1991 \ZP {\bf C50} 185

\bibitem{rkno}
Polyakov A M 1971 {\it Sov. Phys. JETP} {\bf 32} 296 and {\bf 33} 850 \\
Koba Z, Nielsen, H B and Olesen 1972 \NP {\bf B40} 317 \\
Golokhvastov A I 1979 {\it Sov. J. Nucl. Phys.} {\bf 30} 128

\bibitem{rnbd}
Carruthers P and Shih C C 1987 {\it Int. J. Mod. Phys.} {\bf A2} 1447

\bibitem{rlnd}
Carius S and Ingelman G 1990 \PL {\bf B252} 647

\bibitem{rmydist}
ALEPH Collaboration, Buskulic D \etal\ 1995 \ZP {\bf C69} 15

\bibitem{rmjdist}
DELPHI Collaboration, Abreu P \etal\ 1991 \ZP {\bf C52} 511

\bibitem{rQmult}
DELPHI Collaboration, Abreu P \etal\ 1995 \PL {\bf B347} 447 \\
OPAL Collaboration, Akers R \etal\ 1995 \PL {\bf B352} 176 \\
SLD Collaboration, Abe K \etal\ 1996 \PL {\bf B386} 475

\bibitem{rbruce}
Mark II Collaboration, Schumm B A \etal\ 1992 \PR {\bf D46} 453

\bibitem{rdead}
Dokshitzer Yu L, Khoze V A and Troyan S I 1991 \jpg {\bf 17} 1602

\bibitem{rbdif}
Schumm B A, Dokshitzer Yu L, Khoze V A and Koetke D S 1992
\PRL {\bf 69} 3025 \\
Petrov V A and Kiselev 1995 \ZP {\bf C 66} 453 \\
Dias de Deus J 1995 \PL {\bf B355} 539

\bibitem{rchrin}
Chrin J 1995 {\it Proc. Int. Conf. on H.E.P. (Glasgow 1994)} vol~2
eds P J Bussey and I G Knowles (Bristol: Inst. of Physics) p~893

\bibitem{rochs}
Ochs W 1995 {\it Proc. XXIVth Int. Symp. on Multiparticle Dynamics (Vietri
sul Mare 1994)} eds A Giovaninni \etal\ (Singapore: World Scientific) p~243

\bibitem{rprecon}
Amati D and Veneziano G 1979 \PL {\bf B83} 87 \\
Bassetto A, Ciafaloni M and Marchesini G 1980 \NP {\bf B163} 477 \\
Marchesini G, Trentadue L and Veneziano G 1981 \NP {\bf B181} 335

\bibitem{rfif}
Krzywicki A and Petersson B 1972 \PR {\bf D6} 924 \\
Finkelstein J and Peccei R D 1972 \PR {\bf D6} 2606 \\
Niedermayer F 1974 \NP {\bf B79} 355 \\
Casher A, Kogut J and Suskind L 1974 \PR {\bf D10} 732 \\
Hoyer P \etal\ 
1979 \NP {\bf B161} 349 \\
Ali A, Pietarinen E, Kramer G and Willrodt J 1980 \NP {\bf B93} 155

\bibitem{risajet}
Paige F and Protopopescu S 1986 {\it Super Collider Physics} ed D Soper
(Singapore: World Scientific) p~41

\bibitem{rjmb}
Bidulph P and Thompson G 1989 {\it Comp. Phys. comm} {\bf 54} 13

\bibitem{rasdep}
Sj\"ostrand T 1984 \ZP {\bf C26} 93 \\
Bengtsson M, Sj\"ostrand T and van Zijl M 1986 \PL {\bf B179} 164

\bibitem{rmont}
Montevay I 1979 \PL {\bf B84} 331

\bibitem{regge}
Collins P D B 1977 {\it Regge Theory and High Energy Physics} (Cambridge
University Press)

\bibitem{rbag}
Johnson K and Thorn C B 1976 \PR {\bf D13} 1934

\bibitem{rwilson}
Wilson K G 1974 \PR {\bf D10} 2445

\bibitem{ronia}
Appelquist T, Barnett R M and Lane K 1978 {\it Ann. Rev. Nucl. Part. Sci.}
{\bf 28} 387 \\ 
SESAM Collaboration, Gl\"assner \etal\ 1996 \PL {\bf B383} 98

\bibitem{rnambu}
Nambu Y 1970 Chicago preprint EFI 70-07

\bibitem{rsuper}
Nielsen H B and Olesen P 1973 \NP {\bf B61} 45

\bibitem{rartru1}
Artru X 1983 {\it Phys. Rep.} {\bf 97} 147

\bibitem{rgstr}
Andersson B and Gustafson G 1980 \ZP {\bf C3} 33

\bibitem{rgsyst}
Sj\"ostrand T 1984 \NP {\bf B248} 469

\bibitem{rmorris}
Morris D A 1987 \NP {\bf B288} 717

\bibitem{rQstr}
Chodos A and Thorn C B 1974 {\bf B72} 509 \\
Bardeen W A, Bars I, Hanson A J and Peccei R D 1976 \PR {\bf D13} 2364

\bibitem{rgsw}
Green M B, Schwarz J H and Witten E 1987 {\it Superstring theory} (Cambridge:
Cambridge University Press) 

\bibitem{rduality}
Ferrara S 1996 preprint archive hep-th/9610085

\bibitem{rpop}
Andersson B, Gustafson and Sj\"ostrand T 1985 {\it Physica Scripta}
{\bf 32} 574

\bibitem{rartru2}
Artru X and Mennessier G 1974 \NP {\bf B70} 93

\bibitem{rctech2}
Gottschalk T D and Morris D A 1987 \NP {\bf B288} 729

\bibitem{rucla}
Buchanan C D and Chun S B UCLA-HEP-95-2 and  1993 \PL {\bf B308} 153 \\
URL: http://www.physics.ucla.edu/{\char'176}chuns

\bibitem{rctech}
Gottschalk T D 1984 \NP {\bf B239} 325

\bibitem{rlufrg}
Andersson B, Gustafson G and Peterson C 1979 \ZP {\bf C1} 105

\bibitem{rbowl}
Bowler M G 1981 \ZP {\bf C11} 169

\bibitem{rlsff}
Andersson B, Gustafson G and S\"oderberg B \ZP 1983 {\bf C20} 317

\bibitem{rlsff2}
Bowler M G 1984 \ZP {\bf C22} 155

\bibitem{rschwin1}
Bohr H and Nielsen H B 1978 Niels Bohr Institute preprint NBI-HE-78-3 \\
Andersson B, Gustafson G and Sj\"ostrand 1980 \ZP {\bf C6} 235

\bibitem{rschwin2}
Heisenberg W and Euler H 1936 \ZP {\bf 98} 36 \\
Schwinger J 1951 \PR {\bf 82} 664 \\
Brezin E and Itzykson C 1970 \PR {\bf D2} 1191

\bibitem{rschwin3}
Casher A, Neuberger H and Nussinov 1979 \PR {\bf D20} 179

\bibitem{rbo}
Andersson B, 1992 \jpg {\bf 17} 1507

\bibitem{rvp}
Andersson B and Gustafson G 1982 Lund University preprint LU TP 82-5

\bibitem{rpicor}
Andersson B, Gustafson G and Samuelsson J 1994 \ZP {\bf C64} 653

\bibitem{rucla2}
Buchanan C D and Chun S B 1987 \PRL {\bf 59} 1997

\bibitem{rspcor}
Andersson B, Gustafson G and Ingelman G 1979 \PL {\bf B85} 417

\bibitem{rstbe}
Andersson B and Hofmann W 1986 \PL {\bf B169} 364 \\
Bowler M G 1987 \PL {\bf B185} 205 and {\bf 276} 237 \\
Bowler M G and Artru X 1992 \ZP {\bf C37} 293

\bibitem{rhanb}
Hanbury-Brown R and Twiss R G 1954 {\it Phil. Mag.} {\bf 45} 663 and
1956 {\it Nature} {\bf 177} 27 
Goldhaber G, Goldhaber S, Lee W and Pais A 1960 \PR {\bf 120} 300

\bibitem{rbegeom}
Gyulassy M, Kauffman S K and Wilson L W 1979 \PR {\bf C20} 2267

\bibitem{rberes1}
Bowler M G 1988 \ZP {\bf C39} 81

\bibitem{rberes2}
Bowler M G 1990 {\bf C46} 305

\bibitem{rbealg}
Sj\"ostrand T 1989 {\it Multiparticle Production} eds R Hwa \etal\
(Singapore: World scientific) p~237 \\
Zajc W A 1987 \PR {\bf D35} 3396

\bibitem{WWBEC}
L\"onblad L and Sj\"ostrand T 1995 \PL {\bf B351} 293

\bibitem{rdiq}
Anselmino M \etal\
1993 {\it Rev. Mod. Phys} {\bf 65} 1199

\bibitem{rstdiq}
Andersson B, Gustafson and Sj\"ostrand 1982 \NP {\bf B197} 45 \\
Meyer T 1982 \ZP {\bf C12} 77

\bibitem{rbpop}
Bowler M G 1981 Oxford preprint OUNP 76/81

\bibitem{rpop2}
Ed\'en P and Gustafson G 1996 preprint archive hep-ph/9606454,
to be published in \ZP {\bf C} \\
Ed\'en P 1996 preprint archive hep-ph/9610246

\bibitem{rbpfrg}
Bowler M G, Burrows P N and Saxon D H 1989 \PL {\bf B221} 415 \\
Bowler M G and Artru X 1991 \PL {\bf B256} 557

\bibitem{rpopex}
DELPHI Collboration, Abreu P \etal 1997 preprint CERN-PPE/97-027,
submitted to \PL {\bf B}

\bibitem{rctech1}
Gottschalk T D 1984 \NP {\bf B239} 349

\bibitem{rboot}
Hagedorn R 1965 \NC {\bf 3} 147 \\
Frautschi S C 1971 \PR {\bf D3} 2821

\bibitem{rmpm}
Berger E L and Fox G C 1973 \PL {\bf B47} 162 \\
Hamer C J and Peierls R F 1973 \PR {\bf D8} 1358

\bibitem{rwolf}
Wolfram S 1980 {\it Proc. XVth Recontre de Moriond (Les Arcs 1980)} vol~2
ed J. Tran Thanh Van (Gif-sur-Yvette: Editions Fronti\`ere) p~549 \\
Field R D and Wolfram S 1983 \NP {\bf B213} 65

\bibitem{rfast}
Bertolini S and Marchesini G 1982 \PL {\bf B117} 449 \\
Webber B R 1984 {\it Proc. XVth Int. Symp. on Multiparticle Dynamics (Lund
1984)} eds G Gustafson and C Peterson (Singapore: World scientific) p~627

\bibitem{rozi}
Okubo S 1963 \PL {\bf 5} 165 \\
Zweig G 1964 CERN reports TH-401,412 \\
Iizuka J, Okada K and Shito O 1966 {\it Rep. Prog. Phys.} {\bf 35} 1061

\bibitem{rlep2ww}
Kunszt Z \etal 1996 {\it Physics at LEP 2} vol~1 yellow report CERN 96-01 p~141

\bibitem{rjhreco}
Gustafson G and Hakkinen J 1994 \ZP {\bf C64} 659

\bibitem{rllreco}
L\"onnblad L 1996 \ZP {\bf C70} 107

\bibitem{rtsreco}
Sj\"ostrand T and Khoze V A 1994 \ZP {\bf C62} 281

\bibitem{rlamb}
Andersson B, Dahlqvist P and Gustafson G 1989 \ZP {\bf C44} 453

\bibitem{rllrap}
Bjorken J D, Brodsky S J and Lu H J 1992 \PL {\bf B286} 153 \\
L\"onnblad L 1996 \jpg {\bf 22} 947 \\
Friberg C, Gustafson G and Hakkinen J 1996 preprint archive hep-ph/9604347 \\
SLD Collaboration, Abe K \etal\ 1995 \PRL {\bf 76} 4890


\bibitem{bib-ALEPH}
ALEPH Collaboration, Buskulic D \etal\ 1990 \NIM {\bf A294} 121 and
1995 \NIM {\bf A360} 481

\bibitem{bib-DELPHI}
DELPHI Collaboration, Abreu P \etal\ 1996 \NIM {\bf A378} 57

\bibitem{bib-L3}
L3 Collaboration, Adeva B \etal\ 1990 \NIM {\bf A289} 35 and
Adriani O \etal\ 1993 {\it Phys. Rep.} {\bf 236} 1

\bibitem{bib-OPAL}
OPAL Collaboration, Ahmet K \etal\ 1990 \NIM {\bf A305} 275 \\
Allport P P \etal\ 1993 \NIM {\bf A324} 34 and 1994 \NIM {\bf A346} 476

\bibitem{bib-SLD}
Junk, T R 1995, PhD Thesis, SLAC-R-96-476

\bibitem{bib-xlw}
Lafferty G D and Wyatt T R 1995 \NIM {\bf A355} 541

\bibitem{bib-Boehrer}
B\"ohrer A 1996 preprint CERN-OPEN/96-021, submitted to {\it Phys. Rep.}

\bibitem{bib-Ocharged}
OPAL Collaboration, Akers R \etal\ 1994 \ZP {\bf C63} 181

\bibitem{rAxpi0}
ALEPH Collaboration, Barate R \etal\ 1996 preprint CERN-PPE/96-168,
submitted to \ZP {\bf C}

\bibitem{bib-DELpi0}
DELPHI Collaboration, Adam W \etal\ 1996 \ZP {\bf C69} 561

\bibitem{bib-L3pi0}
L3 Collaboration, Acciarri M \etal\ 1994 \PL {\bf B328} 223

\bibitem{bib-L3eta}
L3 Collaboration, Acciarri M \etal\ 1996 \PL {\bf B371} 126

\bibitem{bib-AVM}
ALEPH Collaboration, Buskulic D \etal\ 1995 \ZP {\bf C69} 379

\bibitem{bib-Dlight}
DELPHI Collaboration, Abreu P \etal\ 1994 \ZP {\bf C65} 587

\bibitem{bib-L3ometa}
L3 Collaboration, Acciarri M \etal\ 1996 preprint CERN-PPE/96-171,
to be published in \PL {\bf B}

\bibitem{bib-DEL96}
DELPHI Collaboration, Abreu P \etal\ 1996 preprint CERN-PPE/96-077,
submitted  to \ZP {\bf C}

\bibitem{bib-OVTM}
OPAL Collaboration, Akers R \etal\ 1995 \ZP {\bf C68} 1

\bibitem{bib-f2'}
DELPHI Collaboration, Abreu P \etal\ 1996 \PL {\bf B379} 309

\bibitem{bib-Dcharged}
DELPHI Collaboration, Abreu P \etal\ 1995 \NP {\bf B444} 3 

\bibitem{bib-AK0}
ALEPH Collaboration, Buskulic D \etal\ 1994 \ZP {\bf C64} 361

\bibitem{bib-OK0}
OPAL Collaboration, Akers R \etal\ 1995 \ZP {\bf C67} 389

\bibitem{bib-OK*}
OPAL Collaboration, Acton P \etal\ 1993 \PL {\bf B305} 407

\bibitem{bib-AD}
ALEPH Collaboration, Buskulic D \etal\ 1994 \ZP {\bf C62} 1 

\bibitem{bib-DELD}
DELPHI Collaboration, Abreu \etal\ 1993 \ZP {\bf C59} 533
{\it erratum} {\bf C65} 709

\bibitem{bib-Ocharm}
OPAL Collaboration, Alexander G \etal\ 1996 \ZP {\bf C72} 1

\bibitem{bib-DELB*}
DELPHI Collaboration, Abreu P \etal\ 1995 \ZP {\bf C68} 353

\bibitem{bib-DELB**}
DELPHI Collaboration, Abreu P \etal\ 1994 \PL {\bf B345} 598

\bibitem{bib-AJpsi}
ALEPH Collaboration, Buskulic D \etal\ 1992 \PL {\bf B295} 396

\bibitem{bib-DJpsi}
DELPHI Collaboration, Abreu P \etal 1994 \PL {\bf B341} 109

\bibitem{bib-LJpsi}
L3 Collaboration, Adriani O \etal\ 1993 \PL {\bf B317} 467

\bibitem{bib-OJpsi}
OPAL Collaboration, Alexander G \etal\ 1996 \ZP {\bf C70} 197

\bibitem{bib-Oupsilon}
OPAL Collaboration, Alexander G \etal\ 1996 \PL {\bf B370} 185

\bibitem{bib-DDELTA}
DELPHI Collaboration, Abreu P \etal\ 1995 \PL {\bf B361} 207

\bibitem{bib-ODELTA}
OPAL Collaboration, Alexander G \etal\ 1995 \PL {\bf B358} 162

\bibitem{bib-Dlam}
DELPHI Collaboration, Abreu P \etal\ 1993 \PL {\bf B318} 249

\bibitem{bib-OPSB96}
OPAL Collaboration, Alexander G \etal\ 1996 preprint CERN-PPE/96-099, 
to be published in \ZP {\bf C}

\bibitem{bib-OPsigma96}
OPAL Collaboration, Alexander G \etal\ 1996 preprint CERN-PPE/96-100, 
to be published in \ZP {\bf C}

\bibitem{bib-DELB95}
DELPHI Collaboration, Abreu P \etal\ 1995 \ZP {\bf C67} 543

\bibitem{bib-DELB96}
DELPHI Collaboration, Abreu P \etal\ 1996 \ZP {\bf C70} 371

\bibitem{rbj}
Suzuki M 1977 \PL {\bf B71} 139 \\
Bjorken J D 1978 \PR {\bf D17} 171 

\bibitem{rfunny}
Close F E \etal\
1993 \PL {\bf B319} 291

\bibitem{rgribov}
Gribov V N 1991 Lund preprint Lu-TP 91-7

\bibitem{rpom}
Pomeranchuk I and Smorodinsky Ya 1945 {\it J. Fiz, USSR} {\bf 9} 97

\bibitem{bib-ssHERA}
ZEUS Collaboration, Derrick M \etal\ 1995 \ZP {\bf C68} 29 \\
H1 Collaboration, Aid S \etal\ 1996 DESY-96-122 

\bibitem{bib-ssuppress}
Lafferty G D 1995 \PL {\bf B353} 541

\bibitem{SLDss}
SLD collaboration, Abe K \etal\ 1997 preprint SLAC-PUB-7395,
submitted to \PRL

\bibitem{rbstar}
ALEPH collaboration, Buskulic D \etal\ 1996 \ZP {\bf C69} 393

\bibitem{rbstar1}
L3 Collaboration, Acciarri M \etal\ 1995 \PL {\bf B345} 589

\bibitem{bib-Becattini}
Becattini F 1996 \ZP {\bf C69} 485

\bibitem{bib-Chliapnikov}
Chliapnikov P V and Uvarov V A, 1995 \PL {\bf B345} 313

\bibitem{rhof}
Hofmann W, 1987 {\it Ann. Rev. Nucl. Part. Sci.} {\bf 38} 279

\bibitem{bib-YiPei}
Yi-Jin Pei 1996 \ZP {\bf C72} 39

\bibitem{OPALD**}
OPAL Collaboration, Ackerstaff K \etal\ 1997 preprint CERN-PPE/97-035, 
submitted to  \ZP {\bf C}


\bibitem{rAupsilon}
ALEPH Collaboration, McNeil M 1996 submission to {\it ICHEP96 (Warsaw)} 
PA05-066

\bibitem{rjprev}
CLEO Collaboration, Balest R \etal\ 1995 \PR {\bf D52} 2661 and
submission to {\it ICHEP96 (Warsaw)} PA05-074

\bibitem{rqfrag}
Barger V, Cheung K and Keung W Y 1990 \PR {\bf D41} 1541 \\
Braaten E, Cheung K and Yuan T C 1993 \PR {\bf D48} 4230

\bibitem{rgfrag}
Hagiwara K, Martin A D and Stirling W S, 1991 {\bf B267} 527
{\it Erratum} 1993 {\bf B316} 631 \\
Braaten E and Yuan T C 1993 \PRL {\bf 71} 1673

\bibitem{rjprom}
DELPHI Collaboration, Abreu P \etal\ 1996 \ZP {\bf C69} 575 \\
OPAL Collaboration, Alexander G \etal\ 1996 \PL {\bf B384} 343

\bibitem{rcho}
Cho P 1996 \PL {\bf B368} 171 \\
Cheung K, Keung W Y and Yuan T C 1996 \PRL {\bf 76} 877

\bibitem{rgrad}
K\"uhn J H and Schneider H 1981 \ZP {\bf C11} 263 \\
Keung W Y 1981 \PR {\bf D23} 2072 \\
Abraham K J 1989 \ZP {\bf C44} 467

\bibitem{rcdf}
CDF Collaboration, Abe F \etal\ 1995 \PRL {\bf 75} 4358 and 1996
preprint FERMILAB-CONF-96-156

\bibitem{rcdf8}
Cho P and Leibovich A K 1996 \PR {\bf D53} 150, 6203

\bibitem{roctet}
Braaten E and Yuan T C 1994 \PR {\bf D50} 3176 \\
Braaten E, Fleming S and Yuan T C 1996 {\it Ann. Rev. Nucl. Part. Sci.} 
{\bf 46} 197 \\
Yuan F, Qiao C-F and Chao K-T 1997 preprint archive hep-ph/9703438

\bibitem{rocpot}
Schuler G A 1997 preprint archive hep-ph/9702230

\bibitem{rlatt}
Bodwin G T, Kim S and Sinclair D K 1996 \PRL {\bf 77} 2376

\bibitem{rjspn}
Cho P and Wise M B 1995 \PL {\bf B346} 129 \\
Braaten E and Chen Y-Q 1996 \PRL {\bf 76} 730 \\
Beneke M and Rothstein I Z, 1996 \PL {\bf B372} 157 \\
Baek S, Ko P, Lee J and Song H S 1997 preprint archive hep-ph/9701208

\bibitem{rjmc}
Ernstr\"om P and L\"onnblad L 1996 preprint archive hep-ph/9606472,
to be published in \ZP {\bf C} \\
Ernstr\"om P, L\"onnblad L and Vanttinen 1996 preprint archive hep-ph/9612408,
to be published in \ZP {\bf C}


\bibitem{rxi*}
Dokshitzer Yu L, Khoze V A and Troyan S I 1992 {\it J. Mod. Phys.} {\bf A7}
1875

\bibitem{rbrum}
Brummer N C 1995 \ZP {\bf C66} 367

\bibitem{rmasdep}
Dokshitzer Yu L, Khoze V A and Troyan S I 1991 \jpg {\bf 17} 1481
and 1992 \ZP {\bf C55} 107

\bibitem{rAxpi}
ALEPH Collaboration, Buskulic D \etal\ 1995 \ZP {\bf C66} 355

\bibitem{rassym}
Brodsky S J and Ma B-Q 1997 \PL {\bf B392} 452

\bibitem{rsea}
Martin A D, Roberts R G and Stirling W J 1994 \PR {\bf D50} 6734 \\
CTEQ Collaboration, Lai H L \etal 1995 \PR {\bf D51} 4763

\bibitem{repro}
Gribov V N and Lipatov L N 1971 \PL {\bf B37} 78

\bibitem{rlead}
De Angelis A, Cosmo G and Cossutti 1995 {\it Int. J. Mod. Phys.} {\bf C6}
585


\bibitem{rzqlo}
Bigi I, Dokshitzer Yu, K\"uhn J and Zerwas P 1986 \PL {\bf B181} 157

\bibitem{rzqnlo}
Mele B and Nason P 1991 \NP {\bf B361} 626 \\
Dokshitzer Yu L, Khoze V A and Troyan S I 1995 \PR {\bf D55} 89

\bibitem{rpet}
Peterson C, Schlatter D, Schmitt I and Zerwas P M 1983 \PR {\bf D27} 105

\bibitem{rQfComp}
Chrin J 1987 \ZP {\bf C36} 163 and 1988 {\it Annals. N.Y. Acad. Sci.}
{\bf 535} 131 

\bibitem{rQfA}
ALEPH Collaboration: Buskulic D \etal\ 1995 \PL {\bf B357} 699

\bibitem{rQfD}
DELPHI Collaboration, Podobrin O and Feindt M 1995 submission to
{\it EPS-HEP 95 (Brussels)} eps0560

\bibitem{rQfL}
L3 Collaboration, Adeva O 1991 \PL {\bf B261} 177

\bibitem{rQfO}
OPAL Collaboration, Alexander G \etal\ 1995 \PL {\bf B364} 93

\bibitem{rQfS}
Church E 1996 report SLAC-R-0495

\bibitem{rdimc}
Brodsky S J and Ferrar G R 1973 \PRL {\bf 31} 1153

\bibitem{rcol}
Collins P D B and Spiller T P 1985 \jpg {\bf 11} 1289

\bibitem{rkart}
Kartvelishvili V G, Likehoded A K and Petrov V A 1978 \PL {\bf B78} 615 \\
Kartvelishvili V G, Likehoded A K and Slabosnitski\u{\ii} 1983 {\it Sov. J.
Nucl. Phys.} {\bf 38} 952

\bibitem{rmor}
Morris D A 1989 \NP {\bf B313} 634

\bibitem{rshift}
K\" uhn J H \etal\ 1989 {Z Physics at LEP 1} vol~1 ed G Alterelli \etal\
yellow report CERN 89-08 p~269

\bibitem{rpod}
Podobrin O 1995 \NP {\bf B} ({\it Proc. Supp.}) {\bf C39} 373 

\bibitem{rbstar2}
OPAL Collaboration, Akers R \etal\ 1995 \ZP {\bf C66} 19


\bibitem{bib-poltheory}
Gustafson G and H\"akkinen J 1993 \PL {\bf B303} 350

\bibitem{rpoldis}
Saleev V A 1997 preprint archive hep-ph/9702370

\bibitem{bib-ApolLb}
ALEPH Collaboration, Buskulic D \etal\ 1996 \PL {\bf B365} 437

\bibitem{bib-LBmethod}
Bonvicini G and Randall L 1994 \PRL {\bf 73} 392

\bibitem{bib-ApolL}
ALEPH Collaboration, Buskulic D \etal\ 1996 \PL {\bf B374} 319

\bibitem{bib-sdmatrix}
Bourrely C, Leader E and Soffer J 1980 {\it Phys. Rep.} {\bf 59} 95

\bibitem{bib-statmodels}
Bigi I I Y 1977 \NC {\bf A41} 581

\bibitem{bib-donoghue}
Donoghue J F 1979 \PR {\bf D19} 2806

\bibitem{bib-augustin}
Augustin J E and Renard F M 1979 {\it Proc. of the LEP Summer Study} vol~1
yellow report CERN 79-01 p~185

\bibitem{bib-OPALB*} 
OPAL Collaboration, Ackerstaff K \etal\ 1996 preprint CERN-PPE/96-192,
to be published in \ZP {\bf C}

\bibitem{bib-OPALspin} 
OPAL Collaboration, Ackerstaff K \etal\ 1996 preprint CERN-PPE/97-005,
to be published in \ZP {\bf C}

\bibitem{D*CLEOHRS}
HRS Collaboration, Abachi S \etal\ 1987 \PL {\bf B199} 585 \\
CLEO Collaboration, Kubota Y \etal\ 1991 \PR {\bf D44} 593 \\
TPC-2$\gamma$ Collaboration, Aihara H \etal\ 1991 \PR {\bf D43} 29

\bibitem{bib-Anselmino1} 
Anselmino M, Kroll P and Pire B 1985 \ZP {\bf C29} 135 \\
Anselm A, Anselmino M, Murgia F and Ryskin M G 1994 {\it J. Exp. Th. Phys.}
{\bf 60} 496      


\bibitem{bib-Haywood}
Haywood S 1994 Rutherford Appleton Laboratory preprint RAL-94-074

\bibitem{bib-ABE}
ALEPH Collaboration, Buskulic D \etal\ 1992 \ZP {\bf C54} 75

\bibitem{bib-DBE}
DELPHI Collaboration, Abreu P \etal\ 1992 \PL {\bf B286} 201 and
1994 \ZP {\bf C63} 17

\bibitem{bib-OBE}
OPAL Collaboration, Acton P D \etal\ 1991 \PL {\bf B267} 143

\bibitem{bib-D3BE}
DELPHI Collaboration, Abreu P \etal\ 1995 \PL {\bf B355} 415

\bibitem{bib-DKBE}
DELPHI Collaboration, Abreu P \etal\ 1996 \PL {\bf B379} 330

\bibitem{bib-OBEmult}
OPAL Collaboration, Alexander G \etal\ 1996 \ZP {\bf C72} 389

\bibitem{bib-residBE}
Lafferty G D 1993 \ZP {\bf C60} 659

\bibitem{bib-OPALVM1}
OPAL Collaboration, Acton P \etal\ 1992 \ZP {\bf C56} 521

\bibitem{bib-DELWW}
DELPHI Collaboration, Abreu P \etal\ 1997 preprint CERN-PPE/97-030,
submitted to \PL {\bf B}

\bibitem{bib-OPALFD}
OPAL Collaboration, Alexander G \etal\ 1996 \PL {\bf B384} 377 


\bibitem{rbylow}
TPC/$2\gamma$ Collaboration, Aihara H \etal\ 1986 \PRL {\bf 57} 3140

\bibitem{rSbbcor}
SLD Collaboration, Abe K \etal\ 1995 preprint SLAC-PUB-95 6920 \\
Maruyama T 1996 {\it Proc. Int. Europhysics Conf. on H.E.P. (Brussels 1995)}
eds J Lemonne \etal\ (world Scientific: Singapore) p~330

\bibitem{rOlam2}
OPAL Collaboration, Acton P \etal\ 1993 \PL {\bf B305} 415

\bibitem{rbrev}
Bell K W, Foster B, Hart J C, Proudfoot J, Saxon D H and Woodworth P L
1982 Rutherford Appleton Laboratory preprint RL-82-011

\bibitem{rdirdiq}
Ekelin S, Fredriksson S, J\"andel M and Larsson T I 1984 \PR {\bf D30} 2310 

\bibitem{reco}
Das K P and Hwa R C 1977 \PL {\bf B77} 459

\bibitem{recobar}
Migneron R, Jones L M and Lassila K E 1982 \PR {\bf D26} 2235 \\
Eilam G and Zahir M S 1982 \PR {\bf D26} 2991

\bibitem{rtom}
de Grand T A and Miettinen H I 1981 \PR {\bf D23} 1227

\bibitem{rmark2}
MARK II Collaboration, de la Vaissiere C \etal\ 1985 \PRL {\bf 54} 2071

\bibitem{rb2b10}
TASSO Collaboration, Althoff M \etal\ 1984 \ZP {\bf C17} 5

\bibitem{rb2b30}
JADE Collaboration, Bartel W \etal\ 1981 \PL {\bf B104} 325 \\
ARGUS Collaboration, Albrecht H \etal\ 1989 \ZP {\bf C43} 45 

\bibitem{rtpc}
TPC$/2\gamma$ Collaboration, Aihara H \etal\ 1985 \PRL {\bf 55} 1047


\bibitem{bib-intermittency}
van Hove L 1989 {\it Modern Physics Letters} {\bf A4} 1867

\bibitem{bib-LEPinter}
OPAL Collaboration, Akrawy M Z \etal\ 1991 \PL {\bf B262} 351 \\
ALEPH Collaboration, Decamp D \etal\ 1992 \ZP {\bf C53} 21 

\bibitem{bib-SLCinter}
Murray W N, Frey R E and Ogren H O 1993 Oregon (Eugene) preprint OREXP 93-0016 


\bibitem{rqgrat}
Konishi K, Ukawa A and Veneziano G 1978 \PL {\bf B78} 243: see also \\
Brodsky S J and Gunion J 1976 \PRL {\bf 37} 402

\bibitem{rqgprop}
Shizuya K and Tye S-H H 1978 \PRL {\bf 41} 787 \\
Einhorn M and Weeks B G 1978 \NP {\bf B146} 445

\bibitem{rqgex1}
OPAL Collaboration, Alexander G \etal\ 1991 \PL {\bf B265} 462,
Acton P D \etal\ 1993 \ZP {\bf C58} 387 and 
Akers R \etal\ 1995 \ZP {\bf C68} 179 \\
ALEPH Collaboration, Buskulic D \etal\ 1995 \PL {\bf B346} 389 

\bibitem{rqgex4}
SLD Collaboration, quoted by Fuster J and Marti\`{\ii} 1996 {\it Proc.
Int. Europhysics Conf on HEP} ed J Lemonne \etal\ (Singapore:
World Scientific) p~319

\bibitem{rqgex2}
ALEPH Collaboration, Buskulic D \etal\ 1995 \PL {\bf B384} 353 \\
OPAL Collaboration, Alexander G \etal\ 1996 \ZP {\bf C69} 543

\bibitem{rqgex3}
Gary J W 1994 \PR {\bf D49} 4503 \\
OPAL Collaboration, Alexander G \etal\ 1996 \PL {\bf B388} 659 

\bibitem{rtopol}
ALEPH Collaboration, Barate R \etal\ 1997 preprint CERN-PPE/97-003,
submitted to \ZP {\bf C}

\bibitem{rqgcorr}
Gaffney J B and Mueller A H 1985 \NP {\bf B250} 109

\bibitem{rqgengy}
Dremin I M and Hwa R C 1994 \PL {\bf B324} 477 \\
Dremin I M and Nechitailo V A 1994 {\it Mod. Phys. Lett.} {\bf A9} 1471

\bibitem{rgeta}
Peterson C and Walsh T F 1980 \PL {\bf B91} 455

\bibitem{recopi}
Chang V and Hwa R C 1981 \PR {\bf D23} 728

\bibitem{recoeta}
Chang V, Eilam G and Hwa R C 1981 \PR {\bf D24} 1818

\bibitem{rDidqg}
DELPHI Collboration, Abreu P \etal\ 1996 preprint CERN-PPE/96-193,
submitted to \PL {\bf B}

\bibitem{rOidqg}
OPAL Collaboration 1996
submissions to {\it ICHEP96 (Warsaw)} PA02-011, PA02-016

\bibitem{rargeta}
ARGUS Collaboration, Albrecht H \etal\ 1990 \ZP {\bf C46} 15: see also
1989 \ZP {\bf C41} 557

\bibitem{rcrybeta}
Crystal Ball Collaboration, Bieler Ch \etal\ 1991 \ZP {\bf C49} 225

\bibitem{rbary}
DASP-II Collaboration, Albrecht H \etal\ 1981 \PL {\bf B102} 291 \\
CLEO Collaboration, Behrends S \etal\ 1985 \PR {\bf D31} 2161 \\
ARGUS Collaboration, Albrecht H \etal\ 1988 \ZP {\bf C39} 177 

\bibitem{rjadeta}
JADE Collaboration, Bartel W \etal\ 1985 \ZP {\bf C28} 343 and
1983 \PL {\bf B130} 545

\bibitem{rbarcl}
Field R D 1984 \PL {\bf B135} 203

\bibitem{rbarexp}
Scheck H 1989 \PL {\bf B224} 343

\bibitem{rdeutexp}
ARGUS Collaboration, Albrecht H \etal\ 1990 \PL {\bf B236} 102

\bibitem{rdeut}
Gustafson G and H\"akkinen J 1994 \ZP {\bf C61} 683

\end{thebibliography}
\end{document}